\documentclass{jfm}
\usepackage{graphicx}
\usepackage{epstopdf,epsfig}
\usepackage[colorlinks=true,allcolors=blue]{hyperref}
\usepackage{upgreek} 
\usepackage[export]{adjustbox} 
\usepackage{changes}
\usepackage{multirow}
\makeatletter
\@namedef{Changes@AuthorColor}{red}
\colorlet{Changes@Color}{red}
\makeatother

\usepackage{amsmath,mathtools}
\usepackage{amssymb}
\newcommand{\RR}{\mathit{Re}}
\newcommand{\We}{\mathit{We}}
\newcommand{\diff}{\mathrm{d}}
\newcommand{\f}{\frac}
\newcommand{\ov}{\overline}
\newcommand{\bs}{\boldsymbol}

\shorttitle{Bubble statistics from breaking-wave simulations}
\shortauthor{W. H. R. Chan, P. L. Johnson, P. Moin and J. Urzay}

\title{The turbulent bubble break-up cascade. Part 2. Numerical simulations of breaking waves}

\author{Wai Hong Ronald Chan\aff{1},
  Perry L. Johnson\aff{1,2},
  Parviz Moin\corresp{\email{moin@stanford.edu}\aff{1}}
  \and Javier Urzay\aff{1}}

\affiliation{
\aff{1}Center for Turbulence Research (CTR), Stanford University, Stanford, CA 94305, USA
\aff{2}The Henry Samueli School of Engineering, University of California, Irvine, Irvine, CA 92697, USA
}

\begin{document}

\maketitle

\begin{abstract}
Breaking waves generate a distribution of bubble sizes that evolves over time. Knowledge of how this distribution evolves is of practical importance for maritime and climate studies. The analytical framework developed in Part 1 examined how this evolution is governed by the bubble-mass flux from large to small bubble sizes, which depends on the rate of break-up events and the distribution of child bubble sizes. These statistics are measured in Part 2 as ensemble-averaged functions of time by simulating ensembles of breaking waves, and identifying and tracking individual bubbles and their break-up events. The break-up dynamics are seen to be statistically unsteady, and two intervals with distinct characteristics were identified. In the first interval, the dissipation rate and bubble-mass flux are quasi-steady, and the theoretical analysis of Part 1 is supported by all observed statistics, including the expected $-10/3$ power-law exponent for the super-Hinze-scale size distribution. Strong locality is observed in the corresponding bubble-mass flux, supporting the presence of a super-Hinze-scale break-up cascade. In the second interval, the dissipation rate decays, and the bubble-mass flux increases as small- and intermediate-sized bubbles become more populous. This flux remains strongly local with cascade-like behaviour, but the dominant power-law exponent for the size distribution increases to $-8/3$ as small bubbles are also depleted more quickly. This suggests the emergence of different physical mechanisms during different phases of the breaking-wave evolution, although size-local break-up remains a dominant theme. Parts 1 and 2 present an analytical toolkit for population balance analysis in two-phase flows.
\end{abstract}

\section{Introduction}\label{sec:intro}

Breaking waves in oceans generate bubbles of a wide range of sizes~(\citealp{Blanchard1,Medwin1,Johnson1,Trevorrow1,Melville1,Deane2,Vagle2,Deane1}; and others). A key driver in establishing this wide range of bubble sizes is the turbulence that emerges as these waves break~(\citealp{Kanwisher1,Kitaigorodskii3,Rapp1,Agrawal1,Melville2,Melville1,Terray1,Gemmrich1,Drazen1,Deane7,Deane6}; and others). A travelling water wave carries gravitational potential energy from the variation of its surface height, as well as kinetic energy in the coherent motion of the water beneath its surface. As these waves break, a significant proportion of these potential and kinetic energies is either expended in entraining air beneath the surface or converted to turbulent kinetic energy. This turbulent kinetic energy is then cascaded from large to small scales, establishing a wide range of characteristic scales of turbulent motion before being dissipated as thermal energy~\citep{Richardson1,Kolmogorov1,Onsager1}. The resulting turbulence breaks sufficiently large entrained air cavities up into bubbles of various sizes~\citep{Kolmogorov3,Hinze1}. The smaller the size of the bubbles produced by these break-up events, the slower their rise velocity, and the longer their residence time in the ocean~\citep{Garrettson1,Thorpe2,Thorpe1,Trevorrow1}. As they rise towards the ocean surface, these bubbles are known to scavenge surfactants and other microparticles. This delays their coalescence with the atmosphere and with one another, and stabilizes them for a longer period of time in the ocean~(\citealp{Fox1,Turner1,Blanchard2,Johnson2,Weber1,Johnson4,Stefan1,Takagi1,Czerski2}; and references therein). Some of these bubbles eventually burst at the surface to form film and jet drops~(\citealp{Blanchard1,Thorpe1,Deane1,Veron1}; and references therein). On the whole, then, these bubbles impart an enduring influence on various physical phenomena.

Knowledge of the distribution of bubble sizes informs the total interfacial area, as well as size-dependent effects. These are important to quantify bubble- and drop-mediated mass, momentum, and energy transport near the wave surface~(\citealp{Kanwisher1,Atkinson1,Thorpe2,Thorpe1,Thorpe3,Woolf1,Melville1,Wanninkhof1,Emerson1}; and references therein), the reflection and scattering of solar and other electromagnetic radiation by the water-wave surface and surrounding air--water interfaces~(\citealp{Stramski1,Zhang2,Zhang1,Stefan1,Terrill1,Reed1,Stramski3,Seitz1,Twardowski1,Crook1}; and references therein), as well as the generation, scattering, and propagation of acoustic waves beneath the water-wave surface~(\citealp{Strasberg1,Schulkin1,Schulkin2,Medwin1,Farmer1,Prosperetti2,Hall1, Melville1,Deane2,Deane5,Duineveld1,Vagle2,Farmer5,Stanic1,Czerski1}; and references therein). In order to rigorously quantify these effects, one needs to have a good grasp of the initial distribution of bubble sizes generated in the wave-breaking process, as well as how and why this distribution evolves while the resulting turbulence dissipates and the bubbles rise to the surface. 

A number of experiments have measured the bubble size distribution due to breaking waves generated under a variety of conditions. Among these, the experiments by~\citet{Deane1} and~\citet{Blenkinsopp1} were able to resolve bubbles of sizes spanning over two decades (from under 100~$\upmu$m to over 10~mm)---a formidable size range for a single measurement set-up. Crucially, these experiments could characterize the size distribution early in the wave-breaking process, as the optical methods employed are suitable for these high-void-fraction scenarios. While there are differences in the reported distributions due to the different wave parameters and measurement techniques, as well as the precise stages of wave breaking that were characterized, there appear to be two common themes: first, the distribution of bubble sizes has distinct scalings in different bubble-size subranges; second, the distribution evolves noticeably as the wave breaks. As alluded to in Part 1, these observations imply that different physical mechanisms are at play at different length and time-scales in the formation and dynamics of bubbles in these waves. Similar observations were made in the three-dimensional breaking-wave simulations of~\citet{Wang1} and~\citet{Deike1}, which draw heavily on the foundational two-dimensional simulations of~\citet{Chen1} and~\citet{Iafrati1,Iafrati2}. Numerical simulations have fewer limitations on the access to detailed spatial information than experimental studies, which is crucial for characterizing these physical mechanisms. However, the separation of scales inherent in these oceanic systems makes it prohibitively costly to generate large simulation ensembles that resolve the range of bubble sizes accessed in these seminal experiments. A large number of ensemble realizations are necessary to achieve statistical convergence for various bubble statistics in these statistically unsteady waves. As such, mechanisms governing bubble formation and dynamics have not all been straightforward to isolate and remain a subject of active research. A number of candidate mechanisms have been proposed for various bubble-size subranges and wave-breaking stages, including bubble break-up by turbulence~\citep{Kolmogorov3,Hinze1}, air entrainment and microbubble formation due to plunging jets and drops~\citep{Deane1,Kiger1,Chan3,Chan6,Mirjalili1}, buoyant degassing and dissolution~\citep{Garrett1}, and intermittency in the energy dissipation rate~\citep{Garrett1,Gemmrich2}, although this intermittency remains a topic of active investigation~\citep{Deane5,Deane6}. Note that microbubble formation due to plunging jets and drops may entail the direct generation of sub-Hinze-scale bubbles with sub-unity Weber numbers from larger air sheets and films, which may bypass the action of turbulent break-up~\citep{Deane1,Chan3,Chan6,Mirjalili1}. As will be discussed in \S~\ref{sec:ensemble}, the simulations of this work do not have numerical support for sub-Hinze-scale bubbles, and thus these bubbles are out of the scope of the current work.

Break-up by turbulence is often cited as the dominant mechanism for the generation of bubbles of intermediate sizes during the active wave-breaking phase, when air is entrained beneath the wave surface. The sizes of these fragmenting bubbles are typically comparable to or larger than the global Hinze scale, at which forces of capillary and inertial origins are approximately in balance~\citep{Kolmogorov3,Hinze1}. This characteristic length scale was introduced in Part 1 and is dimensionally about a millimetre in most terrestrial oceanic waves~\citep{Deane7}. Because these bubbles are smaller than the integral scales of the system, the large-scale flow geometry should not have a strong influence on the break-up dynamics. \citet{Garrett1} suggested that the break-up of these bubbles by turbulence occurs through a quasi-steady cascade of bubble mass from large to small bubble sizes. As discussed in the introduction of Part 1, prior studies have theorized and observed in various bubbly flows that this mechanism implies a $D^{-2/3}$ power-law scaling for the break-up frequency of bubbles of size $D$~\citep{Hinze1,MartinezBazan1,RodriguezRodriguez1,Chan5}, and a $D^{-10/3}$ power-law scaling for the corresponding bubble size distribution~\citep{Filippov1,Garrett1,Deane1,Wang1,Deike1,Chan5,Chan3}. The mathematical formalism in Part 1, which quantifies the average rate of bubble-mass transfer from large to small bubble sizes due to break-up events, further demonstrates that these power-law scalings are directly compatible with the notions of locality and self-similarity in the bubble-mass transfer process. This compatibility confirms key physical aspects of the bubble break-up cascade phenomenology, and provides a theoretical basis for the dimensional analysis of~\citet{Garrett1}.

This paper (Part 2) aims to determine the extent to which the theoretical results of Part 1 are supported by numerical simulations, by algorithmically embedding this analytical toolkit and directly measuring its key component metrics in the simulations. While previous breaking-wave simulations have observed a $-10/3$ power-law exponent in the bubble size distribution, this work seeks a more direct observation of the bubble break-up cascade. The bubble-mass transfer rate introduced in Part 1 achieves this objective, and is itself dependent on the size distribution, the break-up frequency, as well as the distribution of child bubble sizes. In Part 2, these bubble statistics are measured by averaging over ensembles of breaking-wave simulations. They were computed using novel post-processing algorithms that identify and track individual bubbles, and record the details of individual break-up events. The evolution of these ensemble-averaged statistics with time is studied, and the effect of time averaging is elucidated with reference to the time-averaged statistics of~\citet{Deane1}, \citet{Wang1}, and~\citet{Deike1}. The statistics discussed in Part 2 provide direct support for the existence of the quasi-steady bubble break-up cascade proposed by~\citet{Garrett1} and theoretically examined in Part 1, at least during the early wave-breaking stages. This support lends credence to the usage of this analytical toolkit to examine the break-up dynamics of turbulent two-phase flows in general, and provides an avenue to develop and validate break-up kernels typically used for population balance modelling.

The dynamics of bubble break-up appear to be distinct in the early and late wave-breaking stages, suggesting the emergence of distinct bubble generation and evolution mechanisms. The size distribution was observed to deviate from the aforementioned $D^{-10/3}$ power-law scaling by~\citet{Deane1}, \citet{Tavakolinejad1}, and~\citet{Masnadi1} late in the wave-breaking process. Prior breaking-wave simulations either did not investigate the evolution of the ensemble-averaged size distribution as a function of time, or did not conclusively recover these alternative scalings in their ensemble-averaged size distribution due ostensibly to a lack of statistical convergence in a time-resolved sense. The results of this work indicate that the size distribution and other bubble statistics are indeed strong functions of bubble size and time. This paper systematically identifies bubble-size subranges and times over which the $D^{-10/3}$ scaling is not fully recovered in the size distribution. The characteristics of the bubble-mass transfer rate are examined in these bubble-size subranges and times in order to explore candidate mechanisms for the formation and dynamics of bubbles under these conditions.

This paper is organized as follows. In \S~\ref{sec:bridging}, the key results of Part 1 are recapitulated and reformulated with the inclusion of volume averaging for the ensembles of breaking-wave simulations described in \S~\ref{sec:ensemble}. Features of the algorithms used to identify individual bubbles in these ensembles, and to detect break-up and coalescence events by tracking the lineages of these bubbles, are briefly discussed in \S~\ref{sec:algorithms}. The resulting bubble statistics are examined in \S~\ref{sec:statistics} in relation to the theoretical analysis of Part 1 and \S~\ref{sec:bridging}. Finally, conclusions are drawn in \S~\ref{sec:conclusions}.


\section{Analysis of the bubble break-up cascade for breaking-wave simulations}\label{sec:bridging}

Volume averaging ($\ov{\cdot}$) is commonly used in the computation of bubble statistics in both numerical and experimental studies of breaking waves. As such, this section recasts the essential expressions of the mathematical formulation for bubble statistics in Part 1 using volume averaging in order to facilitate the subsequent analysis of bubble break-up in breaking-wave simulations. References to \S~\ref{sec:algorithms} and \S~\ref{sec:statistics} are also made to indicate how these statistics are computed in the simulations and where they are reported, respectively. The reader is referred to Part 1 for a more detailed development and interpretation of the analysis tools summarized here, as well as references to relevant prior art.

Volume averaging may be interpreted as an averaging procedure over the small, localized regions in which the turbulent energy and bubble-mass cascades are postulated to coexist, as reviewed in \S~2 of Part 1. The volume averaging procedure is further discussed in appendix~\ref{app:averaging}. The order of magnitude of the correspondingly averaged dissipation rate $\ov{\varepsilon}$ may be estimated in a global sense by $u_L^3/L$, where $L$ is the wavelength of the dominant wave, $u_L = (gL)^{1/2}/(2\upi)^{1/2}$ is the corresponding wave phase velocity, and $g$ is the magnitude of standard gravity. This estimate for the characteristic dissipation rate is addressed in more detail in appendix~\ref{app:dissipation}. It corresponds to the following dimensional expressions for the global Kolmogorov and Hinze scales
\begin{gather}
L_\mathrm{K} \sim \left(\f{\mu_l}{\rho_l}\right)^{3/4}\left(\ov{\varepsilon}\right)^{-1/4},\label{eqn:kolmodim}\\
L_\mathrm{H} \sim \left(\f{\sigma}{\rho_l}\right)^{3/5}\left(\ov{\varepsilon}\right)^{-2/5},\label{eqn:hinzedim}
\end{gather}
where $\rho_l$ and $\mu_l$ denote the density and dynamic viscosity of the liquid phase, respectively, and $\sigma$ denotes the air--water surface tension coefficient.

\subsection{The bubble-mass transfer flux}\label{sec:bridging_pbe}

The population balance equation describing the time evolution of the volume-weighted size distribution, $fD^3$, was discussed in \S~3.2 of Part 1. Volume averaging the phase-space-based equation (3.4) in Part 1 yields
\begin{equation}
\f{\partial \left[\ov{f}(D;t) D^3 \right]}{\partial t} + \f{\partial \left[\ov{v_D f}(D;t) D^3 \right]}{\partial D} = \ov{H}(D;t).
\label{eqn:pbe}
\end{equation}
Correspondingly, volume averaging the kernel-based equation (3.7) in Part 1 yields
\begin{equation}
\f{\partial \left[\ov{f}(D;t) D^3 \right]}{\partial t} = \ov{T_b}(D;t) + \ov{T_c}(D;t).
\label{eqn:pbe_kernel}
\end{equation}
As with the derivation of (3.10) in Part 1, these two equations may be compared and reduced by assuming size-local break-up dominates in an intermediate subrange of bubble sizes to yield a simplified evolution equation for $\ov{f}D^3$
\begin{equation}
\f{\partial \left[\ov{f}(D;t) D^3 \right]}{\partial t} = -\f{\partial \left[\ov{v_D f}(D;t) D^3 \right]}{\partial D} = \ov{T_b}(D;t).
\label{eqn:pbe_reduced}
\end{equation}
The size distribution, $\ov{f}$, is computed using the identification algorithm in \S~\ref{sec:ident}, and is further discussed in \S~\ref{sec:sizedistevol}. Note that the atmosphere above the wave is treated as a large gaseous reservoir in this work. As such, entrainment events are not included in the tally of break-up events, and degassing events are not included in the tally of coalescence events. The local break-up flux $\ov{W_b} = -\ov{v_D f} D^3$ may be interpreted as the ensemble-averaged and volume-averaged rate of bubble-mass transfer from all bubble sizes larger than $D$ to all bubble sizes smaller than $D$. One may derive an expression for the analogous flux arising from $\ov{T_b}$, which is obtained by volume averaging (3.11) in Part 1 as follows
\begin{equation}
\ov{T_b}(D;t) = \int_D^\infty \diff D_p \: \check{q}_b\left(D;t|D_p\right) \ov{g_b f}\left(D_p;t\right) D^3 - \ov{g_b f}(D;t) D^3.
\label{eqn:Tb}
\end{equation}
The ensemble-averaged and volume-averaged differential break-up rate $\ov{g_b f}(D_p;t)$ is the expected number of break-up events per unit time, unit domain volume, and unit size for bubbles of size $D_p$ at time $t$, including all events throughout the characteristic wave volume $L^3$ of each ensemble realization. Then, $\check{q}_b\left(D_c;t|D_p\right)$ describes the probability distribution of sizes of children bubbles in these events. Note that each relevant break-up event is weighted equally within each ensemble realization in the computation of $\check{q}_b$. As such, $\check{q}_b$ satisfies the same constraints as the analogous distribution $q_b$ in Part 1. The quantities $\ov{g_b f}$ and $\check{q}_b$ are both computed using the tracking algorithm in \S~\ref{sec:tracking}, and are further discussed in \S~\ref{sec:trac-gbfe} and \S~\ref{sec:trac-qb}, respectively. The corresponding bubble-mass transfer flux, $\ov{W_b}$, may then be expressed as
\begin{equation}
\ov{W_b}(D;t) = \int_0^D \diff D_c \: D_c^3 \int_D^\infty \diff D_p \: \check{q}_b \left(D_c;t|D_p\right) \ov{g_b f}(D_p;t).
\end{equation}
$\ov{W_b}$ may also be directly computed using the tracking algorithm in \S~\ref{sec:tracking}, and is further discussed in \S~\ref{sec:trac-wb}. In particular, its variations with bubble size and time are discussed in \S~\ref{sec:trac-wb-evolution}. In a system where entrainment at large sizes and break-up towards smaller sizes are the dominant physical mechanisms present, the break-up flux $\ov{W_b}$ may be used as a proxy for the entrainment flux, as suggested in figure 9 of Part 1. Note that a similar procedure may be used to derive the bubble coalescence flux $\ov{W_c}$ from an appropriate model coalescence kernel $\ov{T_c}$.

\subsection{Assessing locality}\label{sec:bridging_locality}

The expressions (3.15) and (3.16) derived in Part 1 to quantify infrared and ultraviolet locality may be rewritten as 
\begin{gather}
\ov{I_p}(D_p;t|D) = \int_0^D \diff D_c \: D_c^3 \check{q}_b \left(D_c;t|D_p\right) \ov{g_b f}\left(D_p;t\right),\\
\ov{I_c}(D_c;t|D) = \int_D^\infty \diff D_p \: D_c^3 \check{q}_b \left(D_c;t|D_p\right) \ov{g_b f}\left(D_p;t\right).
\end{gather}
These quantities may also be directly computed using the tracking algorithm in \S~\ref{sec:tracking}, and are further discussed in \S~\ref{sec:trac-wb-locality}.

In this work, the infrared and ultraviolet locality of $\ov{W_b}$ are investigated using ensembles of breaking-wave simulations in two ways. First, the individual scalings of $\check{q}_b\left(D_c;t|D_p\right)$ and $\ov{g_b f}\left(D_p;t\right)$ with $D_c$ and $D_p$ are computed and compared against the scalings derived in Part 1. Second, $\ov{I_p}$ and $\ov{I_c}$ are directly computed by summing the mass transfers from all relevant break-up events in the simulations, as outlined in appendix B of Part 1. The decay rates of $\ov{I_p}(D_p;t|D)$ and $\ov{I_c}(D_c;t|D)$ with increasing $D_p$ and decreasing $D_c$, respectively, are then examined. Before these statistics are discussed in \S~\ref{sec:statistics}, the breaking-wave simulation ensembles are first described in \S~\ref{sec:ensemble}, and the algorithms used to obtain these statistics are introduced in \S~\ref{sec:algorithms}.

\section{The breaking-wave simulation ensembles}\label{sec:ensemble}

\subsection{Flow solver}\label{sec:solver}

In order to investigate the evolution of the bubble size distribution $\ov{f}$, as well as the other bubble statistics introduced in \S~\ref{sec:bridging}, ensembles of numerical simulations of breaking waves were generated using an unstructured, collocated, node-centred, and unsplit geometric volume-of-fluid-based (VoF) flow solver developed for incompressible and immiscible two-phase flows~\citep{Kim2,Ham2,Bravo4}. Among other flow variables, the solver tracks the spatially discretized volume fraction of one of the two phases $\phi$, noting that the local sum of the volume fractions of the two phases is 1 by definition. Without loss of generality, assume that the solver is used to simulate a liquid--gas mixture, and $\phi$ denotes the local volume fraction of the liquid phase. The mixture is assumed to be non-reacting and electrically uncharged, and no mass transfer takes place between the phases. In each interfacial computational median dual cell, the interface is represented using the conventional piecewise-linear interface calculation (PLIC) scheme, while the corresponding interface normal vector is computed via a reconstructed distance function~\citep{Cummins1}, and the corresponding interface curvature is estimated using the second-order direct front curvature method~\citep{Herrmann1}. Note that the PLIC scheme is known to minimize jetsam and flotsam, and thus suppresses spurious bubble break-up, compared to the traditional simple line interface calculation (SLIC) scheme with which these structures are commonly associated~\citep{Scardovelli1}. (See also a related discussion in \S~\ref{sec:ident}.) The density and viscosity of the fluid in these interfacial cells are defined as
\begin{gather}
\rho = \phi \rho_l + (1-\phi) \rho_g,\\
\mu = \phi \mu_l + (1-\phi) \mu_g,
\end{gather}
where $\rho_g$ and $\mu_g$ denote the density and dynamic viscosity of the gaseous phase, respectively, and $0 < \phi < 1$. Mass and momentum are consistently advected at the interface~\citep{Mirjalili3} using a variant of the non-intersecting flux polyhedron advection algorithm~\citep{Ivey1} to maintain numerical stability for large-density-ratio flows, while the viscous terms in the momentum equation are implicitly time-advanced with the second-order Crank-Nicolson scheme. The fractional-step method is used in order to include a pressure-projection step to maintain a divergence-free velocity field. Within this method, the surface tension force is treated using a balanced-force algorithm~\citep{Francois1,Herrmann1} involving the usual continuum surface force representation to minimize spurious currents of capillary origin. Surface tension gradients are assumed to be negligible. The solver has been used to simulate a number of liquid jet break-up problems involving primary atomization, including laminar, transitional, and turbulent jets in quiescent gas~\citep{Bravo1,Bravo2,Bravo4,Bravo5}, jets in cross flows~\citep{Bravo3}, as well as jets from swirling injectors~\citep{Kim2,Ham2}.

\subsection{Description of set-up of ensembles}\label{sec:setup}

\begin{table}
  \begin{center}
  \begin{tabular}{ccc}
      Quantity & Dimensional notation & Non-dimensional notation \\[3pt]
       Lengths/co-ordinates & $\hat{x},\hat{y},\hat{z},\bs{\hat{x}},\hat{\eta}$ & $x,y,z,\bs{x},\eta = \{\hat{x},\hat{y},\hat{z},\bs{\hat{x}},\hat{\eta}\}/L$ \\
       Bubble size & $D$ & $D/L$ \\
       Gradient operator & $\hat{\nabla}$ & $\nabla = L\hat{\nabla}$ \\
       Velocity components & $\hat{u},\hat{v},\hat{w}$ & $u,v,w = \{\hat{u},\hat{v},\hat{w}\}/\sqrt{(gL)/(2\upi)} = \{\hat{u},\hat{v},\hat{w}\}/u_L$ \\
       Time & $\hat{t}$ & $t = \hat{t}/\sqrt{L/g}$ \\
       Energies & $\hat{E_k},\hat{E_p},\hat{E_s},\hat{E_t}$ & $E_k,E_p,E_s,E_t$ \\
  \end{tabular}
  \caption{Characteristic scales used for non-dimensionalization. The characteristic time-scale was selected to be equal to the one used by~\citet{Wang1}. With this time-scale, the first wave surface impact after the wave overturns occurs shortly after $t = 1$, while a single wave period corresponds to 2.5 characteristic times.}
  \label{tab:nondim}
  \end{center}
\end{table}

In this work, a baseline ensemble of 30 numerical simulations of breaking third-order Stokes water waves in air was generated. Specifically, the density and viscosity ratios of the two phases are equal to those of an air--water mixture. These waves have the dimensionless integral-scale parameters $\We_L = 1.6 \times 10^3$ and $\RR_L = 1.8 \times 10^5$, which match those of a 27-cm-long water wave at atmospheric conditions, and are similar to those selected by~\citet{Wang1}. Here, $\RR_L$ is the integral-scale Reynolds number $\RR_L = \rho_l u_L L / \mu_l$, and $\We_L$ is the integral-scale Weber number $\We_L = \rho_l u_L^2 L / \sigma$. The initial conditions employed are similar to those adopted by~\citet{Chen1},~\citet{Iafrati1,Iafrati2}, and~\citet{Wang1}: the initial dimensionless wave surface height $\eta$ was initialized in terms of the non-dimensional streamwise co-ordinate $x$ using
\begin{equation}
\eta(x,t=0) = \frac{1}{2\upi} \left[ S_1 \cos(2\upi x) + \f{1}{2} S_1^2 \cos(4\upi x) + \f{3}{8} S_1^3 \cos(6\upi x) \right],
\end{equation}
where $S_1=a_1k_1=0.55$ is the slope of the fundamental wave component, $a_1$ and $k_1 = 2\upi/L$ are respectively the corresponding dimensional amplitude and wavenumber, and the wave propagates in the $x$-direction. Note that the lengths in the expression for $\eta$ have been non-dimensionalized by the wavelength of the fundamental wave component, $L$. A list of characteristic scales used for non-dimensionalization is provided in Table~\ref{tab:nondim}.

\begin{figure}
  \centerline{
(a)
\includegraphics[width=0.425\linewidth,valign=t]{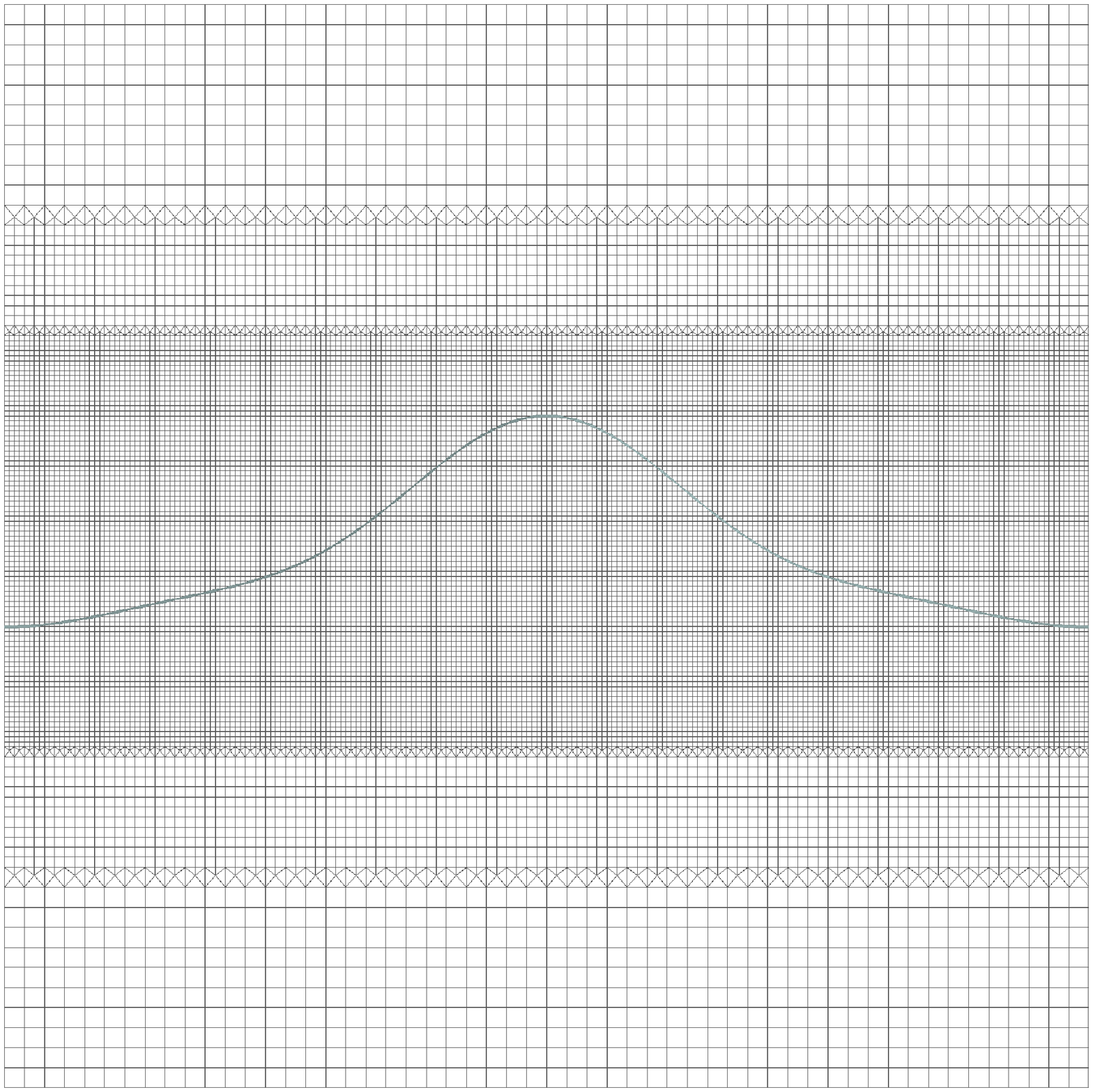}
\quad
(b)
\includegraphics[width=0.425\linewidth,valign=t]{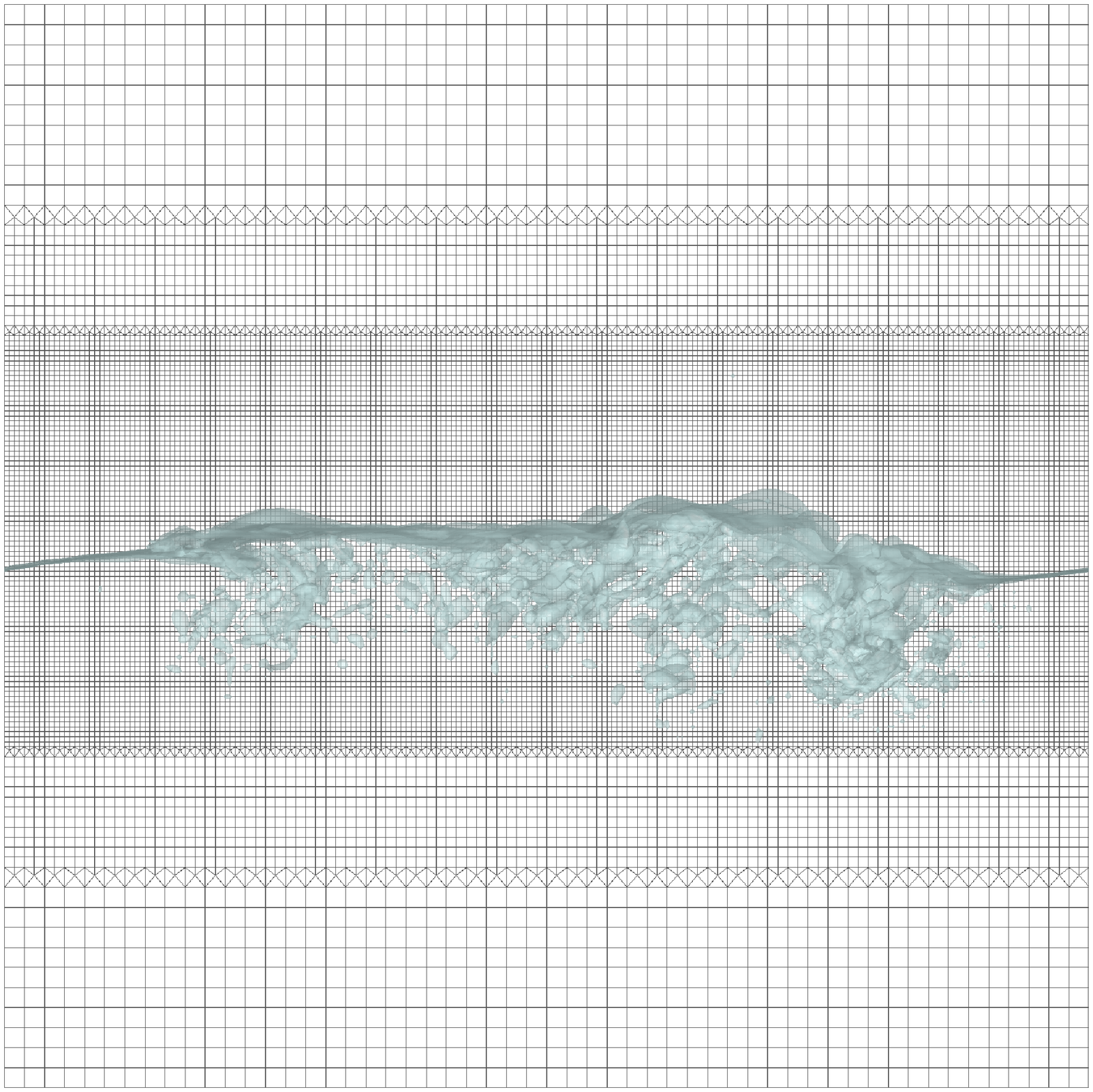}
}
  \caption{Snapshots of the spanwise cross-section of the $\phi=0.5$ isosurface of one of the baseline ensemble realizations corresponding to (a) the initial waveform ($t = 0$), and (b) the wave sometime after breaking has occurred ($t=4.03$). The grey lines in the background depict the local mesh configuration. The nodal mesh volumes in the uniform-resolution regions are isotropic.}
\label{fig:mesh}
\end{figure}

In order to generate an ensemble of statistically independent but similar realizations, the free surface was further perturbed by a set of random displacements $\Delta \eta$ smaller than the local grid spacing, such that every interfacial mesh node with the same $z$ (spanwise) co-ordinate within the same realization has the same $\Delta \eta$. While the shift $\Delta \eta$ is not explicitly resolved by the mesh, it perturbs the volume fraction in these interfacial nodes and provides an implicit disturbance to the original interface. This choice of perturbation preserves the modal content of the wave profile in the streamwise direction since the mesh employed in this work is Cartesian. In other words, the initial conditions are effectively two-dimensional except for a perturbation of spanwise modes. As in the aforecited works, the air above the wave surface was initialized at rest, while the water below the surface was initialized with the following dimensionless velocity field~\citep{Iafrati1,Iafrati2}
\begin{gather}
u(x,y,t=0) = S_1\sqrt{1 + S_1^2}\cos(2\upi x)\exp(2\upi y),\\
v(x,y,t=0) = S_1\sqrt{1 + S_1^2}\sin(2\upi x)\exp(2\upi y),
\end{gather}
where $y$ is the non-dimensional co-ordinate antiparallel to gravity. Periodic boundary conditions were employed in the $x$- and $z$-directions, while free-slip boundary conditions were employed on the two remaining boundary faces of the computational domain, which is a cube of length $L$. In each of the realizations in this ensemble, the computational mesh consists of about 4.2 million mesh nodes with a non-dimensional minimum grid spacing of $1/216$. This is equivalent to a dimensional grid spacing of 1.25 mm for a water wave in air at atmospheric conditions with the aforementioned dimensionless parameters. To put this in context, the dimensional Hinze scale~\eqref{eqn:hinzedim} takes a value on the order of $3\text{ mm}$. As evidenced in figures~\ref{fig:energy}, \ref{fig:bsd_compare_powerlaw}, \ref{fig:bsd_timeavg}, and \ref{fig:bsd_powerlaw_premult_time}, mesh insensitivity is observed in the energetics and bubble statistics at this mesh resolution. For the latter, this insensitivity is observed over a subrange of {\textit{super}}-Hinze-scale bubble sizes where turbulent break-up is expected to be dominant. (See also the description of a more resolved ensemble used in this mesh sensitivity study in the ensuing paragraph.) This resolution was selected to permit more ensemble realizations and enable statistical convergence in a time-resolved sense. The mesh is non-uniform and is finer closer to the central region of the domain, such that a large majority of the generated bubbles are resolved using the minimum grid spacing. Snapshots of the initial and post-breaking waveforms, with the computational mesh overlaid, are illustrated for one of the realizations in figure~\ref{fig:mesh}. The non-dimensional time-step adopted is $\Delta t = 6.0 \times 10^{-5}$, leading to a maximum Courant number of about 0.1 throughout the computational domain in the course of the simulations. The relatively small Courant number reduces the shape mismatch between the streak tube and the numerical flux polyhedron in the VoF advection scheme~\citep{Ivey1}, which is important to manage in the case of inadequately resolved mixed-phase regions. The evolution of the waveform for one of these realizations is depicted from two different viewpoints in figure~\ref{fig:wave} to illustrate the interfacial features generated as the wave breaks. 

Besides the baseline case, an ensemble of 3 numerical simulations with the same wave parameters and a higher mesh resolution was also generated for a mesh convergence study of the specifically desired quantities studied in this work, such as the bubble size distribution. In this ensemble, the computational mesh consists of about 32 million mesh nodes with a non-dimensional minimum grid spacing of $1/432$ (dimensionally equivalent to 0.63 mm) and a non-dimensional time-step of $1.2 \times 10^{-5}$. A more detailed rendering of one of the realizations from this ensemble may be found in the video discussed by \citet{Chan6}. Note that even with this higher spatial resolution, the Kolmogorov length scale~\eqref{eqn:kolmodim}, $L_\mathrm{K} \sim 30\text{ }\upmu\text{m}$, remains inaccessible like in many previous breaking-wave simulations. In fact, it may be shown that the smallest scales of turbulent motion near a phase interface could require sub-Kolmogorov resolution~\citep{Dodd3} that is rarely attained. As remarked in the introduction, an ensemble of direct numerical simulations of breaking waves that resolves these dynamics with converged statistics remains a formidable undertaking due to the inherent scale separation in these oceanic systems, and in energetic multiphase flows in general. In view of these limitations, only intermediate-scale quantities where sub-Hinze-scale dynamics have a limited influence are discussed in this work, including the bubble statistics introduced in \S~\ref{sec:bridging}. In other words, the formation of sub-Hinze-scale bubbles, which may be due to mechanisms distinct from a turbulent break-up cascade, is beyond the scope of this work.

\subsection{Time evolution of the breaking wave}\label{sec:waveevol}

\begin{figure}
  \centerline{
(a)
\includegraphics[trim={0pt 20pt 20pt 0pt},clip=true,width=0.425\linewidth,valign=t]{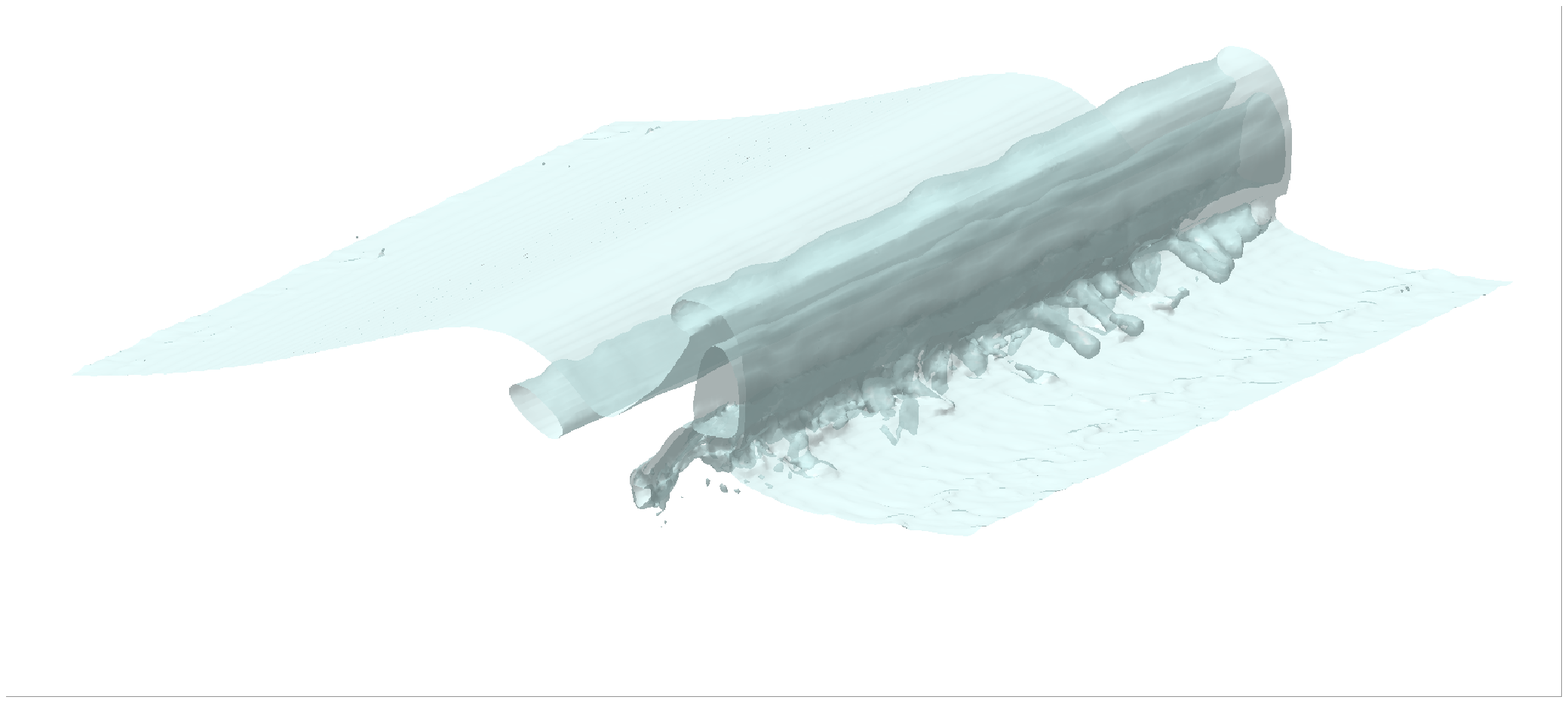}
\quad
(b)
\includegraphics[trim={0pt 20pt 20pt 0pt},clip=true,width=0.425\linewidth,valign=t]{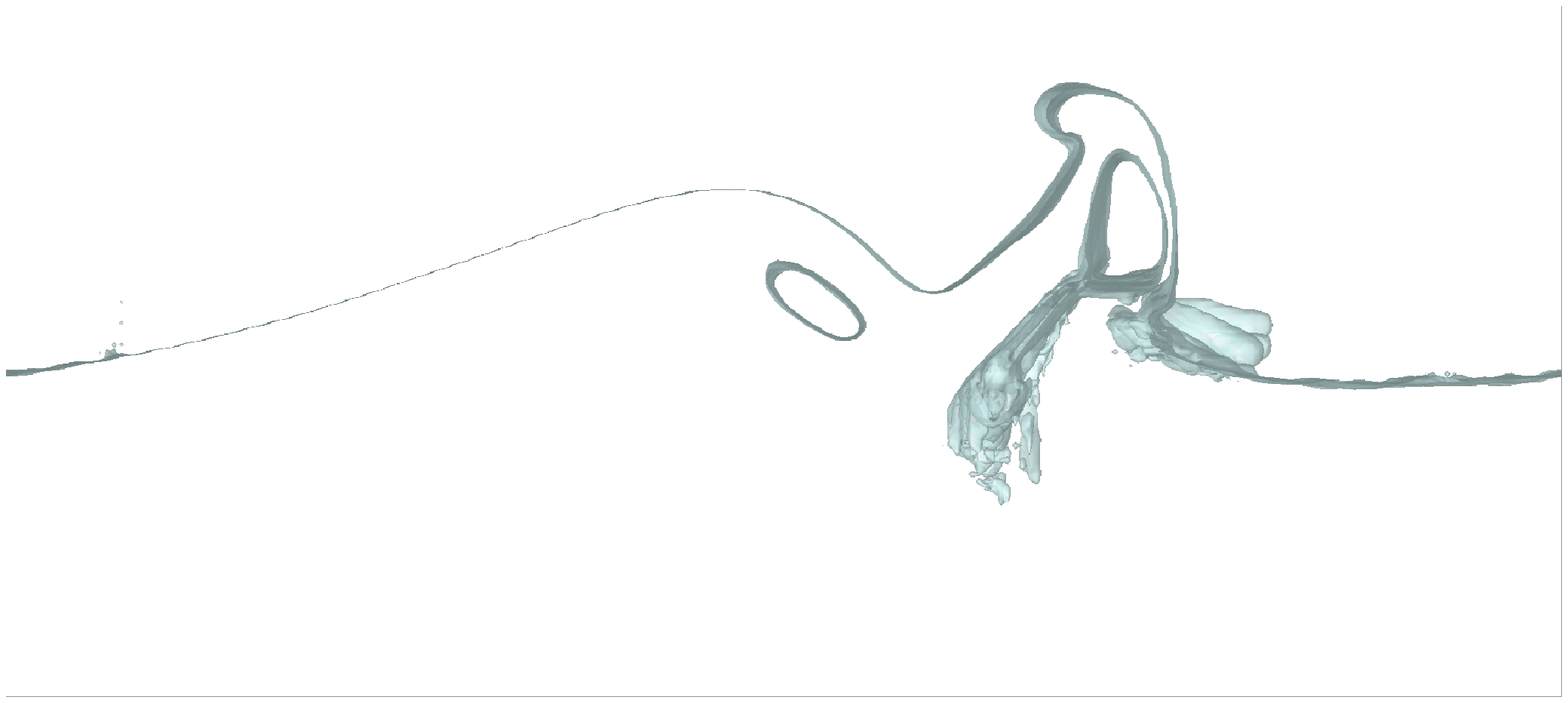}
}
  \centerline{
(c)
\includegraphics[trim={0pt 20pt 20pt 0pt},clip=true,width=0.425\linewidth,valign=t]{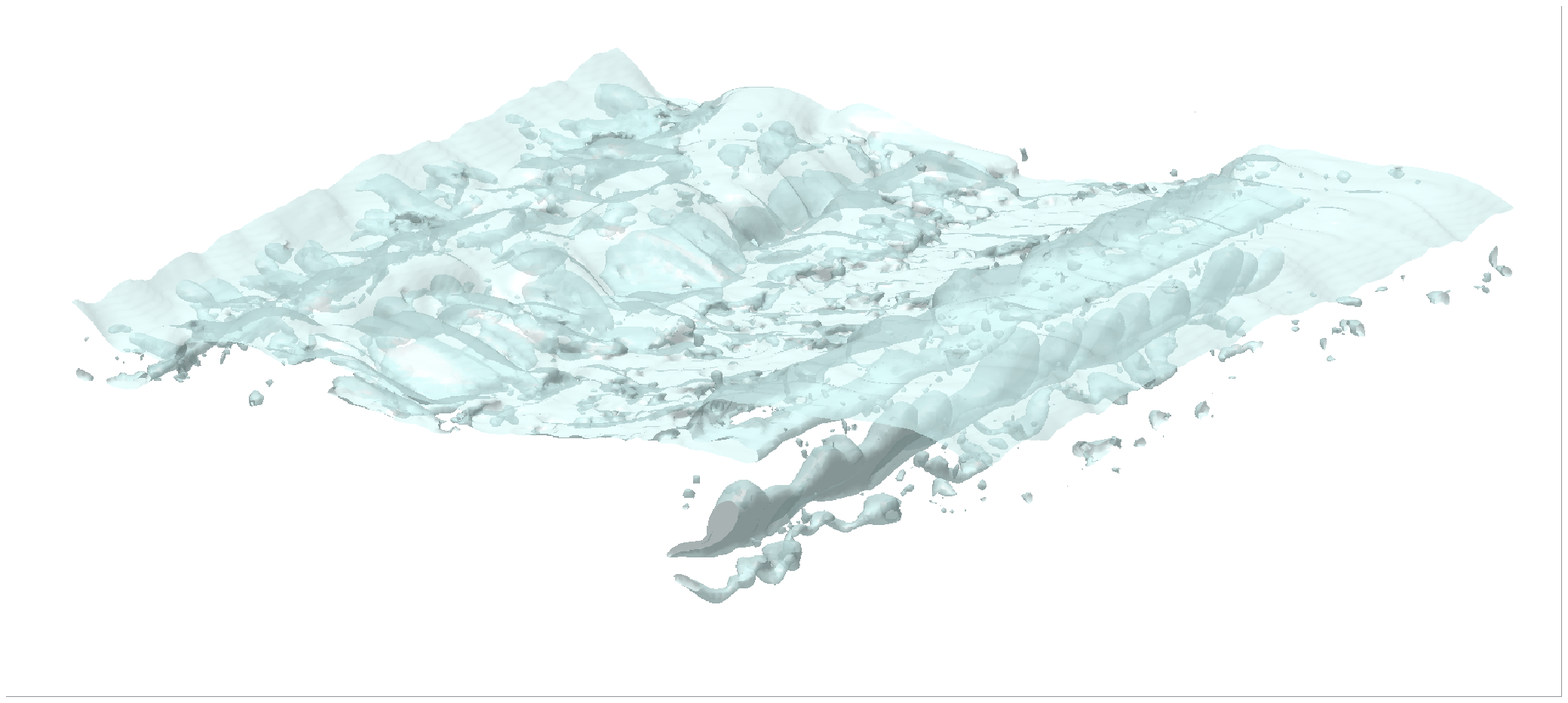}
\quad
(d)
\includegraphics[trim={0pt 20pt 20pt 0pt},clip=true,width=0.425\linewidth,valign=t]{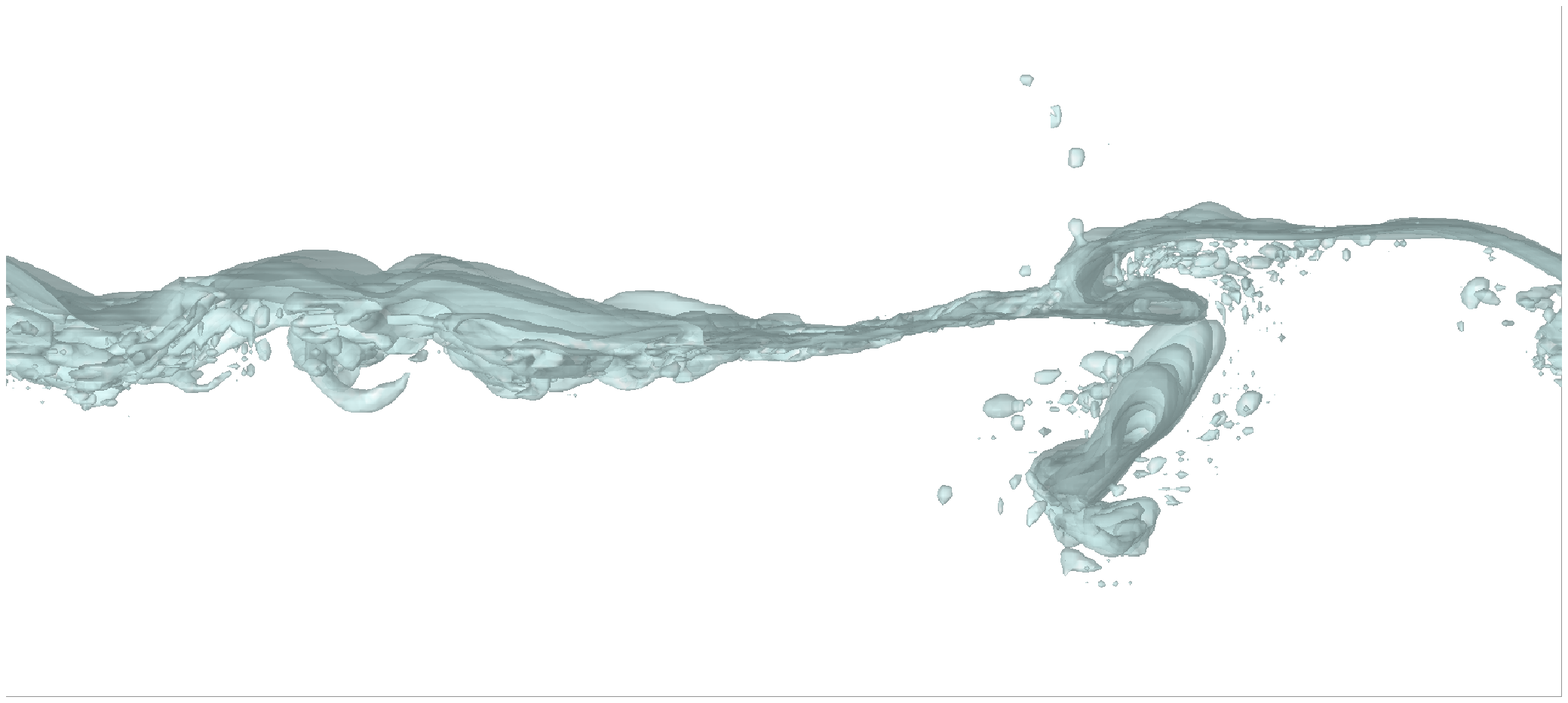}
}
  \centerline{
(e)
\includegraphics[trim={0pt 20pt 20pt 0pt},clip=true,width=0.425\linewidth,valign=t]{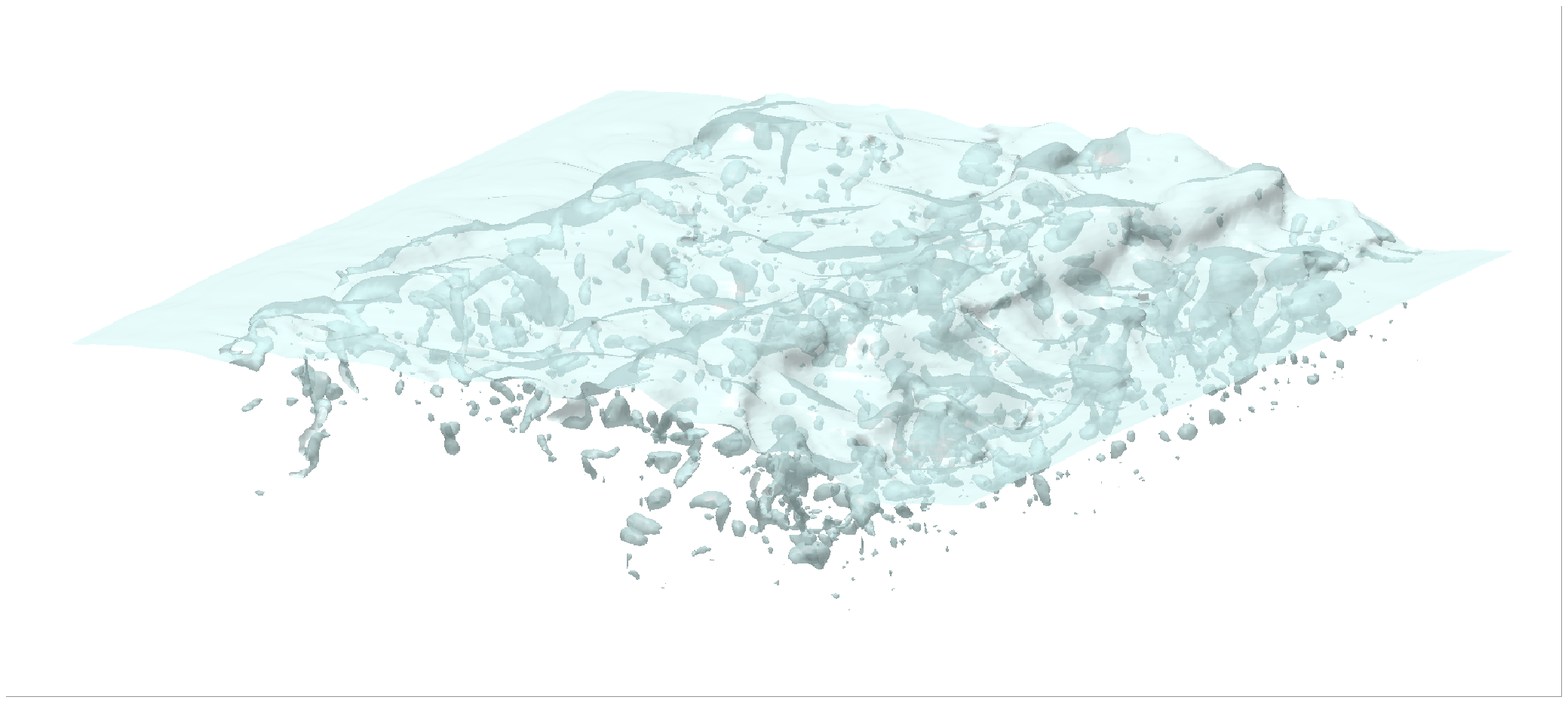}
\quad
(f)
\includegraphics[trim={0pt 20pt 20pt 0pt},clip=true,width=0.425\linewidth,valign=t]{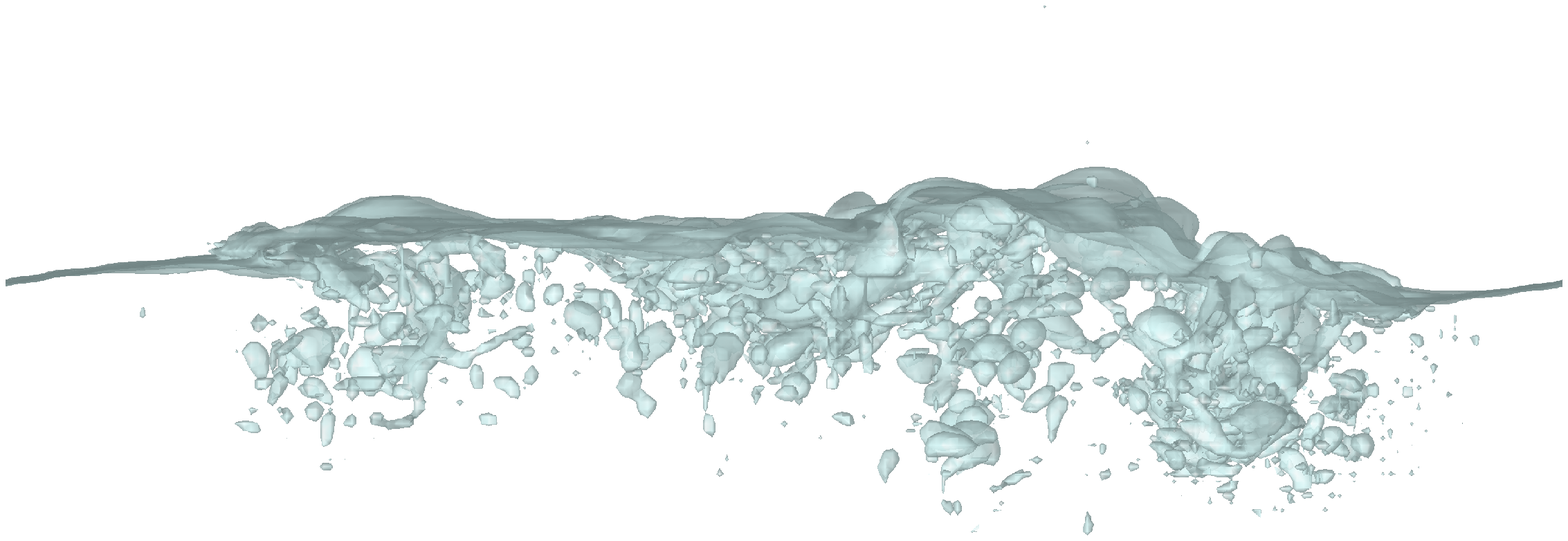}
}
  \centerline{
(g)
\includegraphics[trim={0pt 20pt 20pt 0pt},clip=true,width=0.425\linewidth,valign=t]{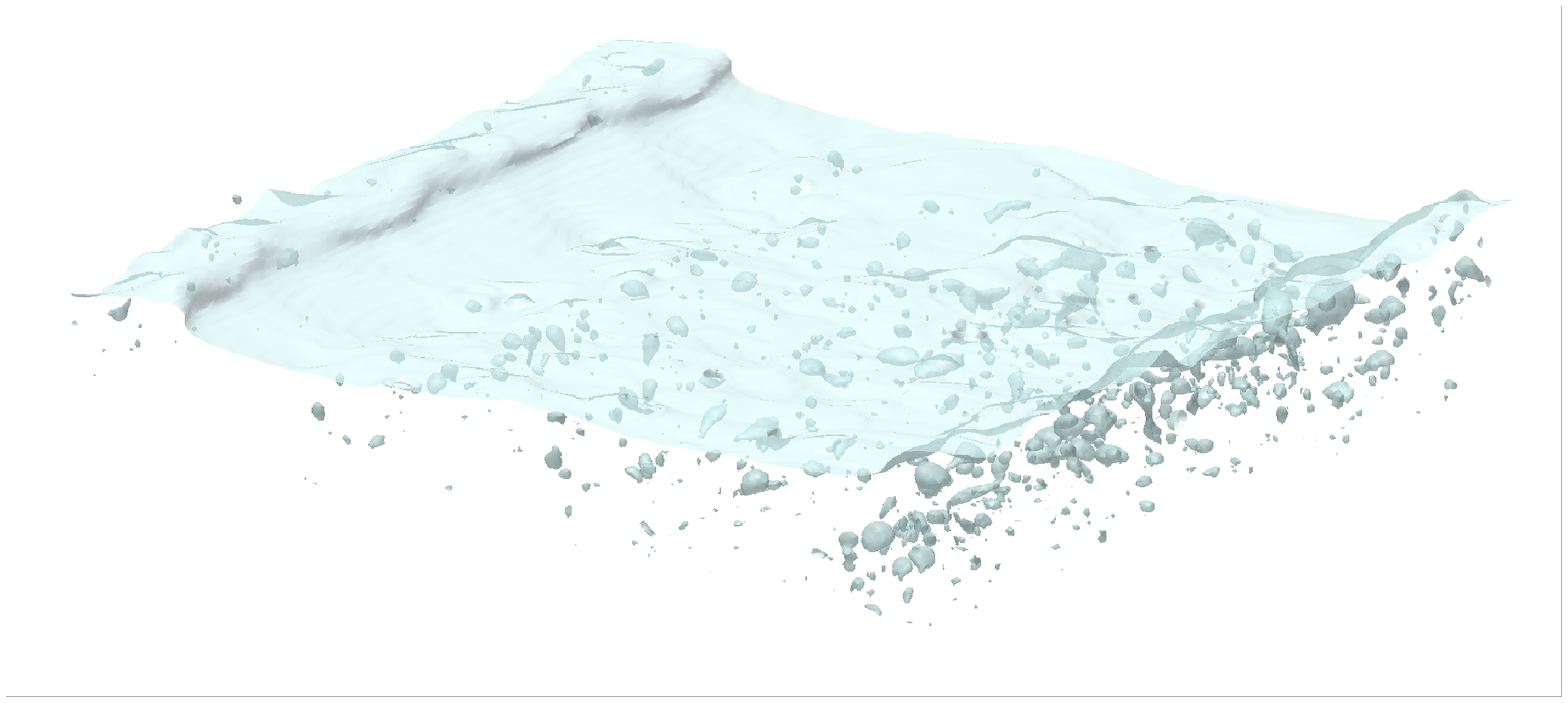}
\quad
(h)
\includegraphics[trim={0pt 20pt 20pt 0pt},clip=true,width=0.425\linewidth,valign=t]{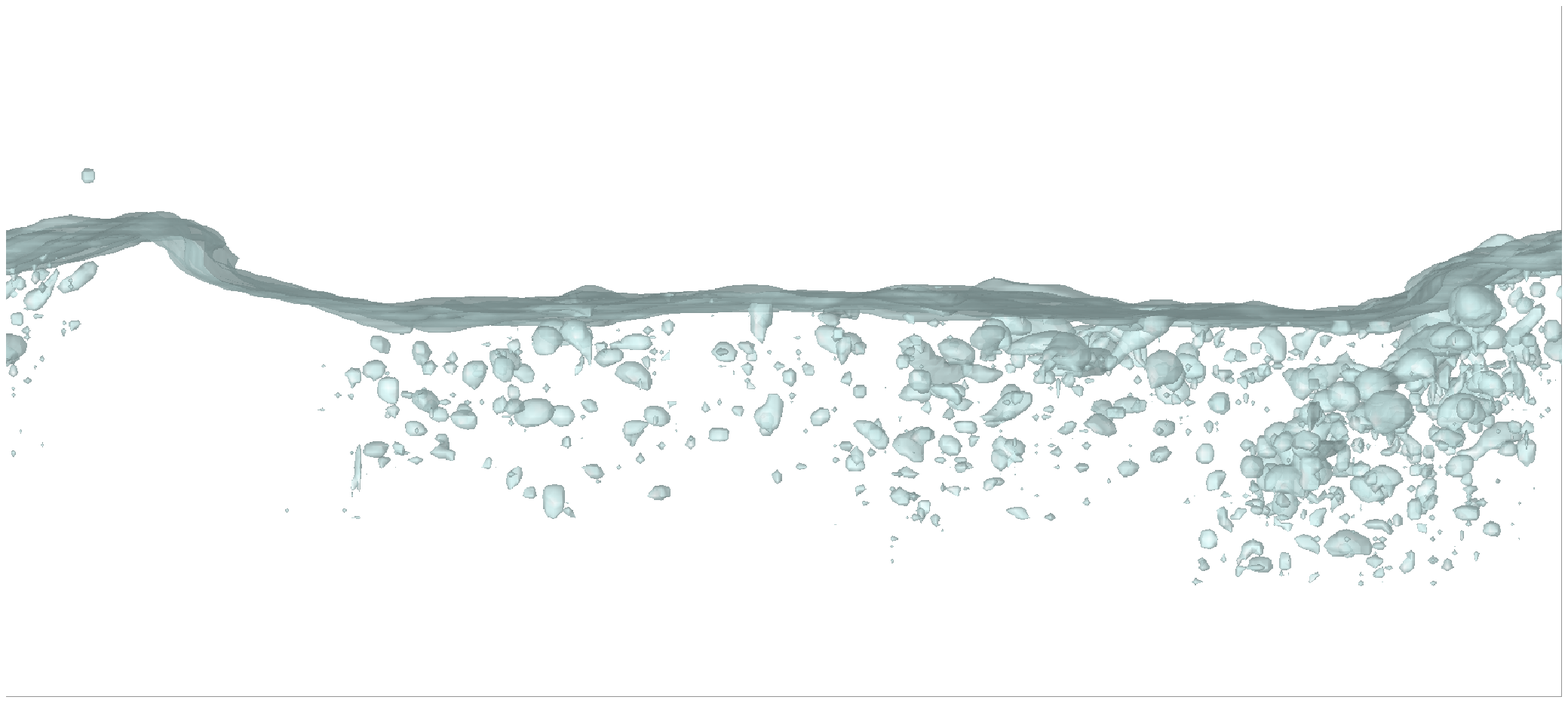}
}
  \caption{Snapshots of (left) an axonometric projection from above the wave and (right) the spanwise cross-section of the $\phi=0.5$ isosurface of one of the baseline ensemble realizations corresponding to various times in the wave-breaking process: (a,b) $t=2.00$, (c,d) $t=3.01$, (e,f) $t=4.03$, and (g,h) $t=5.04$. The snapshots are sampled at an interval of 1.01 characteristic times. The waves are travelling from left to right, and wrap around the domain due to the periodic boundary conditions in the streamwise direction.}
\label{fig:wave}
\end{figure}

From the snapshots of the plunging wave-breaking process in figure~\ref{fig:wave}, it is evident that the wave profile and accompanying distribution of bubbles dramatically evolve in the span of about 1 to 2 wave periods, or 3 to 5 characteristic times. A qualitatively similar evolution was observed in the studies by~\citet{Bonmarin1},~\citet{Chen1},~\citet{Deane1},~\citet{Blenkinsopp2,Blenkinsopp1},~\citet{Iafrati1},~\citet{Rojas2},~\citet{Kiger1}, \citet{Deike2,Deike1},~\citet{Lim1}, and~\citet{Wang1}. In particular, a number of tubular air cavities are formed, and then subsequently deformed and ruptured beneath multiple surface splash-ups. This bubble fragmentation eventually ceases as large bubbles degas more quickly, leaving a plume of smaller, slowly rising bubbles. More details are provided by~\citet{Chan8}.

\begin{figure}
  \centerline{
(a)
\includegraphics[width=0.40\linewidth,valign=t]{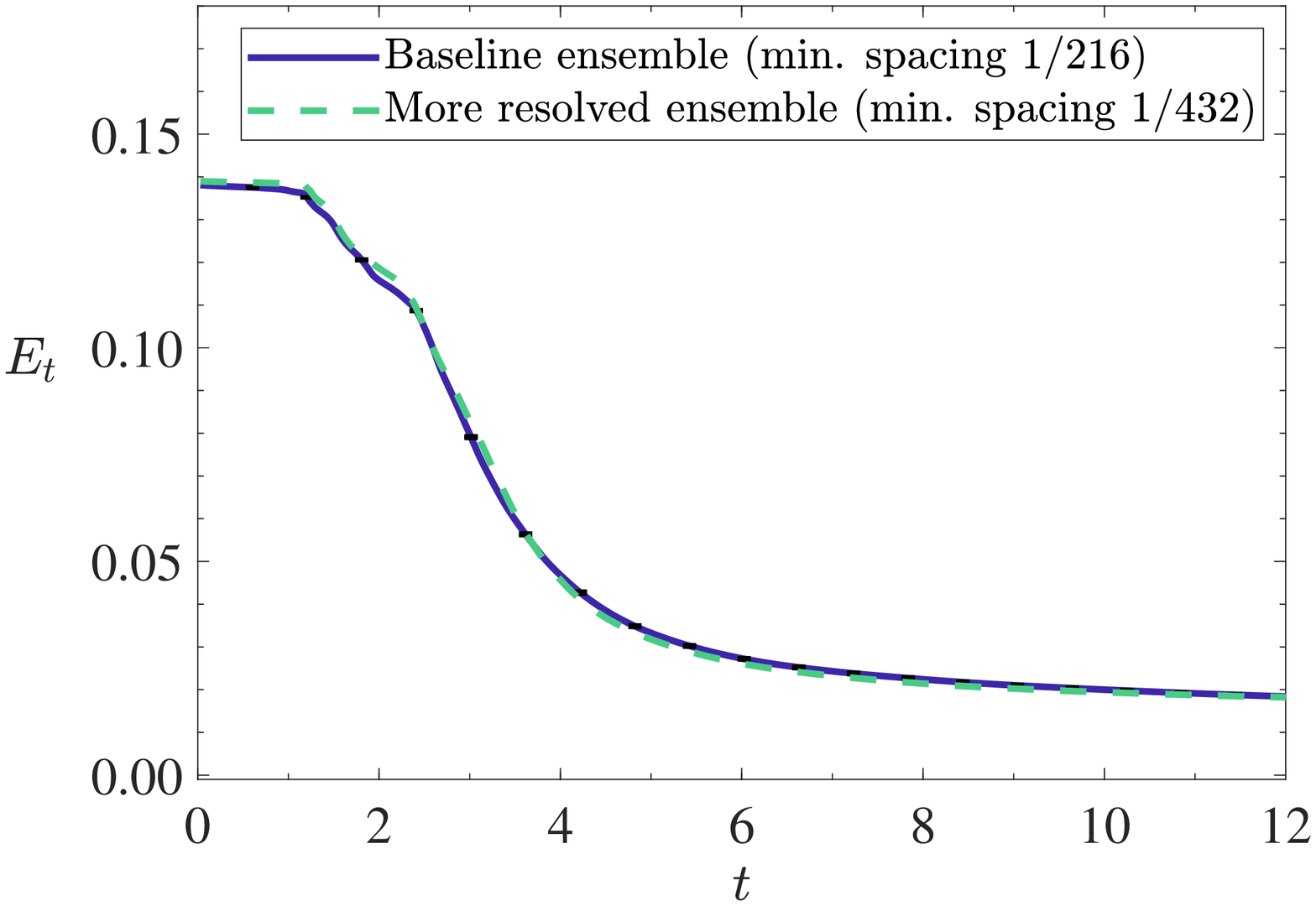}
\quad
(b)
\includegraphics[width=0.45\linewidth,valign=t]{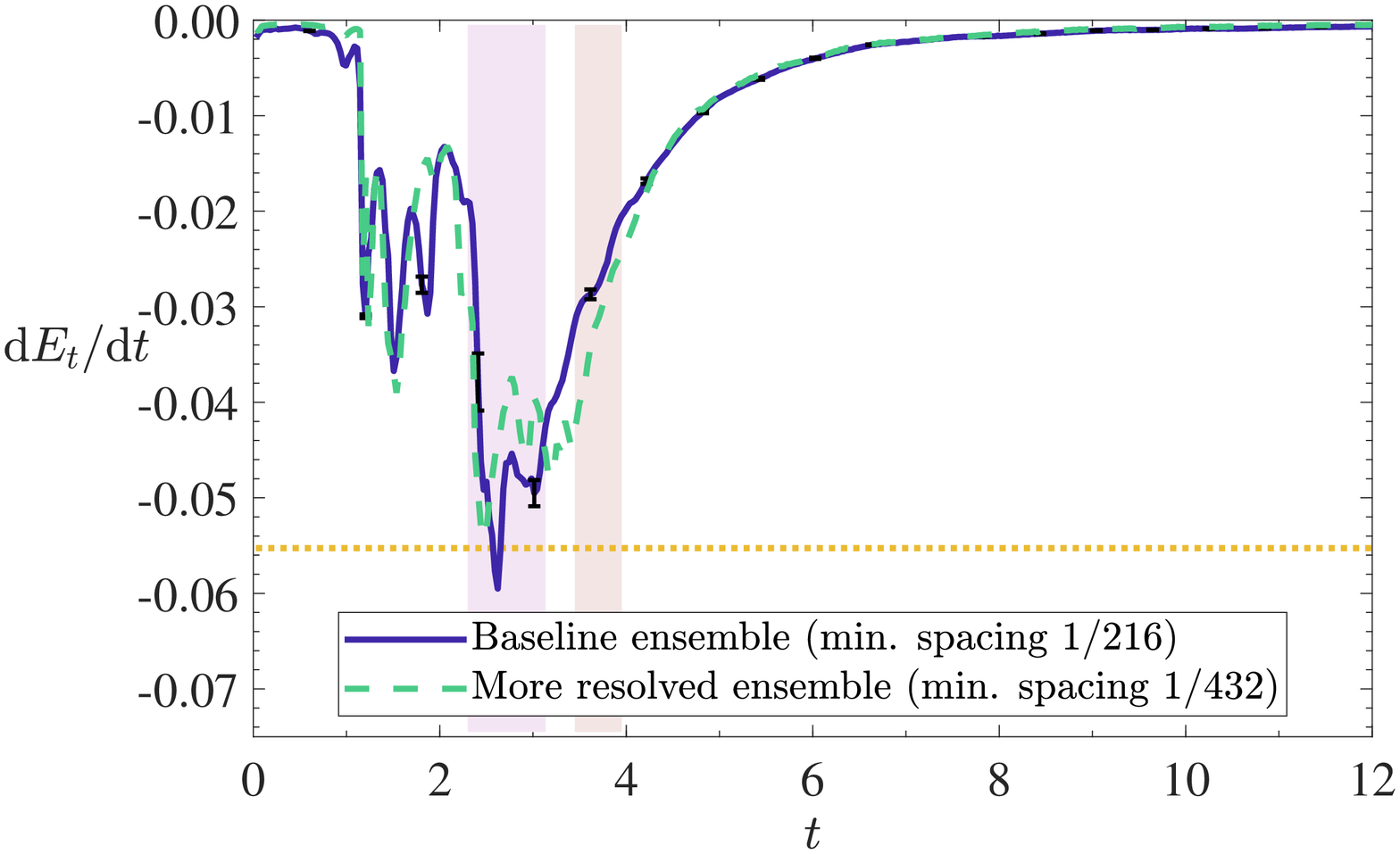}
}

  \caption{(a) Non-dimensional ensemble-averaged total ($E_t$) energy of the waves in the baseline ensemble, as well as the non-dimensional total energy of one of the realizations in the ensemble with increased mesh resolution, as a function of non-dimensional time. (b) Non-dimensional rate of change of the energies in (a), computed by differencing the total energy every 500 time-steps in the baseline ensemble and every 2{,}500 time-steps in the more resolved ensemble. The horizontal dotted line in (b) denotes the nominal rate of change of energy if all the energy initially present in the wave was to be dissipated in one wave period. The shaded region on the left spans a third of a wave period between $t=2.30$ and $t=3.14$, while the shaded region on the right spans a fifth of a wave period between $t=3.45$ and $t=3.95$. In both subfigures, the lines corresponding to the baseline ensemble denote ensemble-averaged quantities, while the error bars span, in each direction, twice the standard error of the energies of each realization with respect to the ensemble average, thus representing the variation over the ensemble.}
\label{fig:energy}
\end{figure}

This evolution in the wave dynamics is also manifested in the energetics of the wave, whose time evolution averaged over the wave ensemble is plotted in figure~\ref{fig:energy}. The volume-integrated total energy plotted in figure~\ref{fig:energy}(a) is defined as follows:
\begin{multline}
E_t = \f{\hat{E}_k + \hat{E}_p + \hat{E}_s}{\hat{E}_n} = \f{\text{kinetic energy}+\text{potential energy}+\text{surface energy}}{\text{reference energy for normalization}} ={}\\
{}= \f{\left[\int_{\Omega_d} \f{1}{2} \rho \left|\bs{\hat{u}}\right|^2 \diff \bs{\hat{x}}\right] + \left[\int_{\Omega_d} \rho g \hat{y} \: \diff \bs{\hat{x}} - \f{1}{8} \left(\rho_g-\rho_l\right) g L^4\right] + \left[\int_{\Omega_d} \sigma \left|\hat{\nabla}\phi\right| \diff \bs{\hat{x}}  - \sigma L^2\right]}{\left(\f{1}{2}\rho_l u_L^2\right)\left(\f{1}{2}L^3\right)},
\end{multline}
where the domain of integration $\Omega_d$ is taken here to be the entire computational domain with volume $\mathcal{V}_d = L^3$. In this domain, the range of $\hat{y}$ is defined as $\hat{y} \in [-L/2,L/2]$. The energy is computed with respect to the reference state of a quiescent air--water system with a flat interface where the water phase occupies the bottom half of the computational domain $\left(y \in [-1/2,0)\right)$ and the air phase occupies the top half $\left(y \in (0,1/2]\right)$. It is normalized by the energy $\hat{E}_n$ of a body of water occupying half the volume of the domain and travelling uniformly at the characteristic speed $u_L = (gL)^{1/2}/(2\upi)^{1/2}$. 

The total energy is observed to decay in time in general agreement with the trends observed by~\citet{Wang1} and~\citet{Deike1}. Figure~\ref{fig:energy}(a) suggests that the total energy is fairly insensitive to the grid spacing, so the baseline mesh should be considered sufficient for the present study. Figure~\ref{fig:energy}(b) generally supports this claim as well, although some differences in the rate of change of the total energy are visible during the active wave-breaking phase due to the difference in the mesh resolution. Note, however, that the standard errors in the baseline ensemble are also noticeably larger during this phase. 

Figure~\ref{fig:energy}(b) suggests that during the large-dissipation phase $t \in (2.3,3.1)$, the system loses energy at a rate comparable to that if the initial energy of the wave was dissipated in one wave period. The extraction of the wave period as a defining time-scale is compatible with the expressions for the global Kolmogorov and Hinze scales, \eqref{eqn:kolmodim} and \eqref{eqn:hinzedim}, and lends support to the dissipation rate estimate in appendix~\ref{app:dissipation}. The rate of energy dissipation also appears to be reasonably approximated as quasi-stationary during this time interval, notwithstanding some temporal fluctuations. 

The time evolution of the dissipation rate in figure~\ref{fig:energy}(b) exhibits two intervals of interest. In the first interval $t \in (2.3,3.1)$ discussed earlier, the dissipation rate remains quasi-steady as it attains its peak magnitude. In the second interval $t \in (3.5,4.0)$, the dissipation rate gradually decays in tandem with the surrounding turbulence. The sequence of events depicted in figure~\ref{fig:wave} illustrates the importance of bubble break-up in both intervals. In \S~\ref{sec:statistics}, the evolution of various bubble statistics in these two intervals will be analyzed, and the characteristics of the bubble-mass transfer dynamics in these intervals will be compared and contrasted.


\section{Algorithms}\label{sec:algorithms}

Two algorithms were used to retrieve bubble statistics from the simulation ensembles described in \S~\ref{sec:ensemble} in order to shed light on bubble-mass transfer between large and small bubble sizes during the active wave-breaking phase. \S~\ref{sec:ident} describes an algorithm used to identify the sizes and locations of bubbles in each simulation snapshot. The algorithm was used to determine the bubble size distribution and the total interfacial area. The algorithm in \S~\ref{sec:tracking} tracks these bubbles over successive simulation snapshots in order to detect break-up and coalescence events, which drive the time evolution of the bubble size distribution as discussed in \S~\ref{sec:bridging_pbe} and \S~\ref{sec:bridging_locality}. More details on these algorithms are discussed by \citet{Chan4}, and a brief overview is offered in this section.

\subsection{Bubble identification algorithm}\label{sec:ident}

An existing bubble identification algorithm~\citep{Hebert1,Herrmann2,Tomar1} was refined in this work to increase the accuracy in determining the volumes of the identified bubbles. The basis of the algorithm is the identification of connected regions of computational nodes, where each region corresponds to an individual bubble. After identifying these regions, one may then integrate the gaseous volume fraction, $1-\phi$, over the nodes in each of these regions to obtain the total volume of each bubble. A region is assembled by linking node pairs that each satisfy a grouping criterion. The traditional criterion requires that $\phi<1$ in each node of a pair, or that each node contains a non-zero amount of air. This criterion may create bubbles of spurious numerical origin because energetic collisions may create small wisps of air, which are sometimes also referred to as flotsam and jetsam, where $1-\phi \ll 1$ in each of these nodes. These wisps tend to linger in the flow field due to their low buoyancy and slow degassing rate. Wisps may form node pairs that satisfy the criterion and become spuriously linked together. Clipping small values of $1-\phi$ mitigates this issue at the expense of the volume accuracy of resolved bubbles. Instead, this work uses the additional criterion that at least one node in a pair satisfies $1-\phi > \phi_{c,m}$, so that nodes with small $1-\phi$ are linked only if they neighbour nodes with large $1-\phi$. In this work, $\phi_{c,m}=0.5$ was selected. 

The algorithm yields a list of bubbles in every simulation snapshot, and can simultaneously yield the volume and centre-of-mass (centroid) location of each identified bubble, among other statistics. Accuracy of the determined volume and centroid location is crucial to the performance of the bubble tracking algorithm to be discussed in \S~\ref{sec:tracking}. Volume accuracy also impacts the trends in the bubble size distribution to be discussed in \S~\ref{sec:sizedistevol}. A comparison of the errors incurred from various grouping criteria has been detailed in other works, such as~\citet{Chan4}. This involved systematically quantifying the incurred error for a number of simple test cases, as summarized in appendix~\ref{app:bubbleident}.

\subsection{An event detection algorithm to track bubbles}\label{sec:tracking}

A bubble tracking algorithm was developed in this work to detect break-up and coalescence events by maintaining a tally of all bubbles over time and tracing the lineage of each bubble. To construct these lineages, lists of bubbles with their sizes and locations are compared between successive simulation snapshots. These snapshots need not arise from consecutive time-steps. This is because the principle of mass conservation and the Courant-Friedrichs-Lewy (CFL) condition are used to determine if a bubble has simply been advected between the two snapshots without any change in topology, or if a break-up or coalescence event has occurred. These two constraints are discussed in appendix~\ref{app:eventdetect} in relation to the errors in the identification algorithm discussed in appendix~\ref{app:bubbleident}. As discussed in \S~\ref{sec:bridging_pbe}, disconnections and reconnections with the atmosphere, which is treated as a large gaseous reservoir, are not included in tallies of break-up and coalescence events, respectively. This is to prevent frequent topology changes involving the atmosphere and large, non-spherical bubbles with convoluted shapes near the wave surface from obscuring the remaining break-up and coalescence statistics.

The time interval between successive snapshots is now discussed in relation to other characteristic time-scales for the baseline ensemble introduced in \S~\ref{sec:ensemble}. The non-dimensional snapshot interval for a set of snapshots available for all 30 realizations is $\Delta t_{s,30} = 3.0\times10^{-2}$. A set of more-closely-spaced snapshots is available for 20 of these realizations with a time separation of $\Delta t_{s,20} = 6.0\times10^{-3}$. From \eqref{eqn:hinzedim} and the inertial subrange scaling $u_{L_n} \sim (\ov{\varepsilon} L_n)^{1/3}$, one may estimate the characteristic mean time interval between break-up events for Hinze-scale bubbles to be $\Delta t_\mathrm{H} \sim 10^{-1}$. This scaling for $u_{L_n}$ also suggests that the corresponding time interval for super-Hinze-scale bubbles will exceed $\Delta t_\mathrm{H}$. Since the snapshot interval is shorter than the mean super-Hinze-scale break-up time, the resulting statistics should provide a realistic picture of at least super-Hinze-scale turbulent break-up. In particular, there are at least $O(10)$ snapshots in the 20-realization dataset between two super-Hinze-scale break-up events on average. 

Two additional remarks are in order here. First, a snapshot interval that exceeds the simulation time-step prevents the algorithm from identifying every break-up event of interest. Second, the discussion in appendix~\ref{app:bubbleident} suggests that the identification algorithm in \S~\ref{sec:ident} cannot effectively discern the separation between two bubbles spaced less than a grid cell apart. Bubbles that break up but do not separate quickly enough may then register repeated break-up and coalescence events. An excessively small snapshot interval, such as one that is much smaller than the integral time-scale, may capture some of these confounding events. This limitation has also been observed in other event detection algorithms~\citep{Rubel1} and appears to be inherent in event detection in turbulent flows. An ideal snapshot interval should resolve the characteristic mean break-up times of most bubbles while being sufficiently long to skip over these confounding events. The impacts of these algorithmic limitations will be discussed in \S~\ref{sec:trac-gbfe}.

\section{Bubble statistics}\label{sec:statistics}

\subsection{The bubble size distribution $\ov{f}$}\label{sec:sizedistevol}

\begin{figure}
  \centerline{
(a)
\includegraphics[width=0.42\linewidth,valign=t]{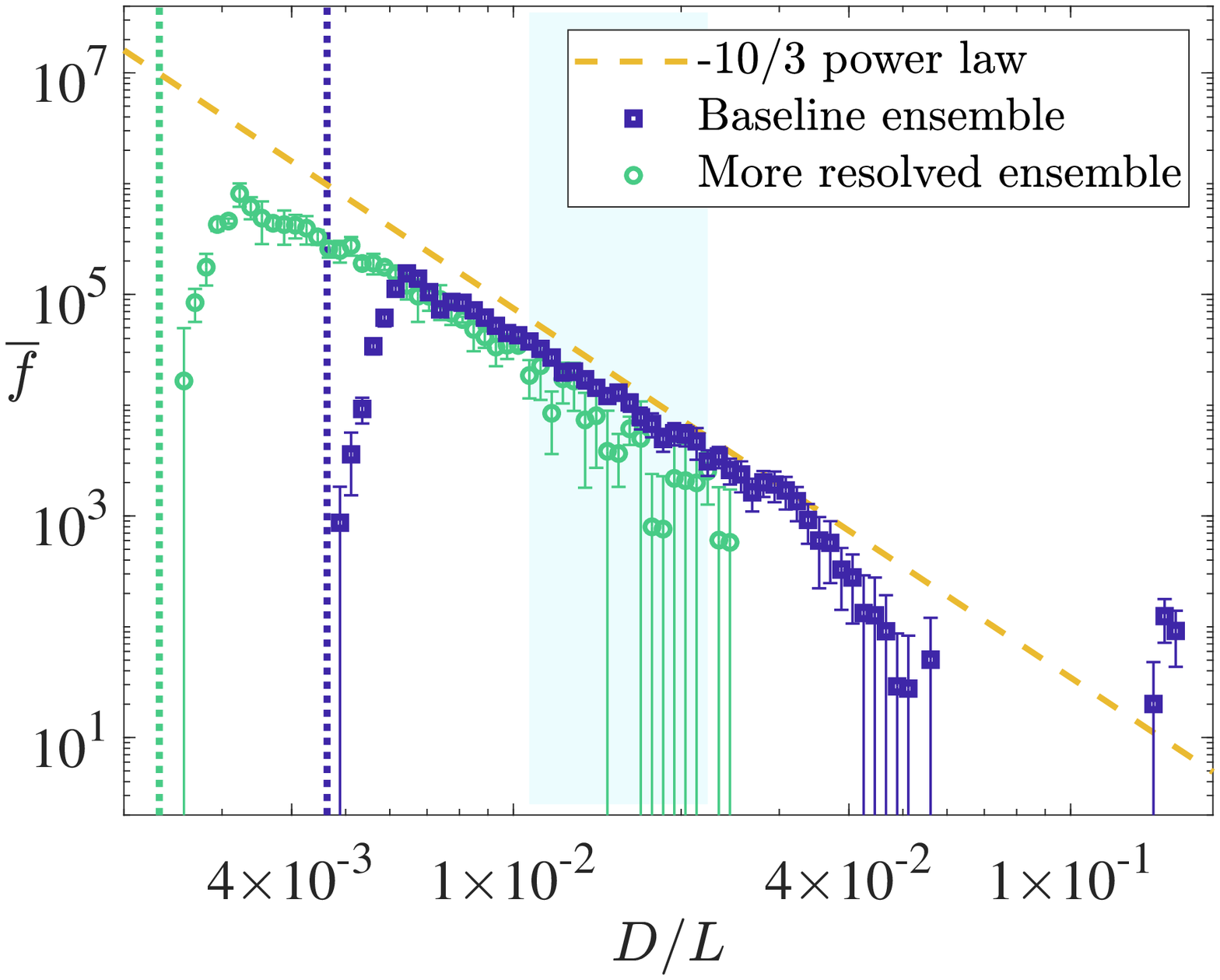}
\quad
(b)
\includegraphics[width=0.42\linewidth,valign=t]{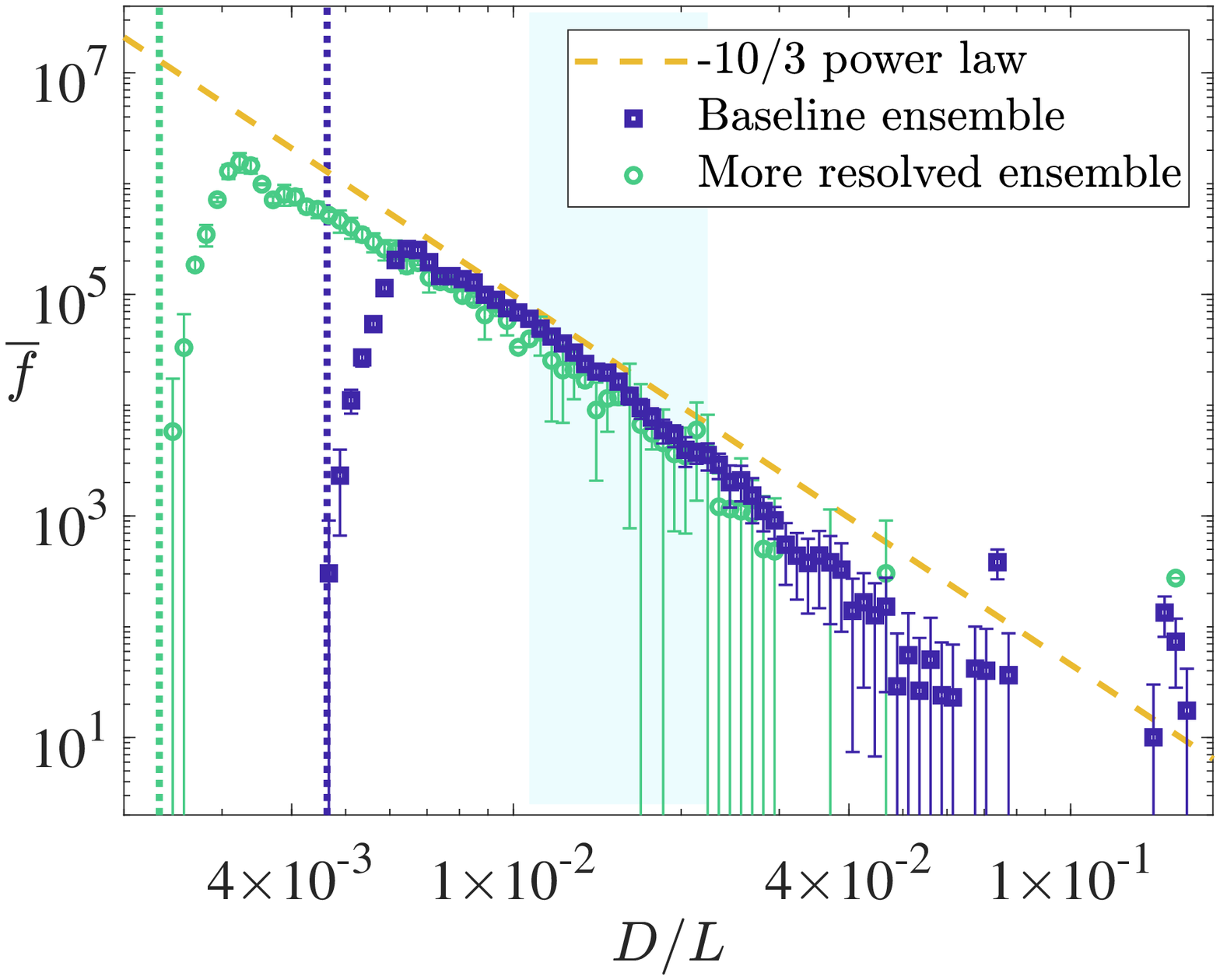}
}
  \centerline{
(c)
\includegraphics[width=0.42\linewidth,valign=t]{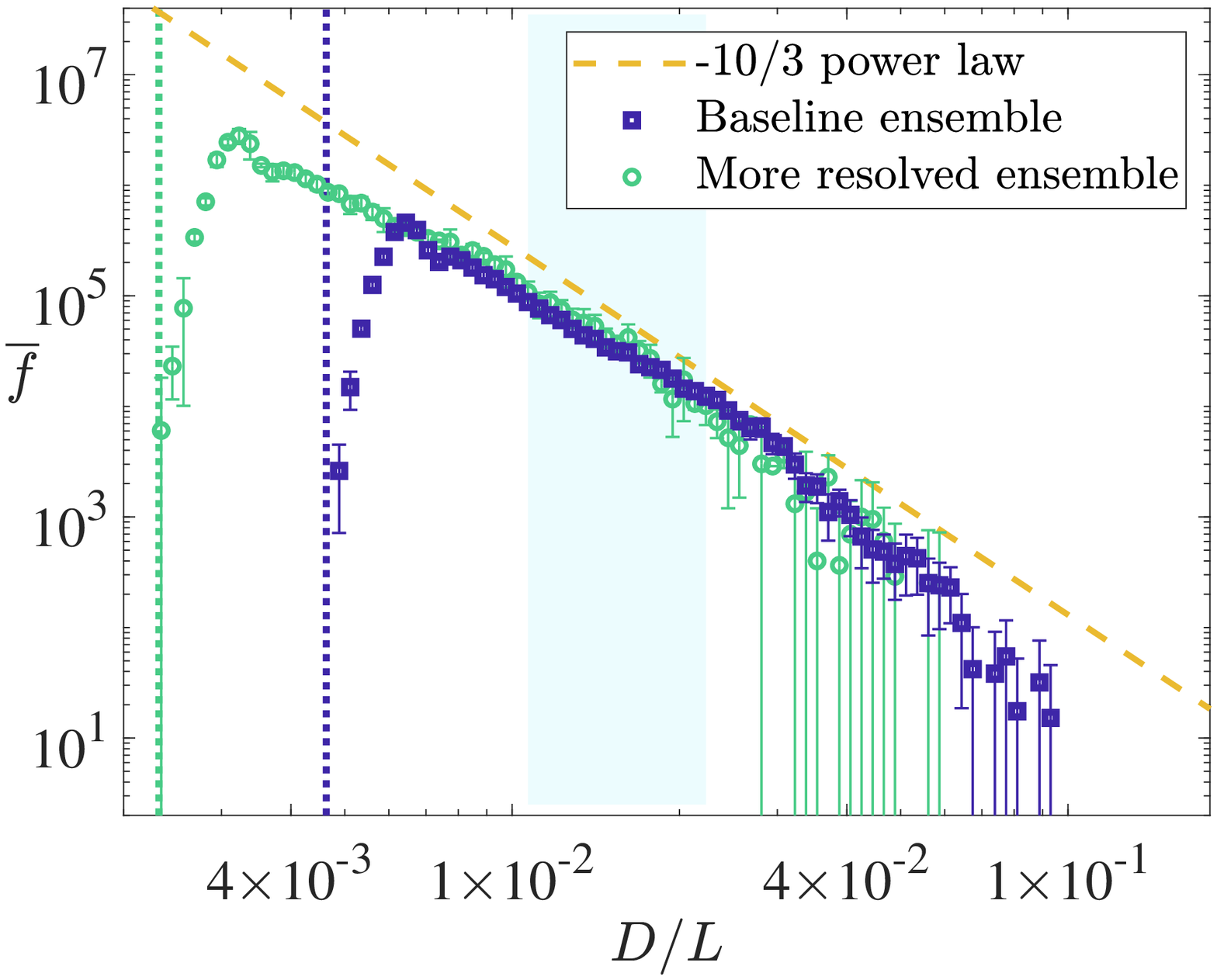}
\quad
(d)
\includegraphics[width=0.42\linewidth,valign=t]{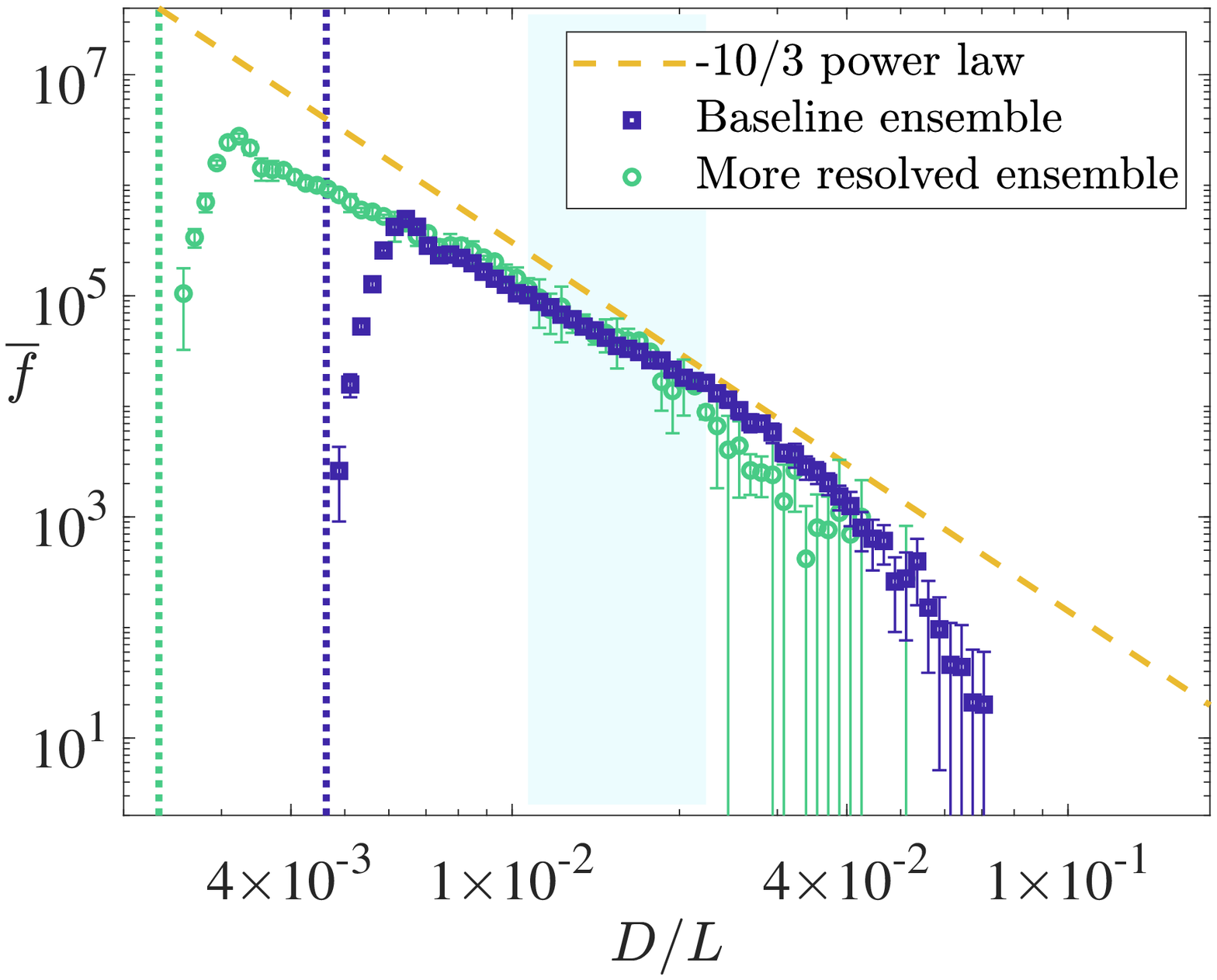}
}
  \caption{The non-dimensional bubble size distribution $\ov{f}(D_j/L;t)$ against non-dimensional size $D/L$ for the baseline and more resolved ensembles at (a) $t=2.50$, (b) $t=2.74$, (c) $t=3.71$, and (d) $t=3.95$. The dotted vertical lines in the same colour/shade as the respective distributions indicate the mesh resolution of the respective ensembles. The dashed sloped lines indicate the $D^{-10/3}$ power-law scaling for an idealized turbulent bubbly flow. The error bars span, in each direction, twice the standard error of the distribution in each realization with respect to the ensemble average. The shaded regions represent the bubble-size subrange over which the power-law fit exponents in figures~\ref{fig:bsd_powerlaw_time} and \ref{fig:bsd_timeavg} were obtained. More snapshots in time of the bubble size distribution are provided by~\citet{Chan8}.}
\label{fig:bsd_compare_powerlaw}
\end{figure}

Figure~\ref{fig:bsd_compare_powerlaw} plots the bubble size distribution $\ov{f}$ at selected time instances for the ensembles introduced in \S~\ref{sec:ensemble}. The bubble size, $D_i = \left[(6\mathcal{V}_i)/\upi\right]^{1/3}$, is the diameter of a sphere with the same volume, $\mathcal{V}_i$. The size distribution has dimensions $(\text{length})^{-4}$, is non-dimensionalized by the wavelength ($L^4$), and was computed by histogram binning with $N_\text{bin}$ bins of equal logarithmic spacing, where the smallest bin is two orders of magnitude larger than the diameter error in appendix~\ref{app:bubbleident}. The non-dimensional discrete distribution satisfies
\begin{equation}
\left\langle N_b(t) \right\rangle = \sum_{j=1}^{N_\text{bin}} \ov{f}(D_j/L;t) \Delta(D_j/L),
\end{equation}
where $\ov{f}(D_j/L;t)$ includes bubbles of sizes between $D_j$ and $D_{j+1}$, and $\Delta(D_j/L) = (D_{j+1}-D_j)/L$. The distributions from the two ensembles reasonably overlap at intermediate bubble sizes, suggesting that the distribution is fairly mesh insensitive at these sizes. While the large-bubble distribution of the more resolved ensemble has significant standard errors due to the small ensemble size and the small number of large bubbles, these large bubbles are well resolved in both ensembles.

One may link the evolution of the size distribution in figure~\ref{fig:bsd_compare_powerlaw} to the wave-breaking process in figure~\ref{fig:wave}. As the cylindrical air cavities deform and rupture between $t=2$ and $t=3$ [figures~\ref{fig:wave}(a)--(d)], the size distribution displays a power-law trend [figures~\ref{fig:bsd_compare_powerlaw}(a)/(b)] with an exponent near the $-10/3$ value postulated by~\citet{Garrett1}, and observed by~\citet{Deane1} and~\citet{Deike1}, suggesting that the turbulent break-up cascade examined in Part 1 dominates the bubble-mass transfer during these early wave-breaking stages. As evidenced in figure~\ref{fig:energy}(b), the dissipation rate is also large. In addition, the smallest resolvable bubbles rapidly appear, as also observed by~\citet{Deane1} and revisited in \S~\ref{sec:trac-qb}. While bubbles continue to deform, fragment, and interact between $t=3$ and $t=4$ [figures~\ref{fig:wave}(c)--(f)], the power-law scaling becomes shallower at intermediate sizes [figures~\ref{fig:bsd_compare_powerlaw}(c)/(d)]. As evidenced in figure~\ref{fig:energy}(b), the dissipation rate also begins to decay. Figures~\ref{fig:wave}(e)--(h) suggest that large-scale air entrainment has mostly ceased by this time, even as intermediate-scale bubble break-up continues, as revisited in~\S~\ref{sec:trac-wb}. 

\begin{figure}
  \centerline{
\includegraphics[width=0.52\linewidth]{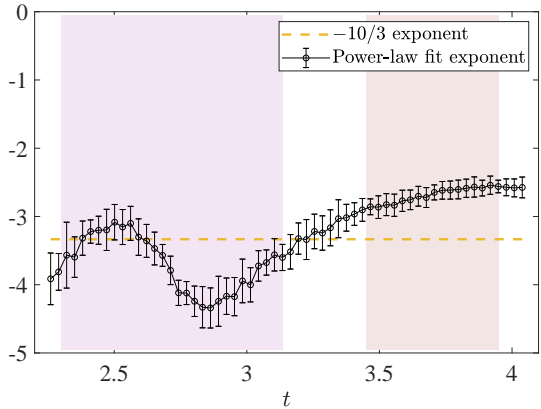}
}
  \caption{The variation of the exponent of a power-law fit over a subset of the baseline distribution, $\ov{f}$, in figure~\ref{fig:bsd_compare_powerlaw} with non-dimensional time. The fit was performed over bubbles with non-dimensional diameters between $1.07\times10^{-2}$ and $2.23\times10^{-2}$. This bubble-size subrange is marked in the shaded regions in figure~\ref{fig:bsd_compare_powerlaw}. The error bars denote twice the standard error in the fit exponent over the baseline ensemble. The shaded regions span the same time intervals as the corresponding regions in figure~\ref{fig:energy}(b).} 
\label{fig:bsd_powerlaw_time}
\end{figure}

\begin{figure}
  \centerline{
(a)
\includegraphics[width=0.42\linewidth,valign=t]{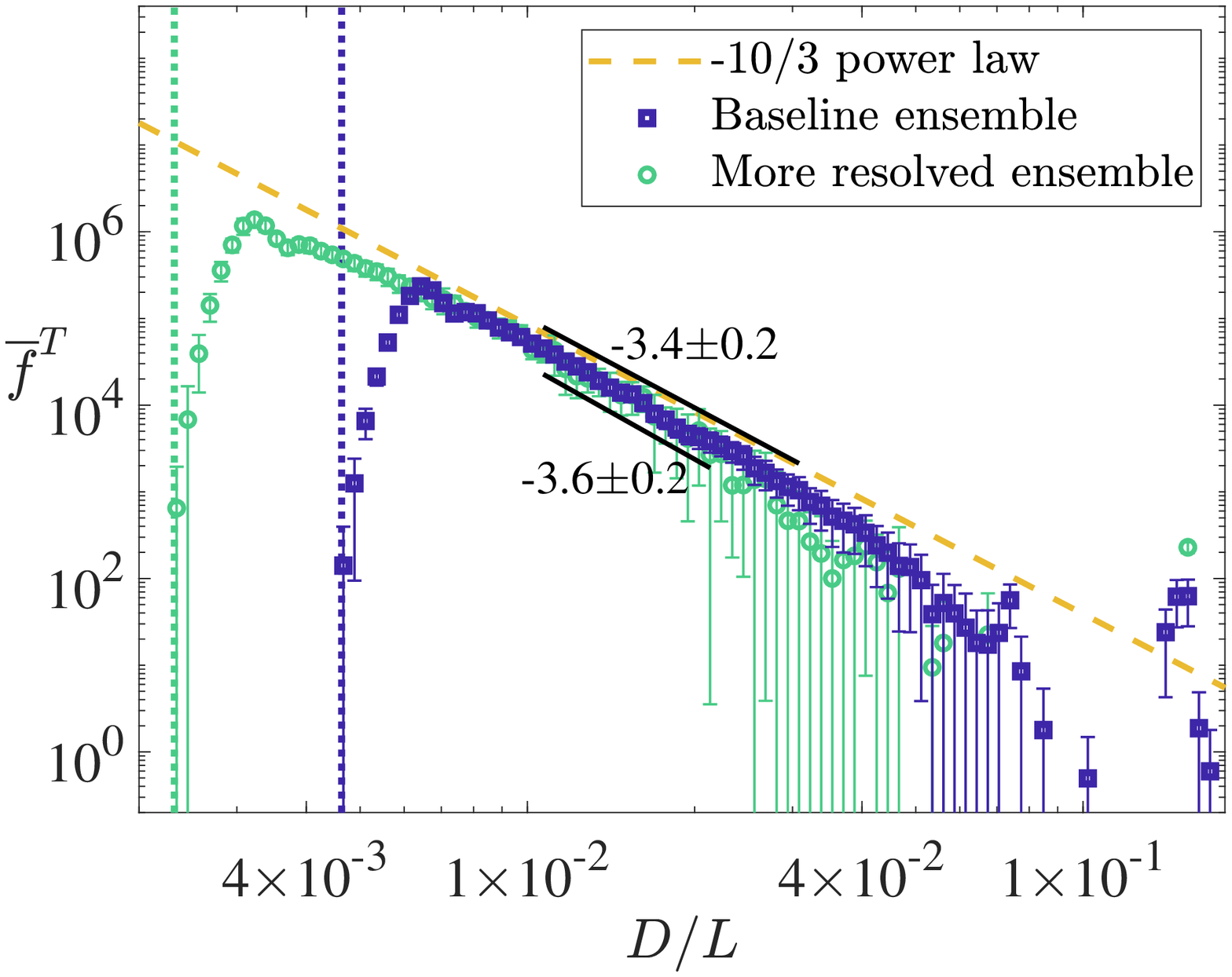}
\quad
(b)
\includegraphics[width=0.42\linewidth,valign=t]{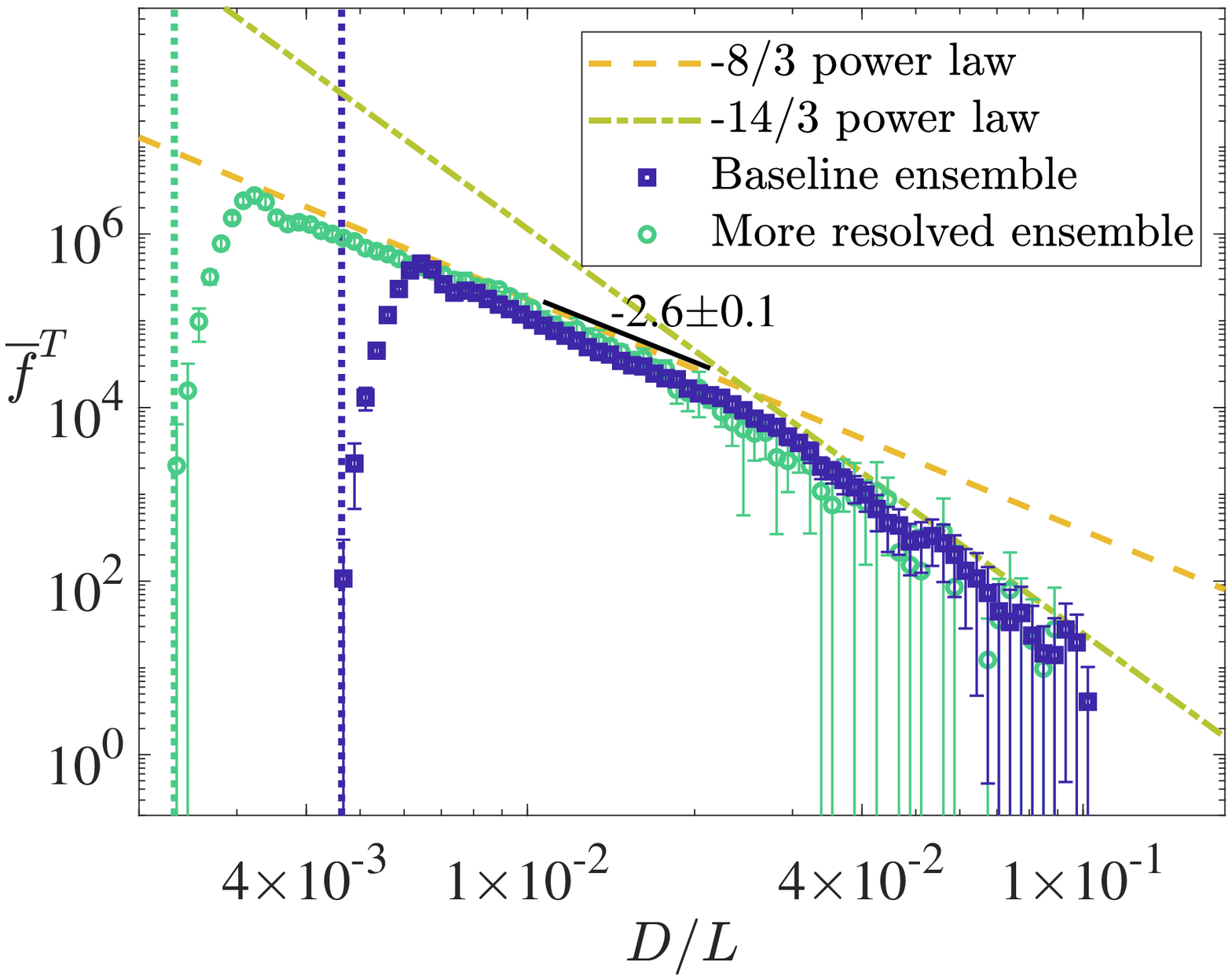}
}
  \caption{The non-dimensional time-averaged size distribution $\ov{f}^{T}(D_j/L)$ against non-dimensional size $D/L$ for both ensembles. The time-averaging interval is (a) between $t=2.30$ and $t=3.14$, corresponding to the left shaded regions in figures~\ref{fig:energy}(b) and~\ref{fig:bsd_powerlaw_time}, and (b) between $t=3.45$ and $t=3.95$, corresponding to the right shaded regions in the same figures. For a description of the vertical lines, dashed sloped line in (a), and error bars, refer to the caption of figure \ref{fig:bsd_compare_powerlaw}. The bubble-size subrange over which the bottom power-law fit in (a) and the power-law fit in (b) were performed is the same subrange highlighted in figure~\ref{fig:bsd_compare_powerlaw} and used in figure~\ref{fig:bsd_powerlaw_time}. The subrange for the top power-law fit in (a) is $[1.07\times10^{-2},3.23\times10^{-2})$. The fit exponent uncertainties denote twice the standard error over the baseline ensemble.} 
\label{fig:bsd_timeavg}
\end{figure}

\begin{figure}
  \centerline{
(a)
\includegraphics[width=0.42\linewidth,valign=t]{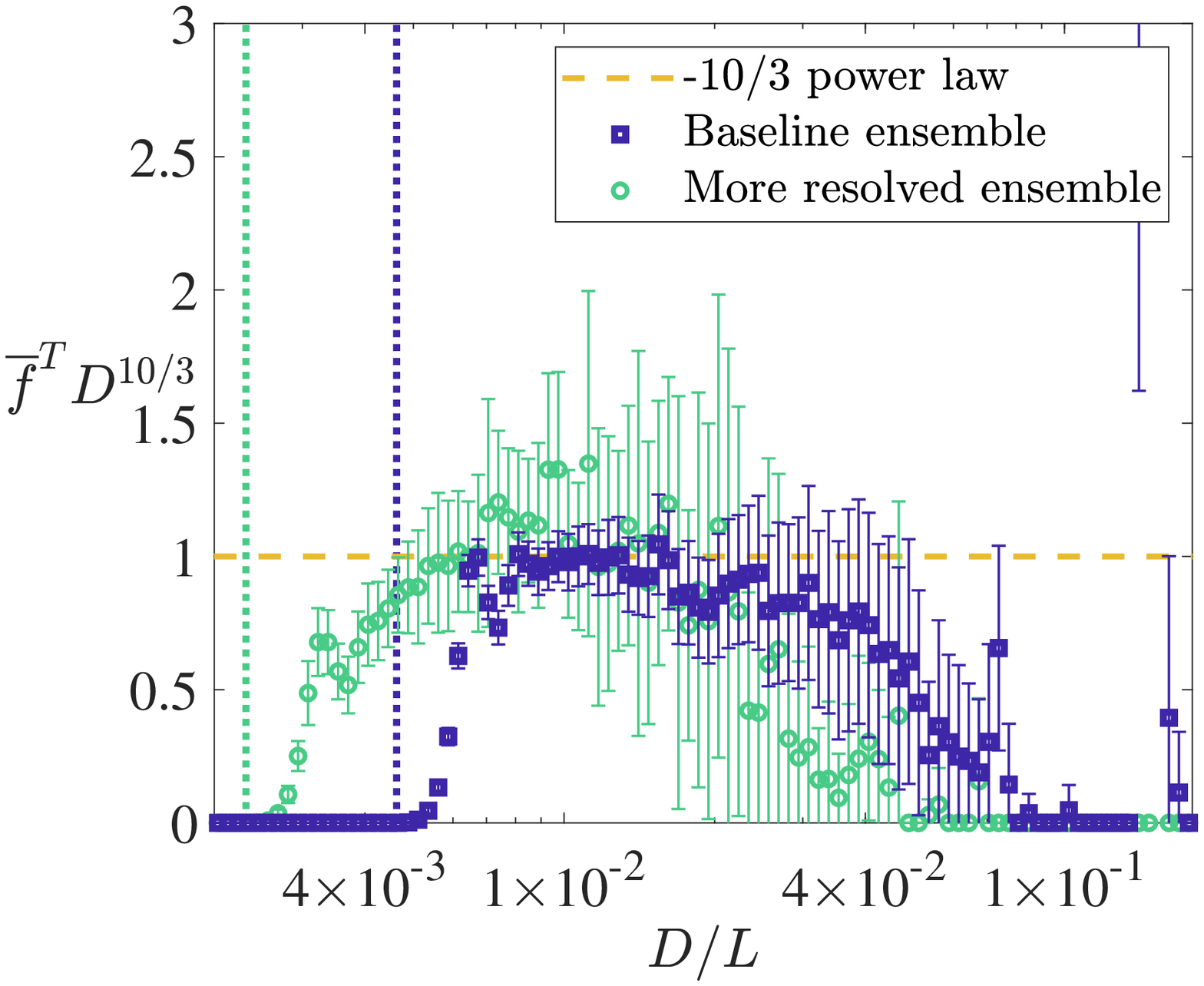}
\quad
(b)
\includegraphics[width=0.42\linewidth,valign=t]{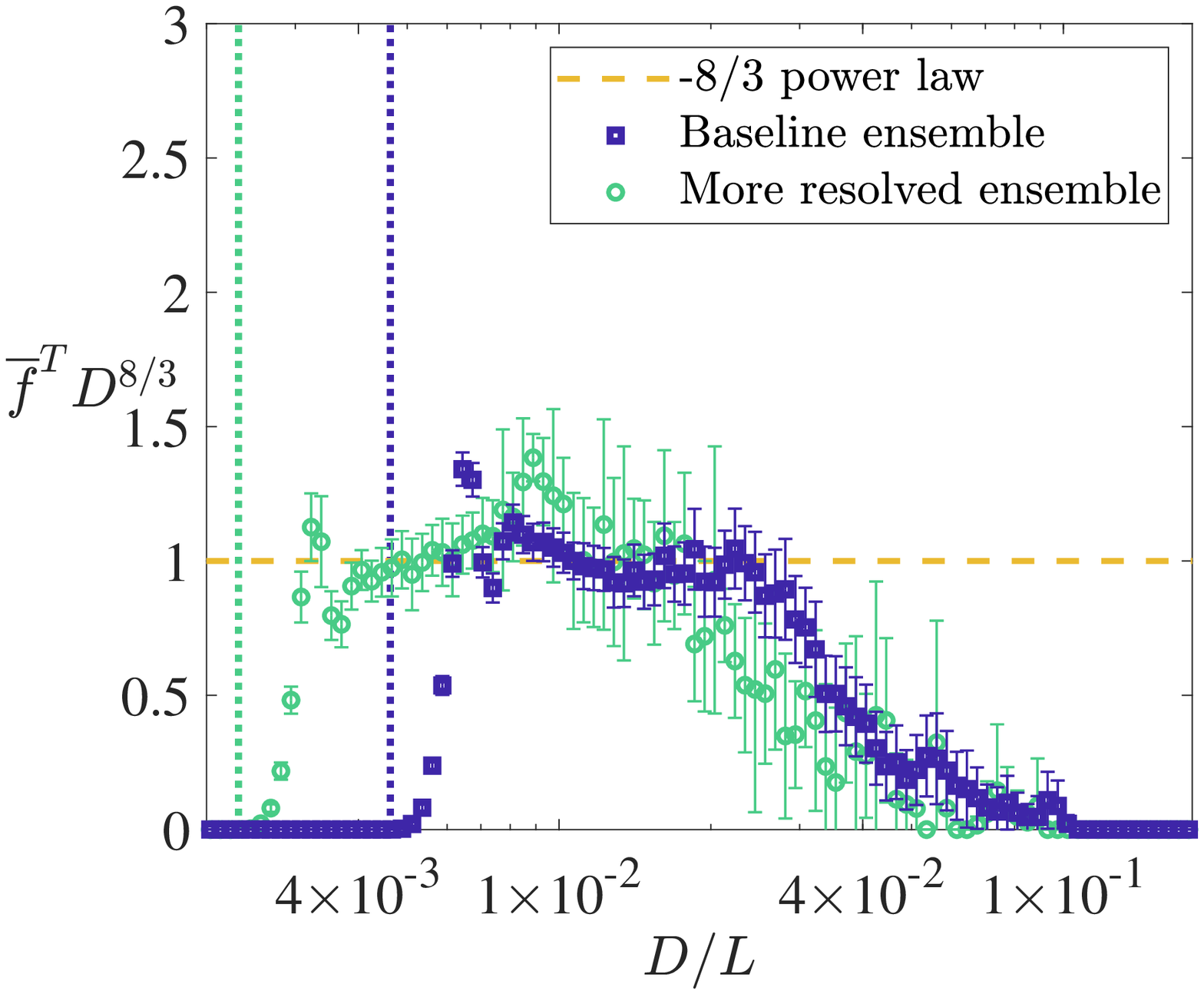}
}
  \caption{The compensated time-averaged size distribution using the same time-averaging intervals as in figure~\ref{fig:bsd_timeavg}: (a) $\ov{f}^{T}(D_j/L) D_j^{10/3}$ from the first interval, premultiplied by the inverse of the $D^{-10/3}$ scaling, and (b) $\ov{f}^{T}(D_j/L) D_j^{8/3}$ from the second interval, premultiplied by the inverse of the observed $D^{-8/3}$ scaling, against non-dimensional size $D/L$ for both simulation ensembles. The compensated distribution is normalized by its value at $D/L = 1.07\times10^{-2}$. For a description of the vertical lines and error bars, refer to the caption of figure \ref{fig:bsd_compare_powerlaw}.} 
\label{fig:bsd_powerlaw_premult_time}
\end{figure}

This work focuses on the statistics of intermediate-sized (super-Hinze-scale) bubbles with non-dimensional diameters larger than $1.1\times10^{-2}$, or dimensional diameters larger than $2.9\text{ mm}$, for which mesh insensitivity was observed. Figure~\ref{fig:bsd_powerlaw_time} quantifies the evolution of the exponent of a power-law fit to the size distribution in figure~\ref{fig:bsd_compare_powerlaw} over an intermediate size subrange. Observe its oscillatory variation around $-10/3$ at early times, which may be attributed to regularity in the successive wave-surface splash-ups and impingements. This variation suggests that the largest scales may be influencing these intermediate-size statistics through analogous fluctuations in the energy and bubble-mass cascade rates. Indeed, fluctuations manifest in the dissipation rate in figure~\ref{fig:energy}(b), and the bubble-mass transfer rate, $\ov{W_b}$, in \S~\ref{sec:trac-wb}. This variation may also reflect the finite convergence time of the break-up dynamics towards quasi-equilibrium~\citep{Filippov1,Qi1}. Figure~\ref{fig:bsd_timeavg}(a) depicts the size distribution averaged over these early time instances with two power-law fits, which are in general agreement with $D^{-10/3}$, although the fit exponent exhibits slight sensitivity to the size subrange used for fitting. A similar agreement was obtained in the ensemble-averaged and time-averaged statistics of~\citet{Deane1} and~\citet{Deike1}, and the time-averaged statistics of~\citet{Wang1}, which builds confidence in the results of this work. Ensemble averaging improves statistical convergence and reduces data scatter such as that observed by~\citet{Wang1} in their instantaneous distributions. While the instantaneous distribution in this work oscillates between $D^{-3}$ and $D^{-4}$, better agreement with $D^{-10/3}$ is obtained via time averaging over the entire interval, consistent with the suggestion of \citet{Deike1} that the scaling of the time-averaged distribution, $\ov{f}^T$, is sensitive to the time-averaging window, as will be revisited in \S~\ref{sec:trac-wb}. A more rigorous test of the $D^{-10/3}$ scaling is shown in figure~\ref{fig:bsd_powerlaw_premult_time}(a) by premultiplying the time-averaged distribution and using a linear scale on the vertical axis. Time averaging is further explored in appendix~\ref{app:averaging}.

Returning to figure~\ref{fig:bsd_powerlaw_time}, the fit exponent deviates from $-10/3$ at later times, about one wave period after the onset of breaking. Interestingly, the exponent does not oscillate, in seeming correlation with the cessation of large-scale entrainment. Figure~\ref{fig:bsd_timeavg}(b) depicts the size distribution averaged over these late time instances. Its increased magnitude at most resolved sizes relative to (a) (by about 3--5 times; see also figure~\ref{fig:breakup_w_t}) suggests that there are more small and intermediate-sized bubbles at later times, and the break-up flux from larger sizes consistently outweighs mass loss from degassing. Two distinct power-law scalings emerge in two size subranges in agreement with the measurements of~\citet{Tavakolinejad1} and~\citet{Masnadi1}, which builds further confidence in the results of this work [see also~\citet{Castro2}]. The size distribution is reasonably described by a $D^{-8/3}$ scaling at intermediate sizes, which is shallower than $D^{-10/3}$. This is supported by the compensated time-averaged distribution in figure~\ref{fig:bsd_powerlaw_premult_time}(b). At large sizes, the power-law scaling is steeper than $D^{-8/3}$ by a factor of $D^{-2}$ in agreement with the increased importance of buoyant degassing postulated by~\citet{Garrett1} and observed by~\citet{Deike1}, as well as in figures~\ref{fig:wave}(e)--(h). For comparison, \citet{Tavakolinejad1} obtained an exponent between $-2.85$ and $-2.58$ at small sizes, and between $-6.57$ and $-3.85$ at large sizes.

Figures~\ref{fig:energy}(b) and \ref{fig:bsd_compare_powerlaw}--\ref{fig:bsd_powerlaw_premult_time} reflect the unsteadiness in the wave dynamics and demonstrate the emergence of two time intervals with distinct characteristics, suggesting the presence of distinct bubble generation and evolution mechanisms. In the first interval, the dissipation rate appears to be quasi-steady, and the $D^{-10/3}$ power-law scaling in the time-averaged size distribution supports the presence of a quasi-steady bubble break-up cascade. In the second interval, the dissipation rate is seen to decay, and the time-averaged size distribution deviates from $D^{-10/3}$. Yet, the robustness of an alternative power-law scaling with a negative exponent is indicative of size-invariant cascade-like behaviour, as Part 1 suggested that size-local break-up may still occur in the absence of the $D^{-10/3}$ scaling. These observations motivate a closer look at other bubble statistics during these two intervals to test the theoretical analysis developed in Part 1 for the break-up cascade in the first interval, and to extend this analysis to the possible cascade-like behaviour in the second interval. Note that these conclusions were drawn from statistics over a limited size range. Future ensembles with larger scale separation will be valuable for shedding more light on these conclusions. Volume averaging also precludes the reporting of spatial size distributions. The physical parameters and grid size selected in this work reflect the trade-off between the resolved size range and the ensemble size. As noted in \S~\ref{sec:intro} and \S~\ref{sec:setup}, a larger ensemble increases statistical convergence in a statistically unsteady flow that does not strictly permit time averaging. However, it also necessitates a smaller resolved range from a practical perspective. The mesh insensitivity in figures~\ref{fig:bsd_compare_powerlaw}, \ref{fig:bsd_timeavg}, and \ref{fig:bsd_powerlaw_premult_time}, as well as the agreement of their power-law scalings with prior works, supports these choices in light of this trade-off.

\subsection{The differential break-up rate $\ov{g_b f}$}\label{sec:trac-gbfe}

\begin{figure}
  \centerline{
(a)
\includegraphics[width=0.42\linewidth,valign=t]{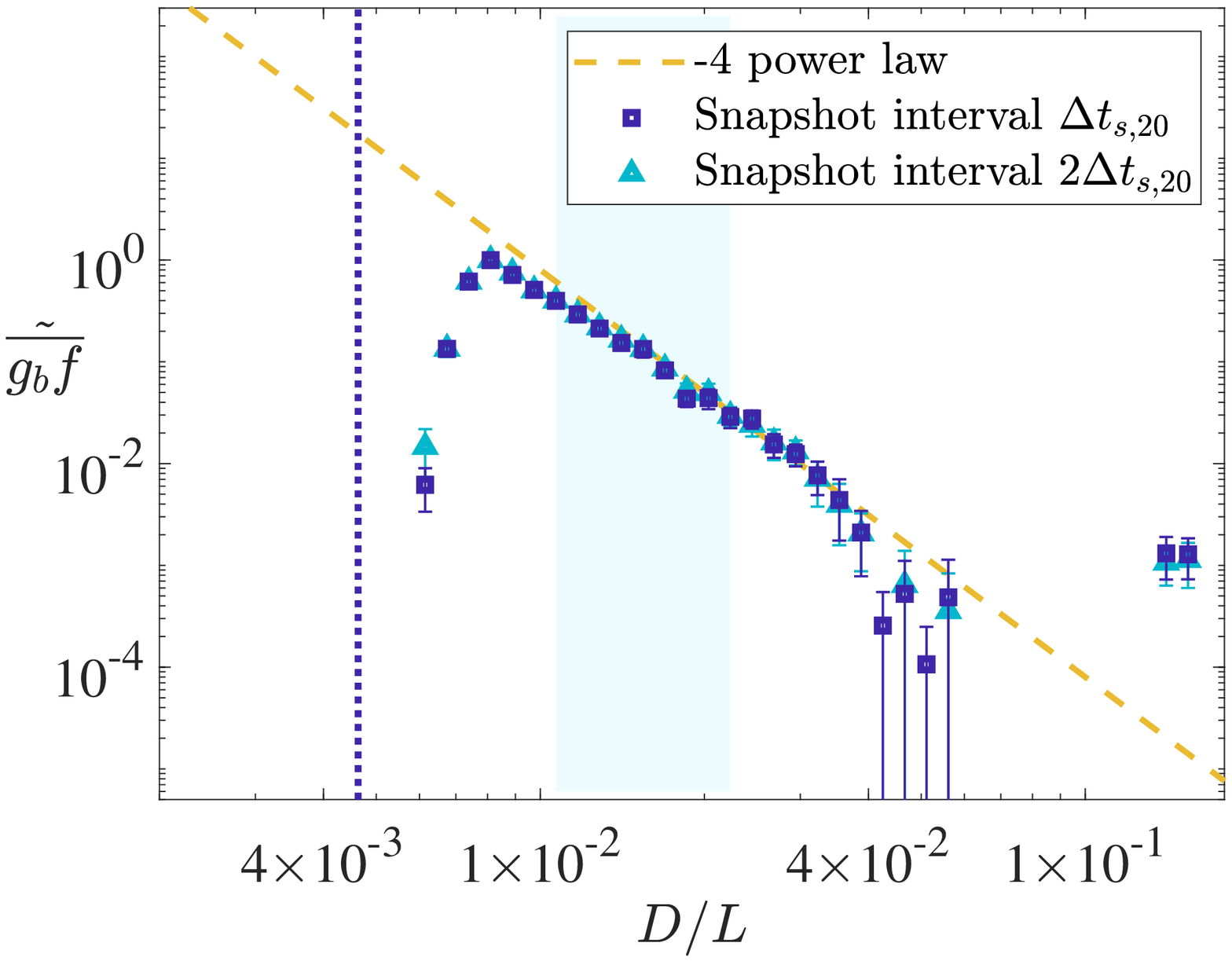}
\quad
(b)
\includegraphics[width=0.42\linewidth,valign=t]{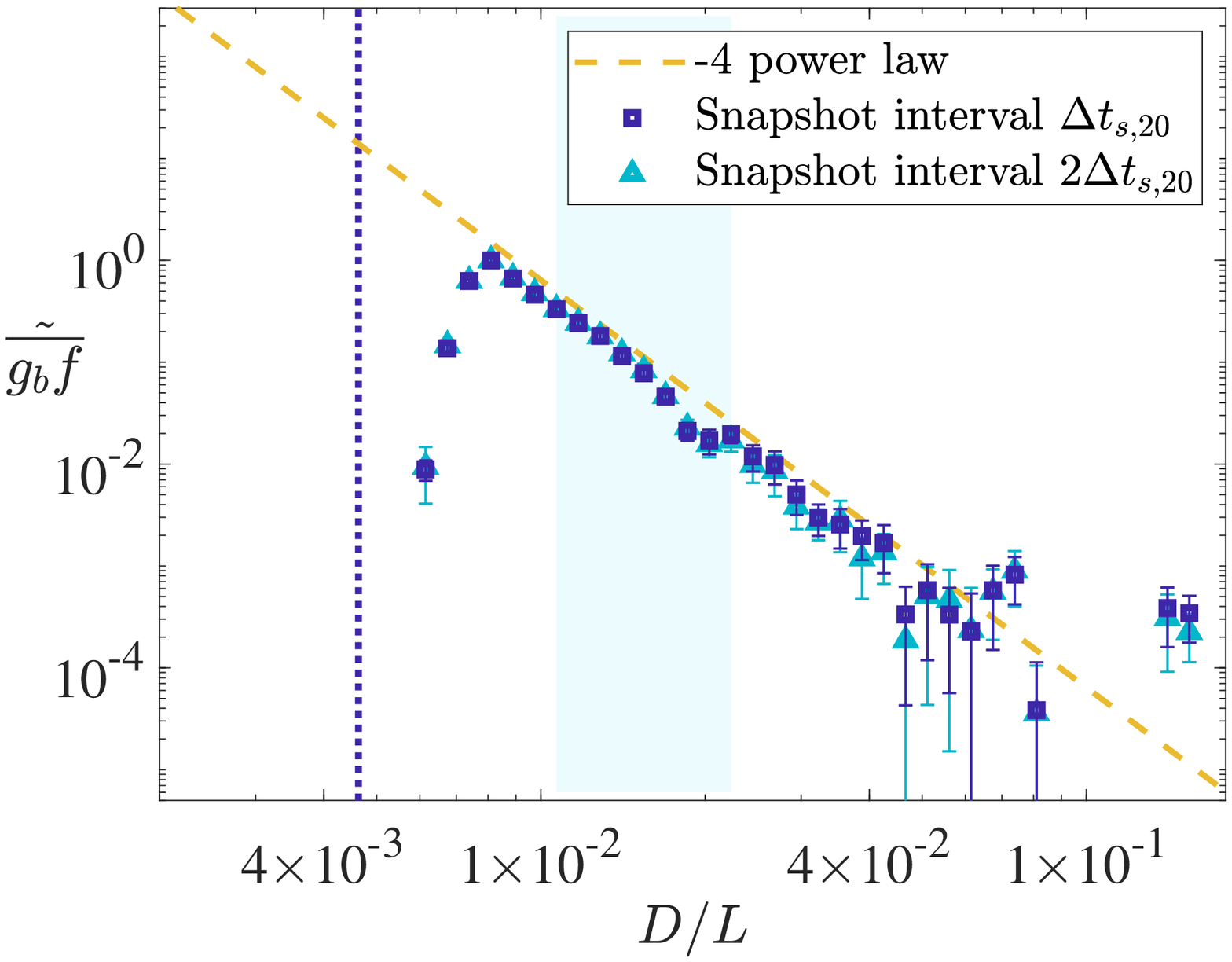}
}
  \centerline{
(c)
\includegraphics[width=0.42\linewidth,valign=t]{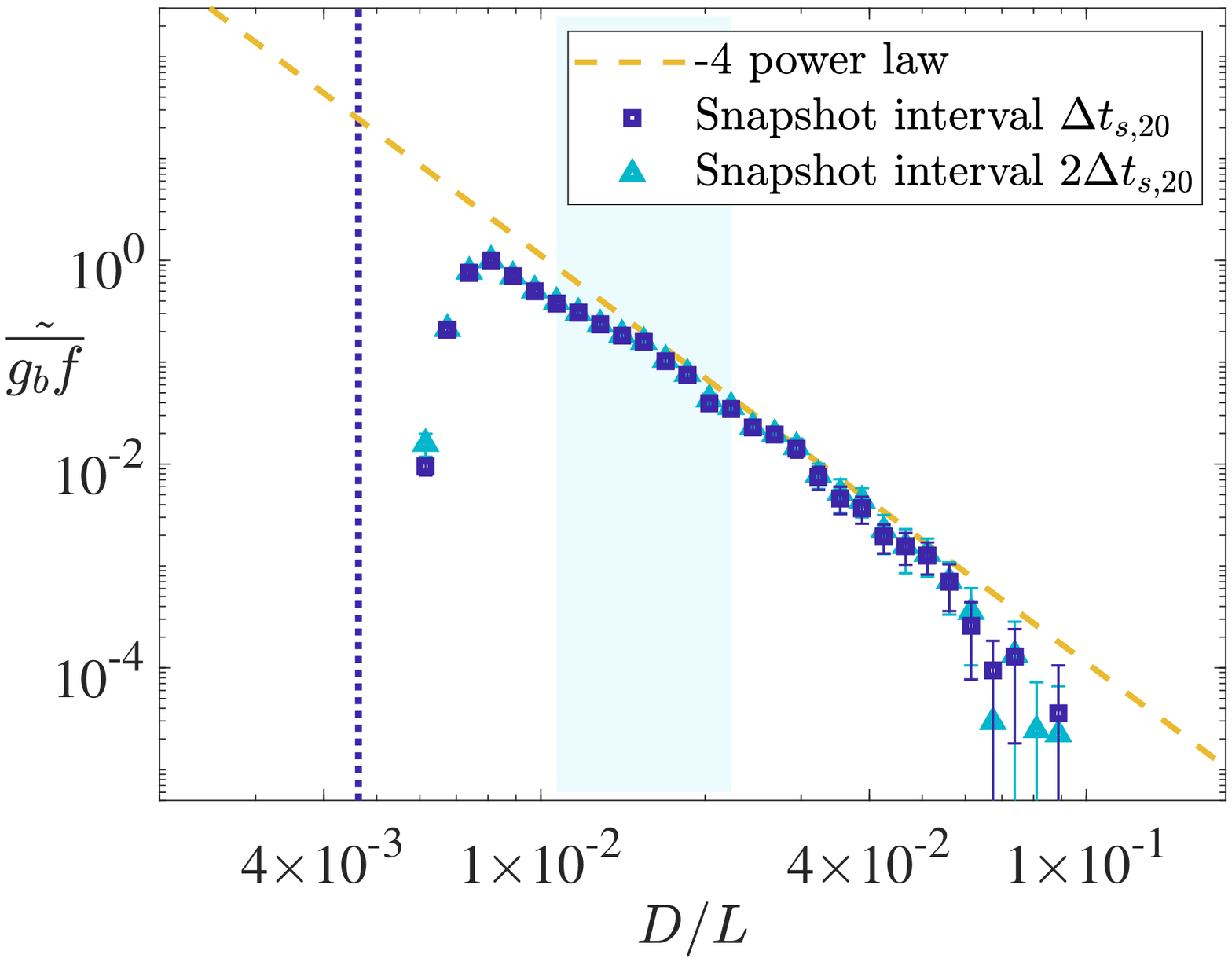}
\quad
(d)
\includegraphics[width=0.42\linewidth,valign=t]{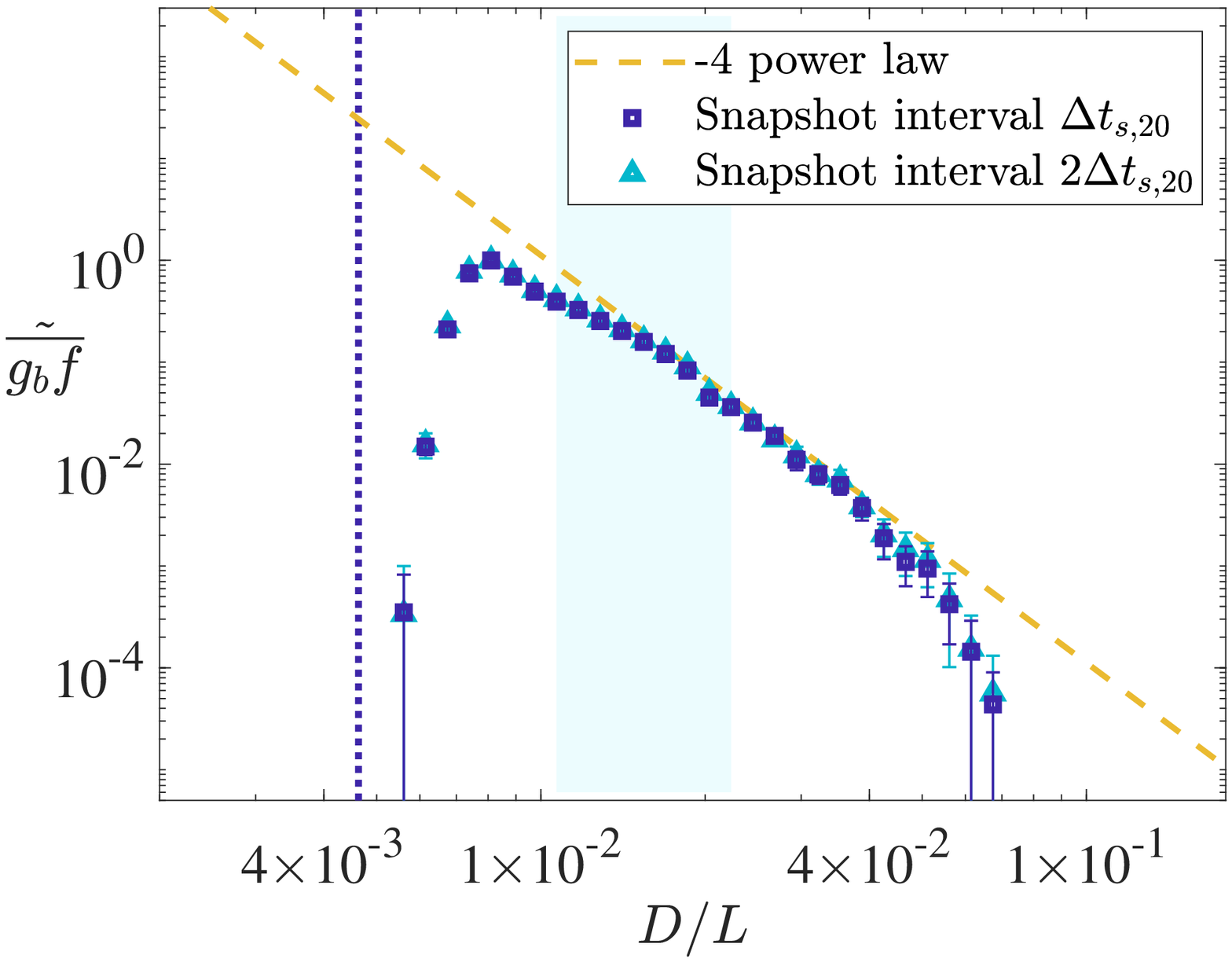}
}
  \caption{The normalized differential break-up rate $\tilde{\ov{g_b f}} (D_j/L;t)$ averaged over the non-dimensional sampling interval $16 \Delta t_{s,20} = 9.6\times10^{-2}$, plotted against non-dimensional size $D/L$ for a 20-realization baseline ensemble subset at (a) $t=2.51$, (b) $t=2.75$, (c) $t=3.71$, and (d) $t=3.95$. The rate is normalized such that the maximum value in each plot is 1, i.e., $\tilde{\ov{g_b f}} = \ov{g_b f}/\left.\ov{g_b f}\right|_\text{max}$. The selected bin sizes are twice those in figure \ref{fig:bsd_compare_powerlaw} in logarithmic space. For a description of the vertical lines and error bars, refer to the caption of figure \ref{fig:bsd_compare_powerlaw}. The dashed sloped lines indicate the theoretical $D^{-4}$ power-law scaling for an idealized turbulent bubbly flow. The shaded regions represent the bubble-size subrange over which the power-law fit exponents in figures~\ref{fig:breakup_gbfe_time} and~\ref{fig:breakup_gbfe_timeavg} were obtained. More snapshots in time are provided by~\citet{Chan8}.}
\label{fig:breakup_gbfe}
\end{figure}

In order to gain more insights into the evolution of the size distribution, $\ov{f}$, as described by \eqref{eqn:pbe_kernel} and \eqref{eqn:Tb}, the differential break-up rate $\ov{g_b f}$ is now analyzed. Break-up events were detected in the baseline ensemble using the algorithm in \S~\ref{sec:tracking}. The ensemble-averaged event rate per unit domain volume and unit size is averaged over the non-dimensional sampling interval $16 \Delta t_{s,20} = 9.6\times10^{-2}$ to converge statistics over more events. This interval is kept small enough that temporal variations are not significantly smoothed. The number of histogram bins for $\ov{g_b f}$ is half that for $\ov{f}$ to increase the bin width and improve this convergence. To ensure that the snapshot interval discussed in \S~\ref{sec:tracking} has been reasonably selected, rates obtained using two intervals, $\Delta t_{s,20}$ and 2$\Delta t_{s,20}$, are reported to demonstrate snapshot convergence. Since the variation of $\ov{g_b f}$ with parent bubble size is of interest, $\ov{g_b f}$ is normalized by its maximum value $\left(\tilde{\ov{g_b f}}\right)$ at every time instance to highlight this convergence. This normalization is also carried out in \S~\ref{sec:trac-wb}.

Figure~\ref{fig:breakup_gbfe} shows $\ov{g_b f}$ near the time instances for which $\ov{f}$ was plotted in figure~\ref{fig:bsd_compare_powerlaw}. Signatures of the $D^{-4}$ scaling derived for a quasi-steady turbulent break-up cascade in Part 1 emerge at intermediate sizes. Note that the computation of $\ov{g_b f}$ samples energetic regions more frequently since break-up only occurs with sufficient energy to change the bubble topology. Thus, $\ov{g_b f}$ is susceptible to large-scale inhomogeneities, and is difficult to converge since the scale separation in the simulated waves is not exceedingly large.

\begin{figure}
  \centerline{
\includegraphics[width=0.52\linewidth]{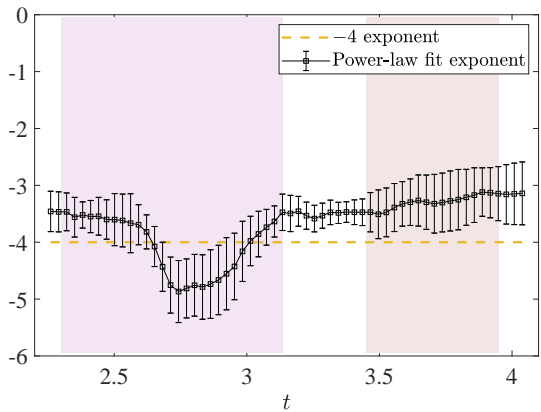}
}
  \caption{The variation of the exponent of a power-law fit over a subset of the differential break-up rate, $\ov{g_b f}$, in figure~\ref{fig:breakup_gbfe} with non-dimensional time. The fit was performed on the data with a snapshot interval of $\Delta t_{s,20}$. For a description of the bubble-size subrange over which the fit was performed, error bars, and shaded regions, refer to the caption of figure~\ref{fig:bsd_powerlaw_time}.} 
\label{fig:breakup_gbfe_time}
\end{figure}

\begin{figure}
  \centerline{
(a)
\includegraphics[width=0.42\linewidth,valign=t]{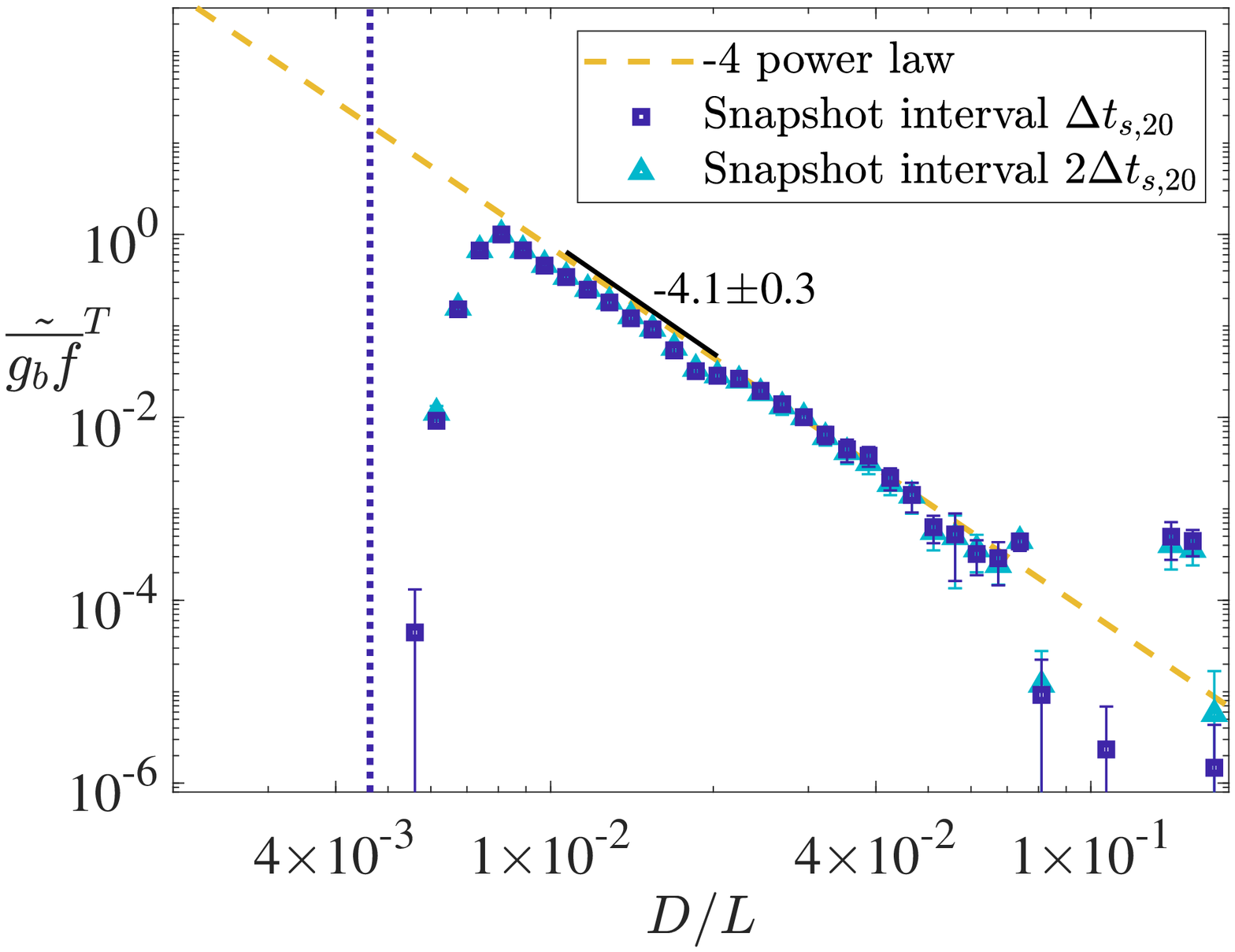}
\quad
(b)
\includegraphics[width=0.42\linewidth,valign=t]{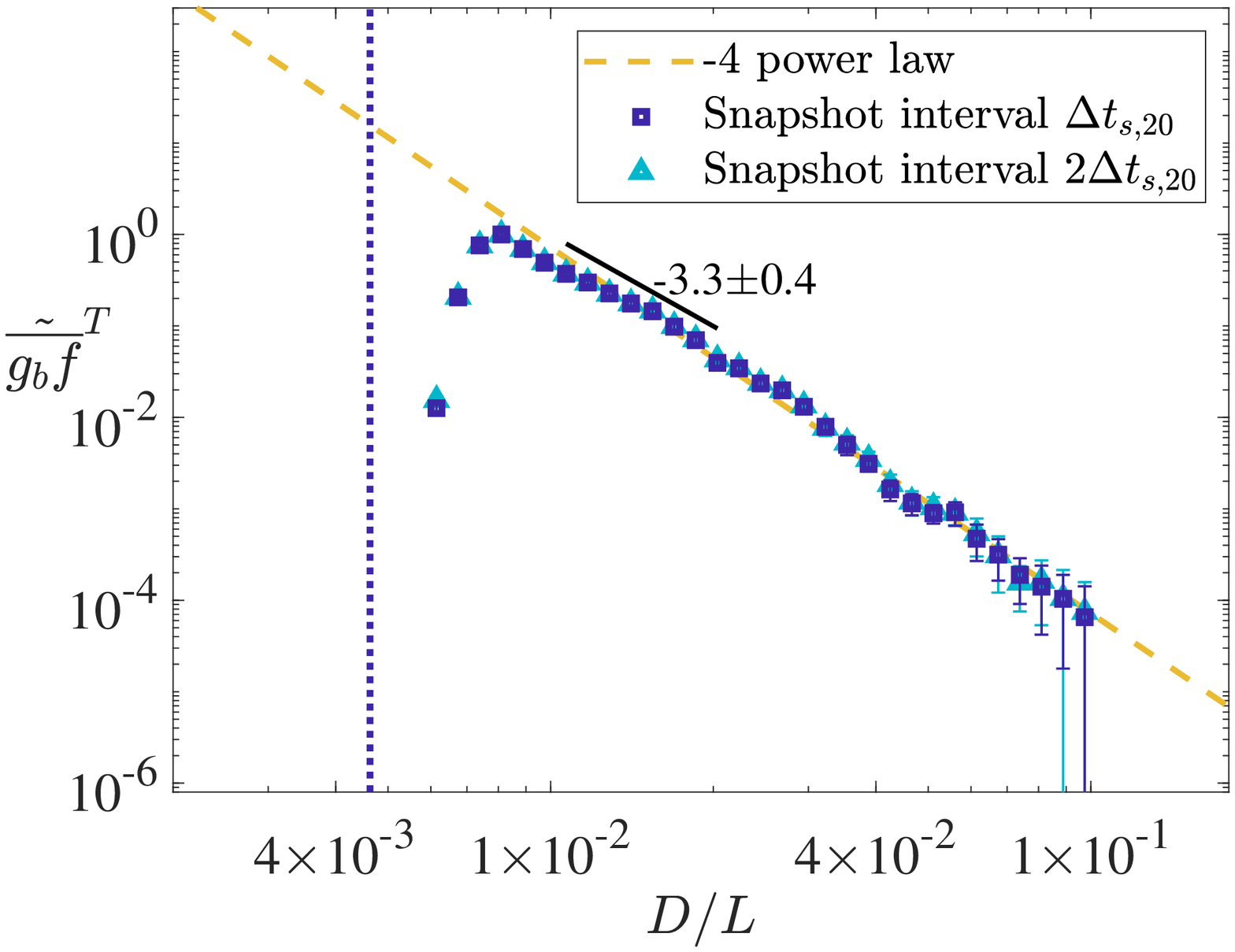}
}
  \caption{The normalized time-averaged differential break-up rate $\tilde{\ov{g_b f}}^T (D_j/L)$ against non-dimensional size $D/L$ for the 20-realization baseline ensemble subset. Subfigures (a) and (b) correspond to the time-averaging intervals in figure~\ref{fig:bsd_timeavg}. For a description of the vertical and sloped lines, error bars, normalization, bubble-size subrange and dataset over which the fits were performed, and fit-exponent uncertainties, refer to the captions of figures~\ref{fig:bsd_compare_powerlaw}, \ref{fig:bsd_timeavg}, \ref{fig:breakup_gbfe}, and~\ref{fig:breakup_gbfe_time}.} 
\label{fig:breakup_gbfe_timeavg}
\end{figure}

Figure~\ref{fig:breakup_gbfe_time} quantifies the evolution of the exponent of a power-law fit on the differential rate in figure~\ref{fig:breakup_gbfe} over a subrange of intermediate sizes, relative to the $-4$ exponent derived in Part 1 using traditional turbulent-flow scalings. The behaviour of the exponent in figure~\ref{fig:breakup_gbfe_time} for $\ov{g_b f}$ loosely tracks the corresponding behaviour in figure~\ref{fig:bsd_powerlaw_time} for $\ov{f}$: in the early wave-breaking stages, the exponent oscillates around the exponent discussed in Part 1; in the late stages, it does not strongly oscillate. As with the size distribution in figure~\ref{fig:bsd_timeavg}, the differential rate may be time averaged over these two intervals, as depicted in figure~\ref{fig:breakup_gbfe_timeavg}. In the first interval, it may be described by a $D^{-4.1\pm0.3}$ power-law fit, in agreement with $D^{-4}$. In the second interval, it may be described by a $D^{-3.3\pm0.4}$ fit, which is about a factor of $D^{-2/3}$ steeper than the $D^{-8/3}$ scaling in the corresponding size distribution, although vestiges of the $D^{-4}$ scaling remain in the differential rate at larger sizes. These power-law scalings at intermediate sizes are largely consistent with the turbulent scaling of the break-up frequency, $g_b \sim D^{-2/3}$, in agreement with the references in \S~\ref{sec:intro} that were used in the theoretical analysis of Part 1. 

In summary, the trends observed in $\ov{g_b f}$ mostly echo those for $\ov{f}$ in \S~\ref{sec:sizedistevol}. The unusual persistence of the $D^{-4}$ scaling may be suggestive of a tendency of the bubble-mass flux to remain quasi-local and quasi-self-similar, since $\ov{W_b} \sim \ov{g_b f} D^4$ in this limit as shown in Part 1. These characteristics of the bubble-mass flux are further addressed in \S~\ref{sec:trac-wb}. Analogous to \S~\ref{sec:sizedistevol}, future ensembles of higher $\RR_L$ and $\We_L$ simulations could bring about more clarity on the robustness of these scalings.

\subsection{The break-up probability distribution $\check{q}_b$}\label{sec:trac-qb}

In addition to $\ov{g_b f}$, the probability distribution of child bubble sizes, $\check{q}_b$, also drives the evolution of $\ov{f}$, as described by \eqref{eqn:pbe_kernel} and \eqref{eqn:Tb}, and may also be computed by the algorithm in \S~\ref{sec:tracking}. The distribution $\check{q}_b\left(D_c^3;t|D_p^3\right)$ is symmetric in bubble-volume space and is presented in this space for a more intuitive interpretation. It satisfies
\begin{equation}
2 = \sum_{j=1}^{N_\text{bin}} \check{q}_b\left(D_{c,j}^3;t|D_p^3\right) \Delta\left(D_{c,j}^3/D_p^3\right).
\end{equation}
The data for each parent bubble size $D_{p,j}$ is averaged over two histogram bins $[D_{p,j},D_{p,j+2})$, where the bins are identical to those in \S~\ref{sec:trac-gbfe}. The reported $\check{q}_b$ is time-averaged over the two time intervals identified in \S~\ref{sec:waveevol}, \S~\ref{sec:sizedistevol}, and \S~\ref{sec:trac-gbfe}.

\begin{figure}
  \centerline{
(a)
\includegraphics[width=0.42\linewidth,valign=t]{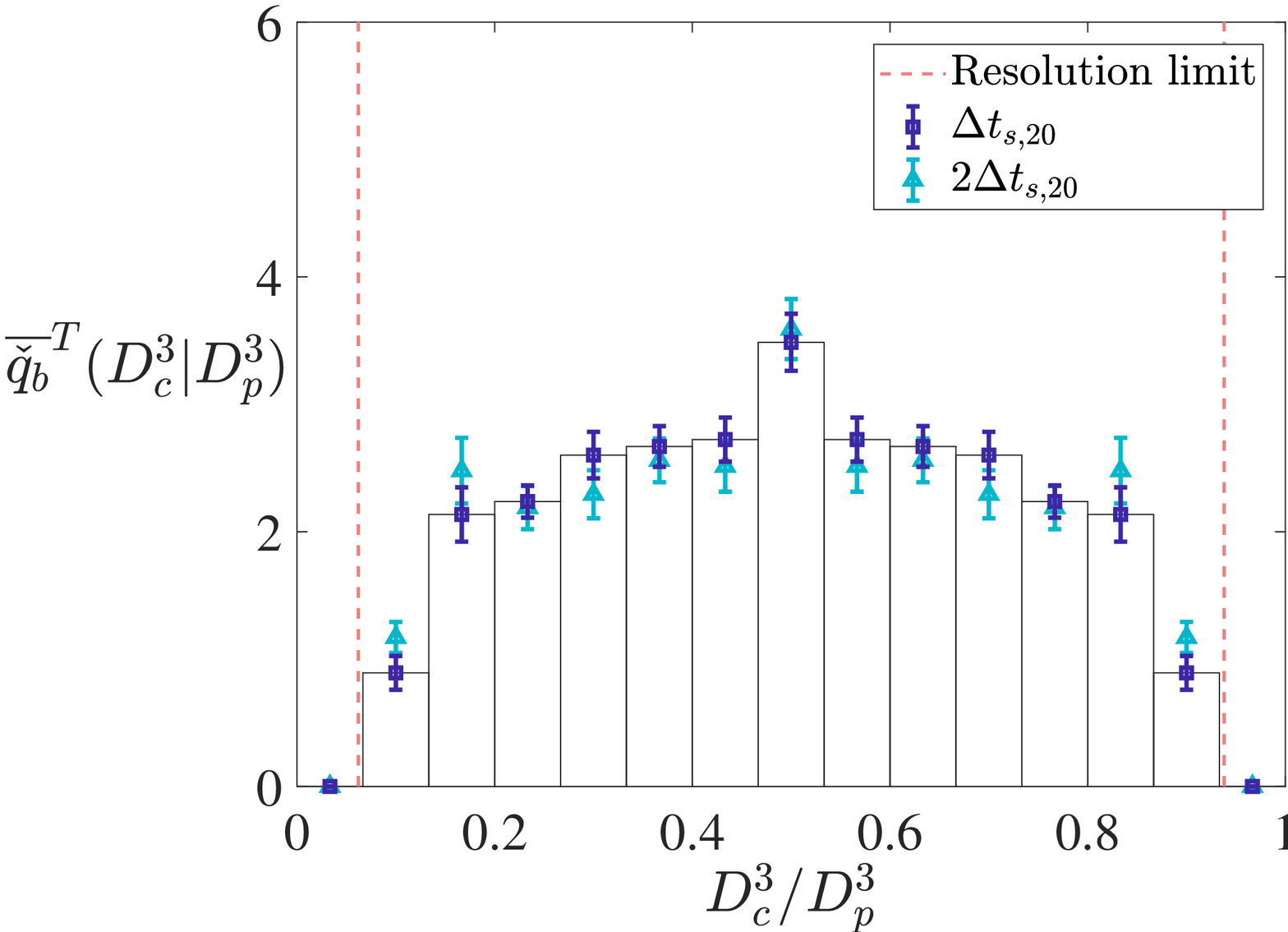}
\quad
(b)
\includegraphics[width=0.42\linewidth,valign=t]{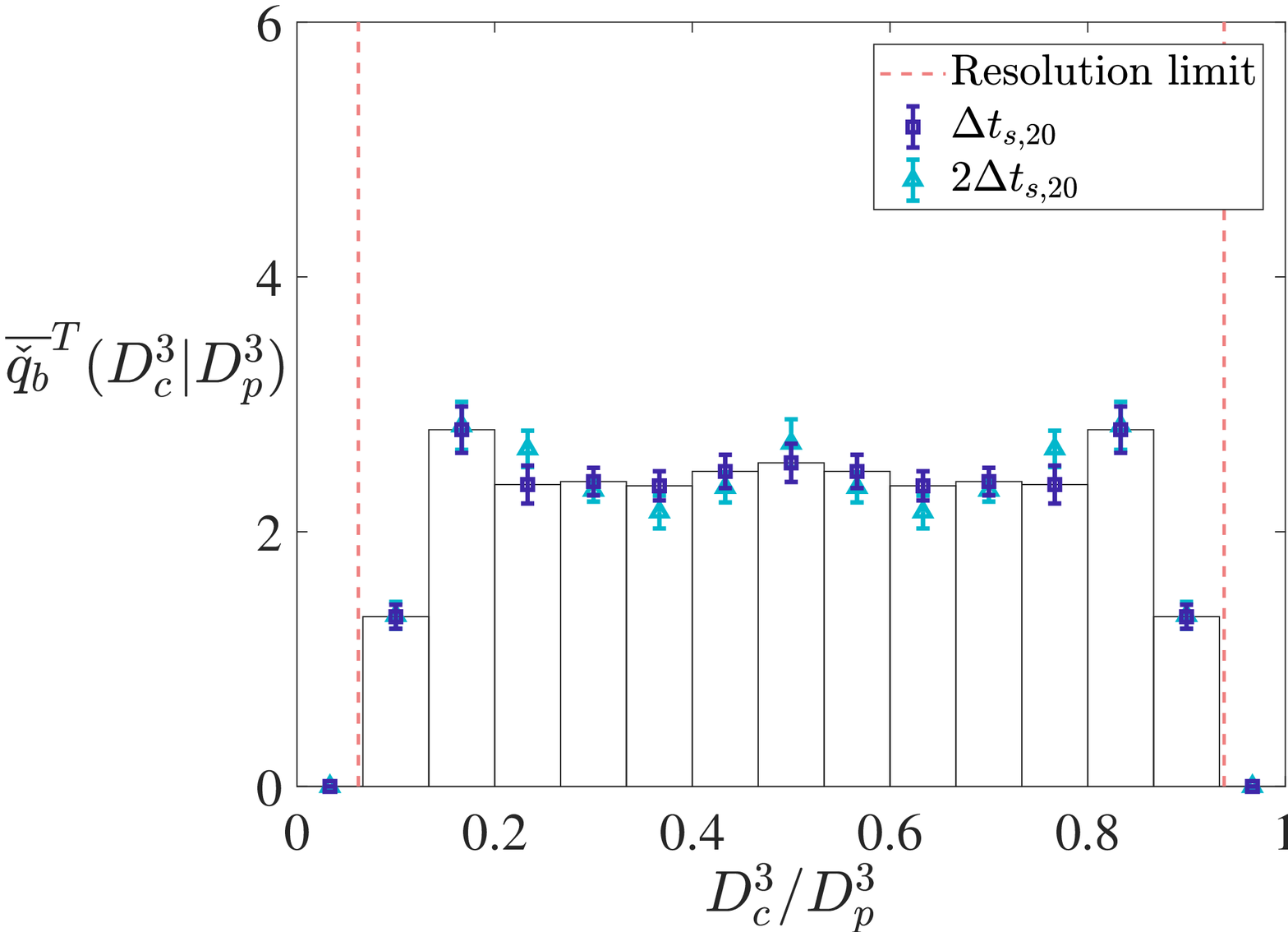}
}
  \centerline{
(c)
\includegraphics[width=0.42\linewidth,valign=t]{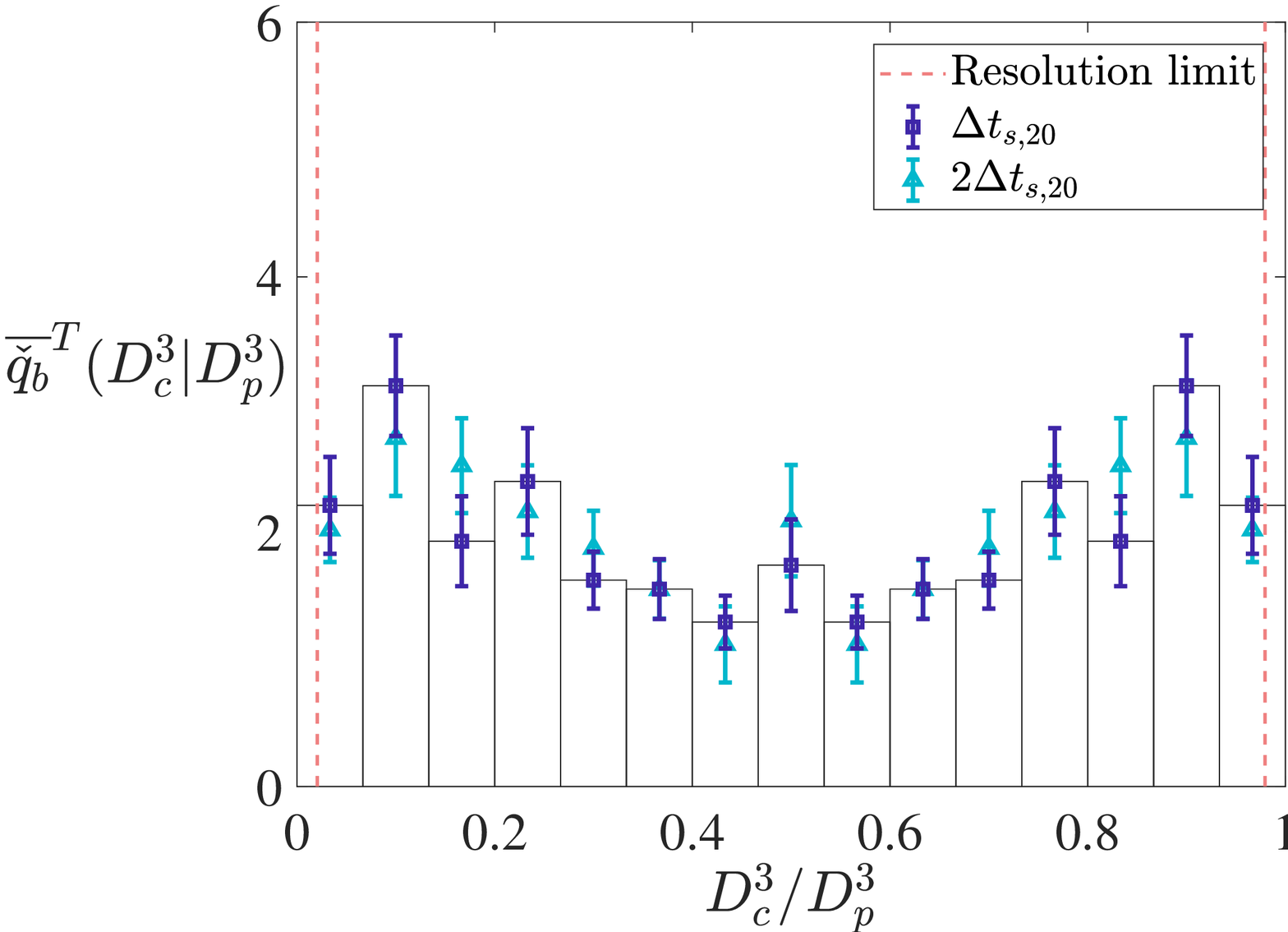}
\quad
(d)
\includegraphics[width=0.42\linewidth,valign=t]{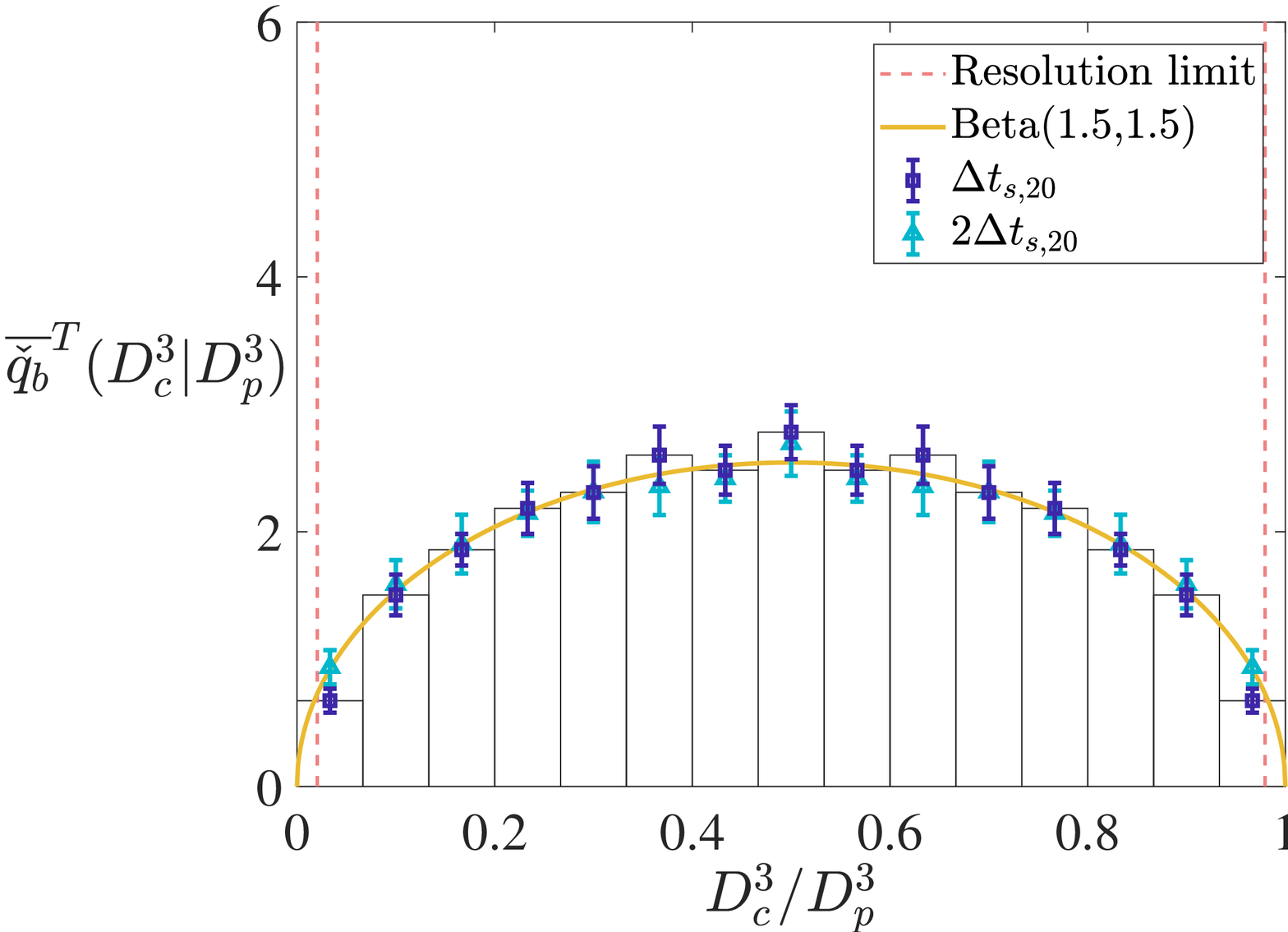}
}
  \centerline{
(e)
\includegraphics[width=0.42\linewidth,valign=t]{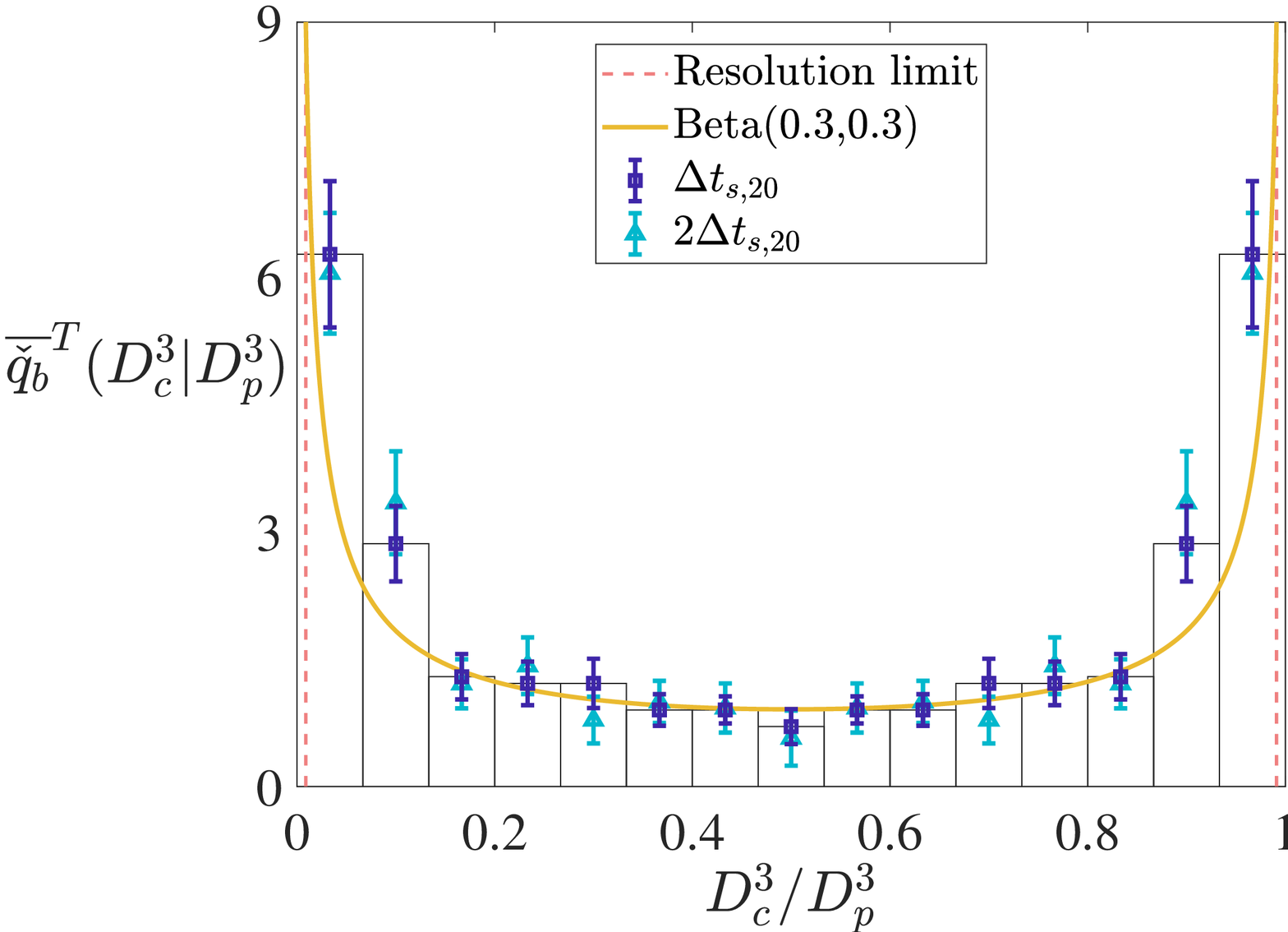}
\quad
(f)
\includegraphics[width=0.42\linewidth,valign=t]{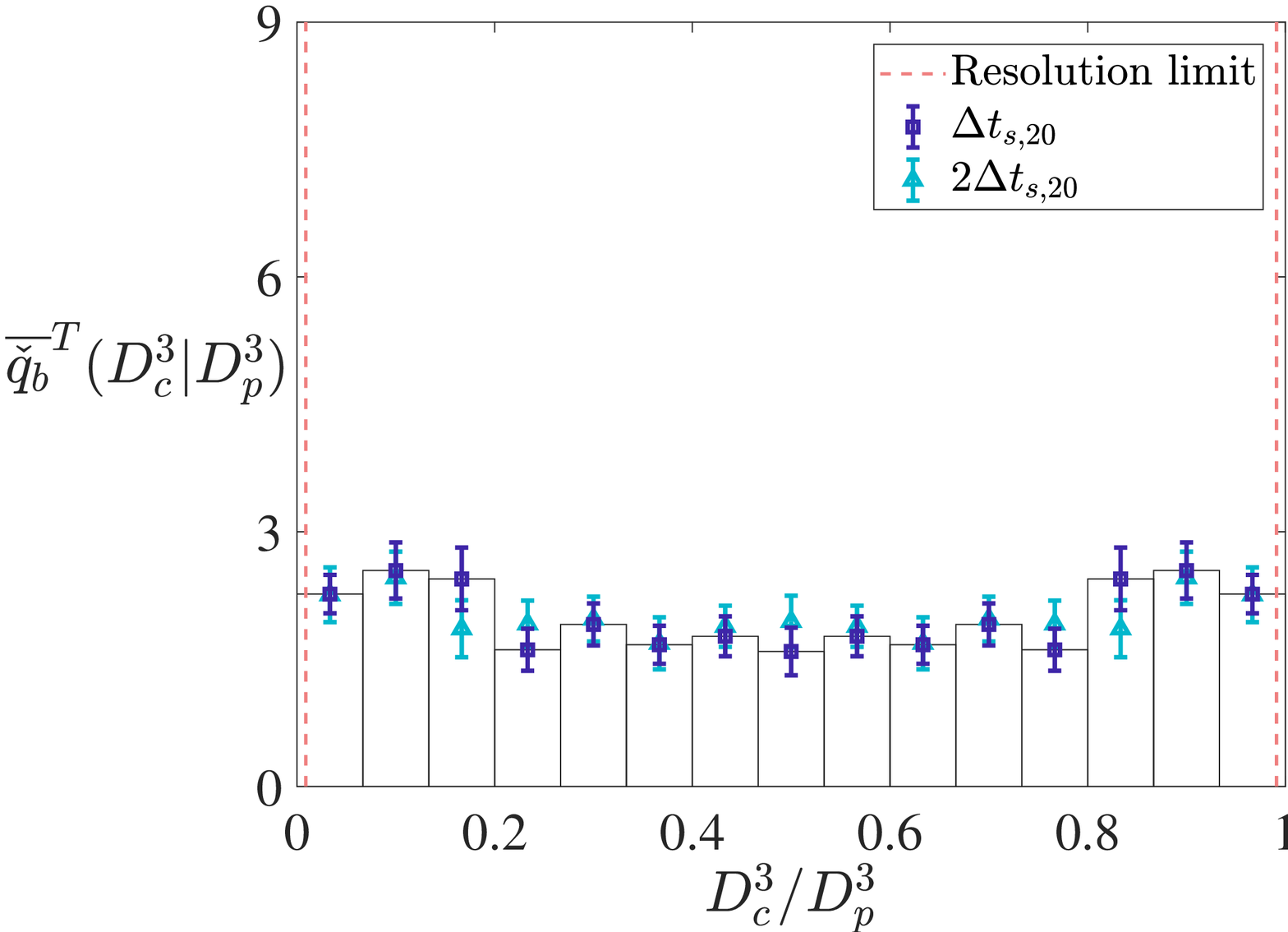}
}
  \caption{The time-averaged break-up probability $\ov{\check{q}_b}^T\left(D_{c,j}^3|D_p^3\right)$ against normalized child bubble volume $D_c^3/D_p^3$ for the 20-realization baseline ensemble subset during the (a,c,e) early ($t=2.30$--$3.14$) and (b,d,f) late ($t=3.45$--$3.95$) wave-breaking stages. The non-dimensional parent bubble sizes considered are (a,b) between $1.86 \times 10^{-2}$ $\left(\We_{D_p} = 18\right)$ and $2.23 \times 10^{-2}$ $\left(\We_{D_p} = 25\right)$, (c,d) between $2.68 \times 10^{-2}$ $\left(\We_{D_p} = 33\right)$ and $3.23 \times 10^{-2}$ $\left(\We_{D_p} = 45\right)$, and (e,f) between $3.54 \times 10^{-2}$ $\left(\We_{D_p} = 53\right)$ and $4.25 \times 10^{-2}$ $\left(\We_{D_p} = 72\right)$. For the definition of $\We_{D_p}$, refer to (2.4) in Part 1. Only children bubbles of radii larger than the mesh resolution are considered. The vertical dashed lines demarcate the child bubble volumes in the break-up event where one of the child bubble radii corresponds to this resolution limit. The error bars denote one standard error. The time intervals in the legends denote the snapshot intervals in the detection algorithm. Fits to the beta distribution have been added in (d) and (e). These fits were performed on the data obtained by taking the snapshot interval to be $\Delta t_{s,20}$, which was also used to plot the histogram bars. The two numbers in the corresponding legend entries are the shape parameters characterizing each fit, which were obtained via the method of moments.}
\label{fig:breakup_qb}
\end{figure}

Figure~\ref{fig:breakup_qb} plots the time-averaged break-up probability distribution, $\ov{\check{q}_b}^T$, for three distinct parent bubble sizes and the two aforementioned time intervals. Each row corresponds to a single parent bubble size, while each column corresponds to a single time interval. Some child bubble sizes are smaller than the mesh resolution and have to be excluded. Under the influence of this exclusion, $\ov{\check{q}_b}^T$ falls off towards large and small child bubble sizes for small parent bubbles. 

In general, the child bubble volume distribution is relatively close to uniform for events within the intermediate bubble-size subrange where power-law fits were earlier obtained for $\ov{f}$ and $\ov{g_b f}$ in \S~\ref{sec:sizedistevol} and \S~\ref{sec:trac-gbfe}. Hence, the uniform distribution, which was analyzed in Part 1, appears to be a suitable, albeit rudimentary, surrogate model for $\ov{\check{q}_b}^T$ in most cases. One exception occurs at large parent bubble sizes outside the aforementioned size subrange in the early wave-breaking stages. Here, large-size-ratio events involving one large child bubble and one small child bubble do frequently occur as evidenced in figure~\ref{fig:breakup_qb}(e), bypassing the break-up cascade and generating small bubbles as observed in \S~\ref{sec:sizedistevol}. These individual non-local contributions to bubble-mass transfer are small in volume and may not necessarily influence the locality of the break-up flux, $\ov{W_b}$, as noted in appendix B of Part 1. Their relative influence will be analyzed in more detail in \S~\ref{sec:trac-wb-locality}. Another exception occurs in the late wave-breaking stages in figure~\ref{fig:breakup_qb}(d). In these two cases, the break-up probability is reasonably described by a beta-distribution fit, which was also discussed in Part 1 as a potential surrogate model. These exceptions indicate that it does not seem appropriate to characterize a turbulent bubbly flow with a single distribution shape without giving more consideration to the bubble-size subrange or flow stage of interest, as many previous studies have done. For example, experimental studies typically report a single distribution of child bubble volumes, agglomerating data from different parent bubble sizes [see figure 1 of~\citet{Qi1}, and references therein]. This has contributed to a multitude of existing models for the child bubble volume distribution, as noted in Part 1. Interestingly, the results of this work are in qualitative agreement with the model of~\citet{Lehr1}, where ``equal sized breakage is preferred for small bubbles \ldots as the size of the parent bubbles increases, unequal breakage is preferred".

\subsection{The break-up flux $\ov{W_b}$}\label{sec:trac-wb}

The results of \S~\ref{sec:sizedistevol} revealed that the bubble size distribution dramatically evolves between $t=2$ and $t=4$. Indirect evidence of a quasi-steady bubble break-up cascade~\citep{Garrett1} was observed in the early wave-breaking stages, although multiple mechanisms could explain the observed $D^{-10/3}$ power-law scaling in the size distribution~\citep[e.g.,][]{Yu1,Yu2}. The present simulations, with the accompanying detection algorithm for break-up events, also provide detailed break-up statistics, which were analyzed in \S~\ref{sec:trac-gbfe} and \S~\ref{sec:trac-qb}. These statistics contribute to the break-up flux, $\ov{W_b}(D;t)$, as discussed in Part 1 and summarized in \S~\ref{sec:bridging_pbe}, and provide further indirect evidence of a break-up cascade. The break-up flux across size $D$ can be directly decomposed into incoming contributions $\ov{I_p}\left(D_p;t|D\right)$ from parent bubbles of sizes $D_p > D$ and outgoing contributions $\ov{I_c}\left(D_c;t|D\right)$ to children bubbles of sizes $D_c < D$, as shown in \S~\ref{sec:bridging_locality}, as well as figures 4 and 5 of Part 1. These complementary decompositions respectively provide information on infrared and ultraviolet locality in the flux. Satisfying these measures of locality is a necessary condition for the presence of a break-up cascade, with $\ov{I_p}$ and $\ov{I_c}$ decaying rapidly away from $D$. Cascade-like behaviour may also be directly characterized by size invariance and quasi-stationarity of the break-up flux at intermediate sizes. These observations are direct consequences of the theoretical analysis performed in Part 1. In this subsection, the simulation results are further leveraged by examining these quantities to obtain more direct observations of cascade-like behaviour.

\subsubsection{Infrared and ultraviolet locality}\label{sec:trac-wb-locality}

\begin{figure}
  \centerline{
(a)
\includegraphics[width=0.42\linewidth,valign=t]{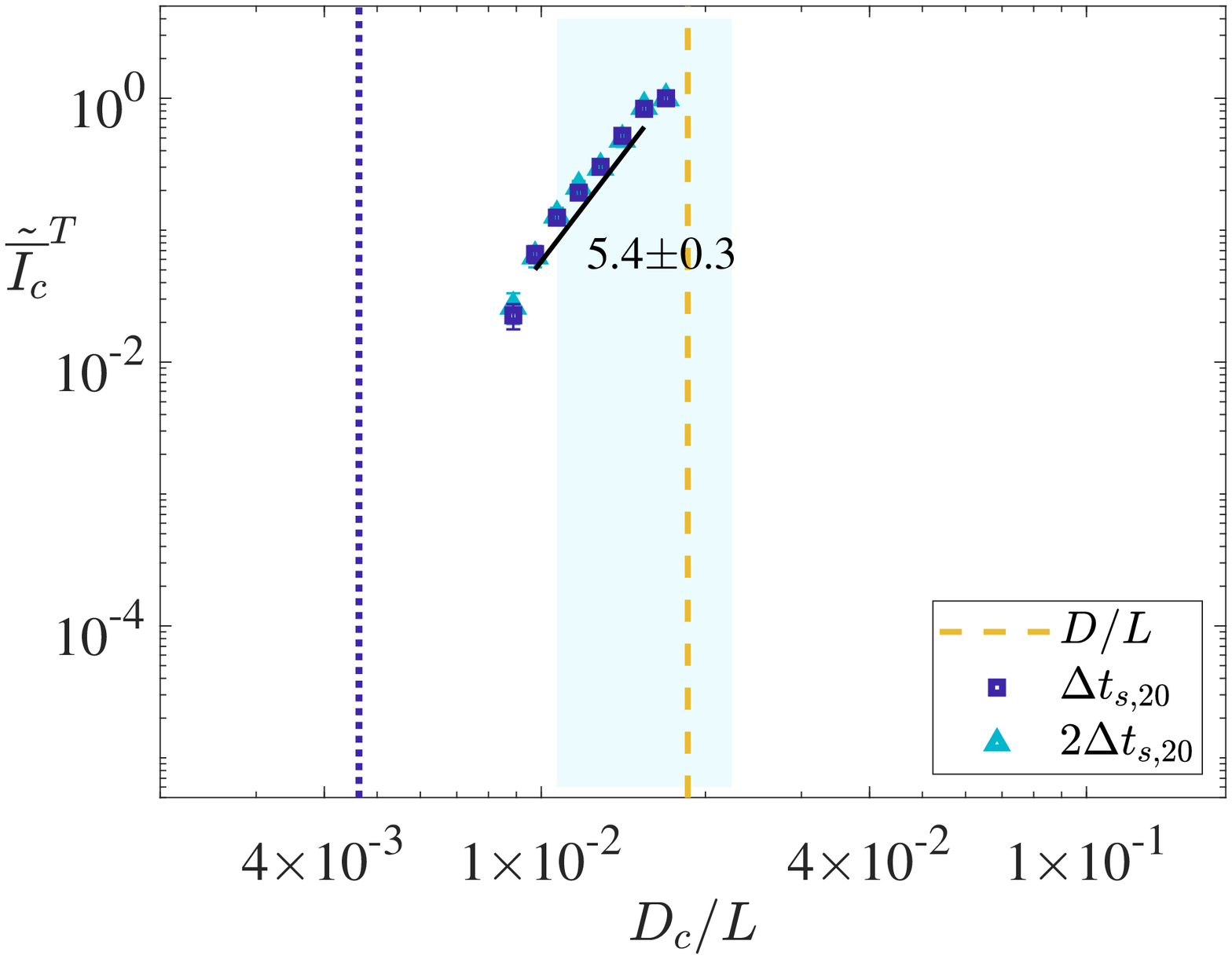}
\quad
(b)
\includegraphics[width=0.42\linewidth,valign=t]{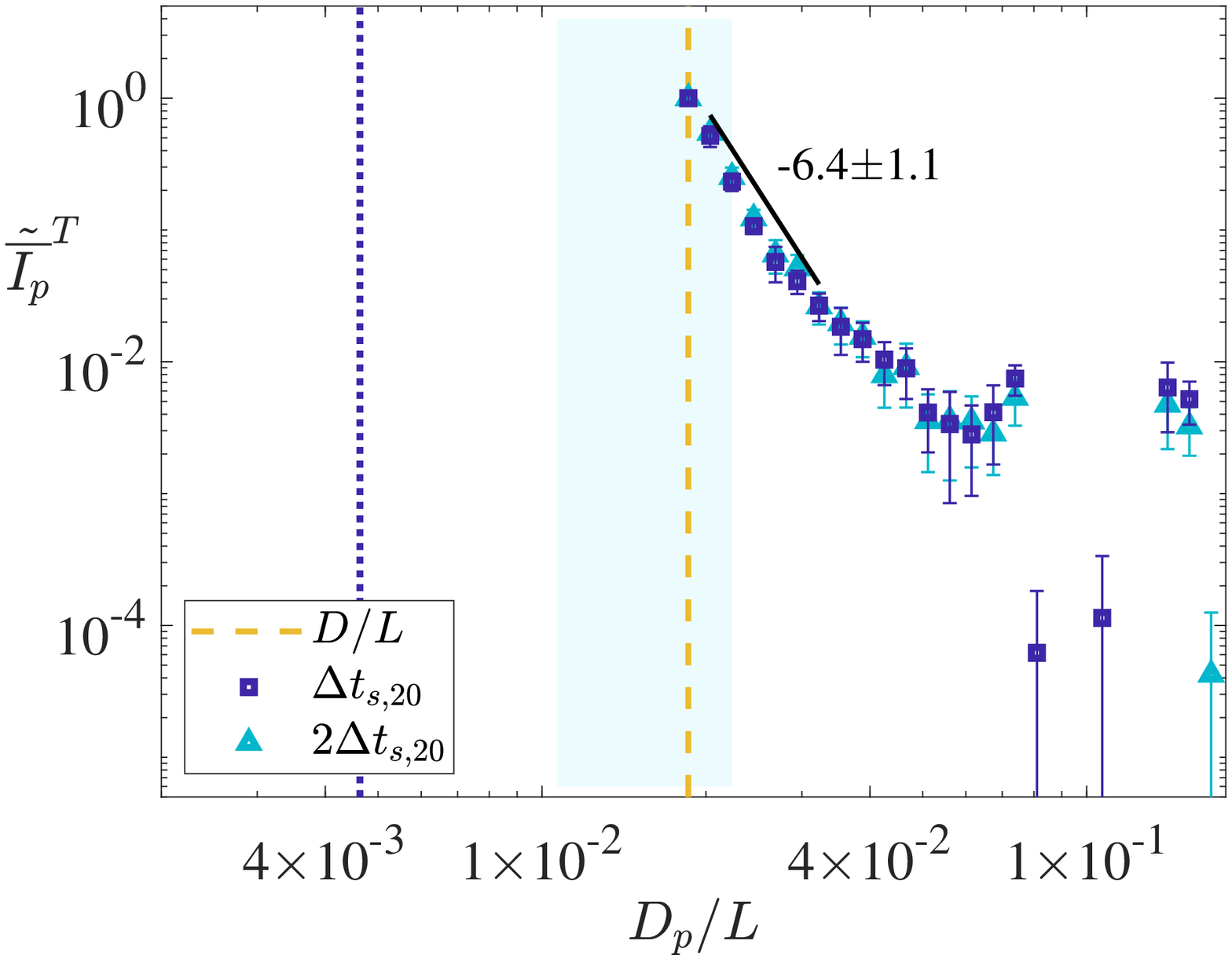}
}
  \centerline{
(c)
\includegraphics[width=0.42\linewidth,valign=t]{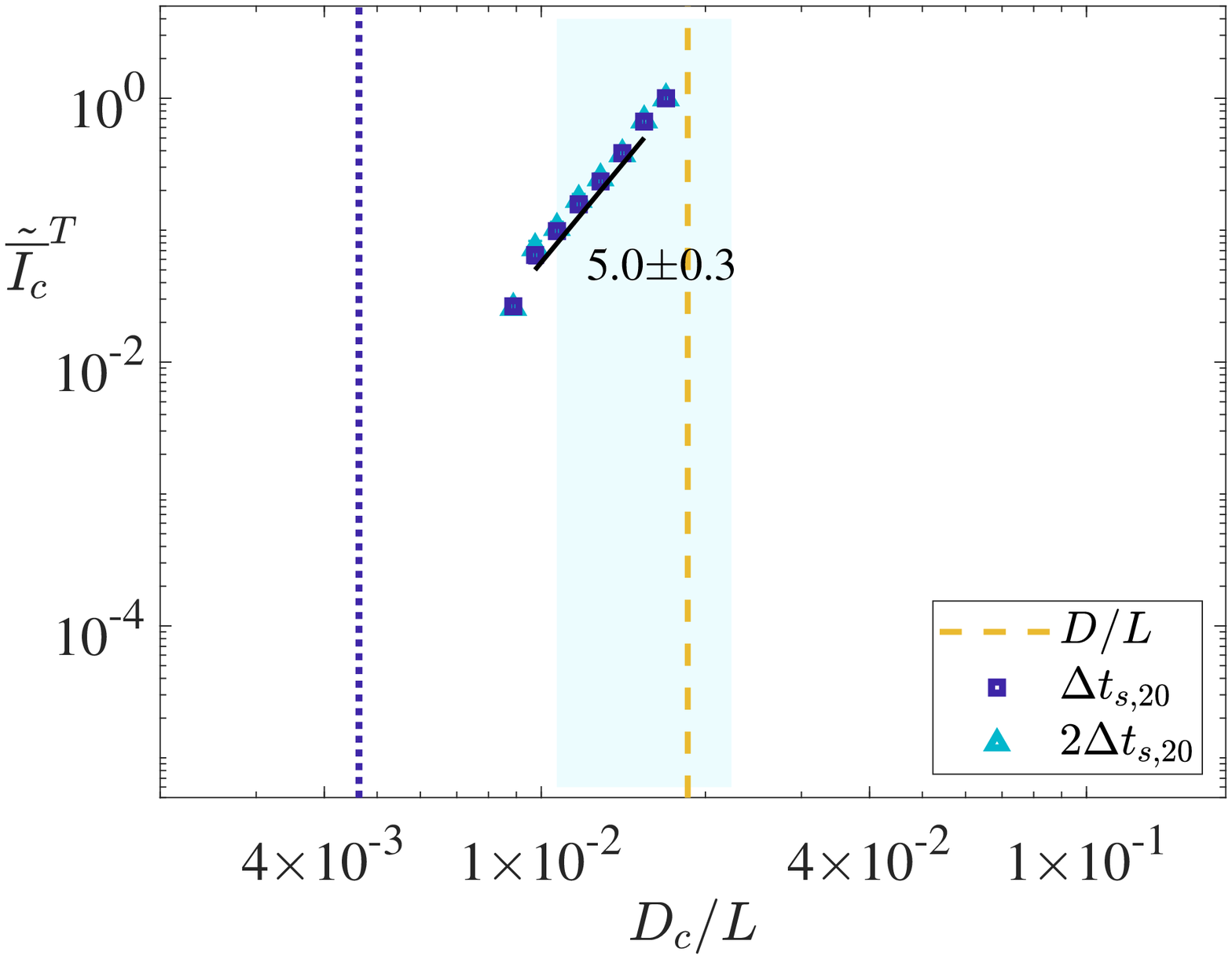}
\quad
(d)
\includegraphics[width=0.42\linewidth,valign=t]{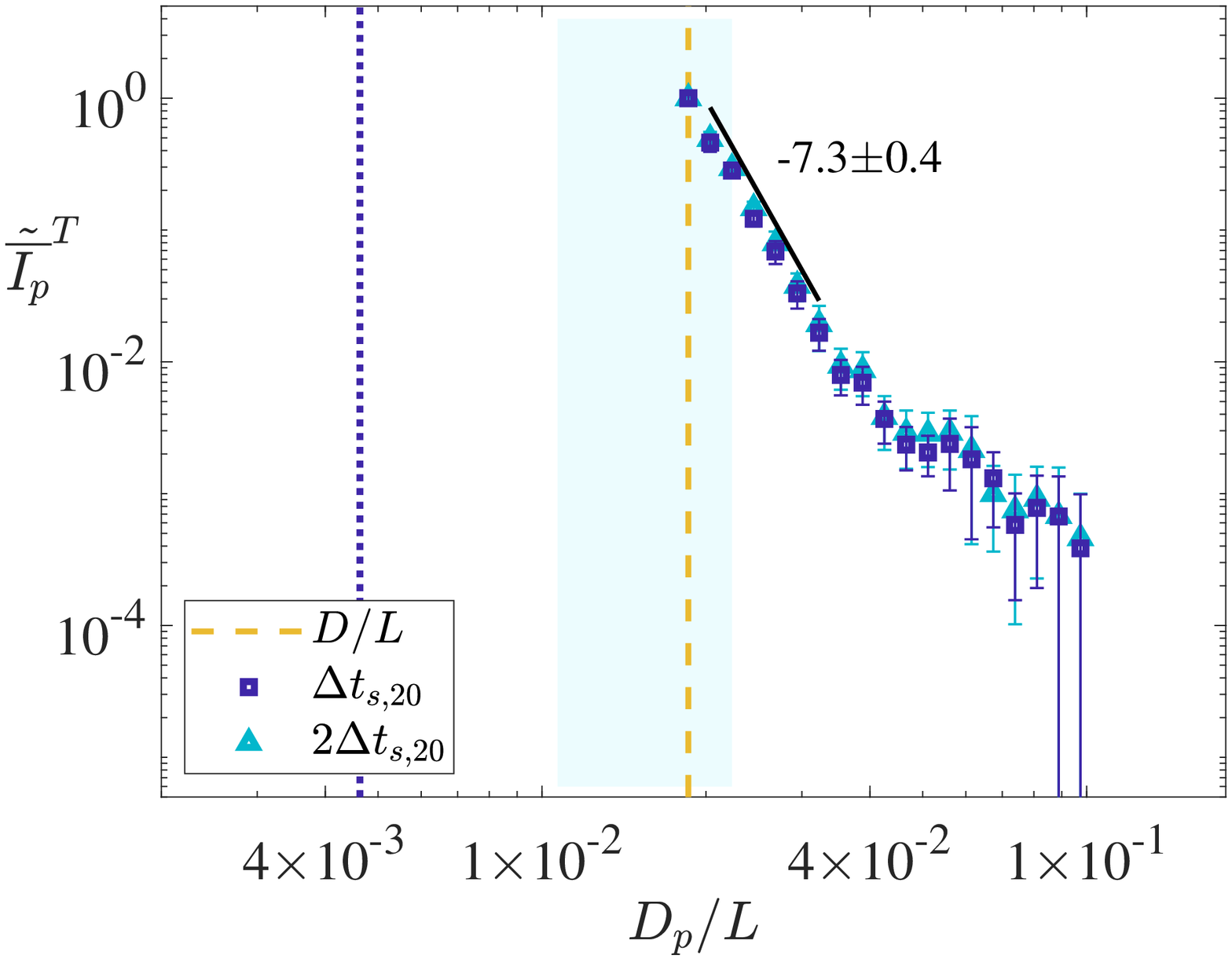}
}
  \caption{(a,c) The normalized time-averaged differential outgoing transfer rate $\tilde{\ov{I_c}}^T\left(D_{c,j}|D\right)$ to children bubbles as a function of non-dimensional child bubble size $D_c/L$, and (b,d) the corresponding differential incoming transfer rate $\tilde{\ov{I_p}}^T\left(D_{p,j}|D\right)$ from parent bubbles as a function of non-dimensional parent bubble size $D_p/L$, for the 20-realization baseline ensemble subset at the non-dimensional cut-off size $D/L = 1.86\times10^{-2}$ marked by the dashed vertical line. These statistics are time averaged over the (a,b) early ($t=2.30$--$3.14$) and (c,d) late ($t=3.45$--$3.95$) wave-breaking stages. The normalization, histogram bins, shaded regions, and dotted vertical lines are identical to those in figure~\ref{fig:breakup_gbfe}. Break-up events involving subgrid children bubbles are excluded. The error bars and fit-exponent uncertainties denote twice the standard error over the ensemble. The power-law fits were performed over bubbles with non-dimensional diameters between $9.74\times10^{-3}$ and $1.69\times10^{-2}$ for $\tilde{\ov{I_c}}^T$, and between $2.04\times10^{-2}$ and $3.54\times10^{-2}$ for $\tilde{\ov{I_p}}^T$, on the dataset with snapshot interval $\Delta t_{s,20}$.}
\label{fig:breakup_i}
\end{figure}

\begin{figure}
  \centerline{
(a)
\includegraphics[width=0.42\linewidth,valign=t]{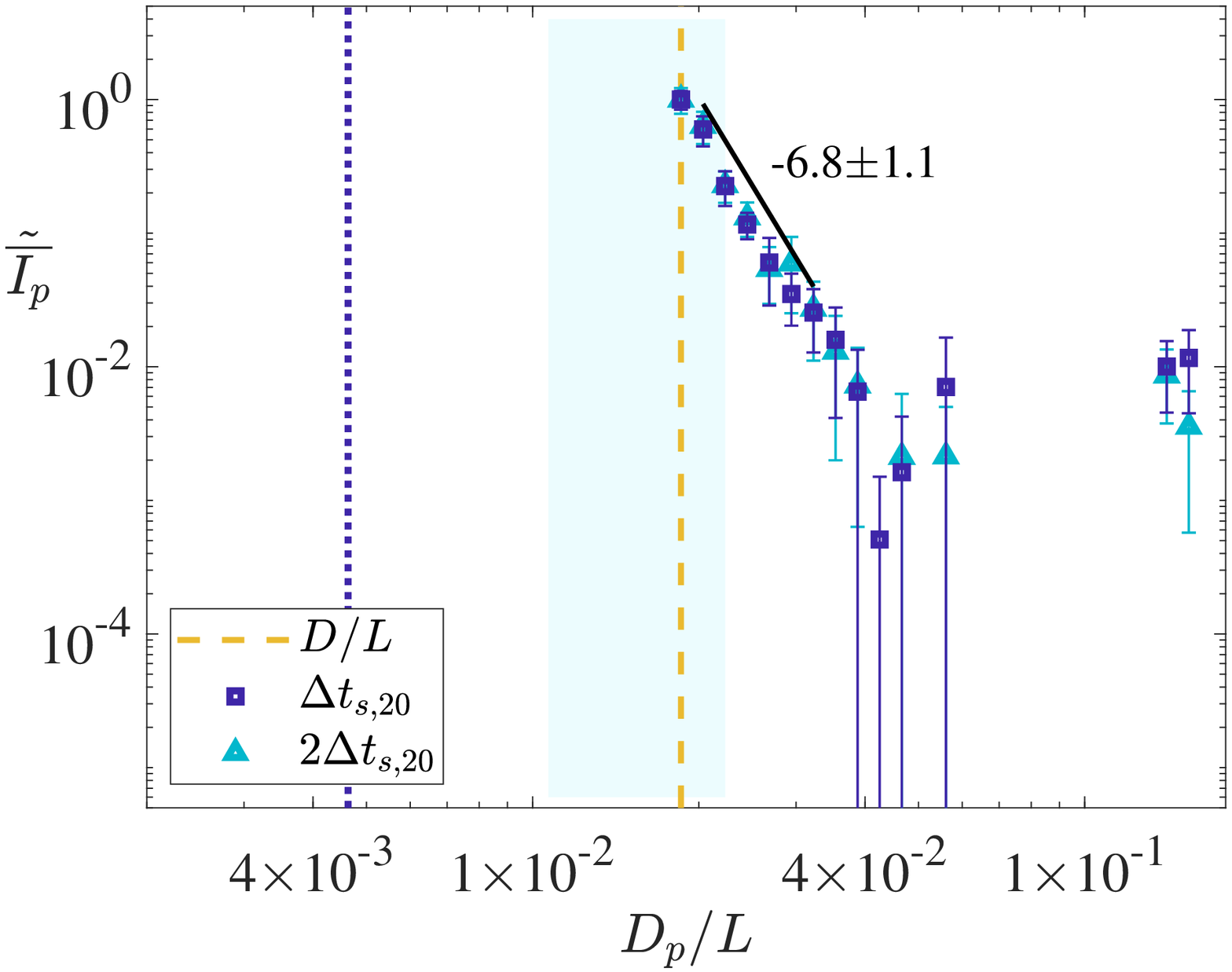}
\quad
(b)
\includegraphics[width=0.42\linewidth,valign=t]{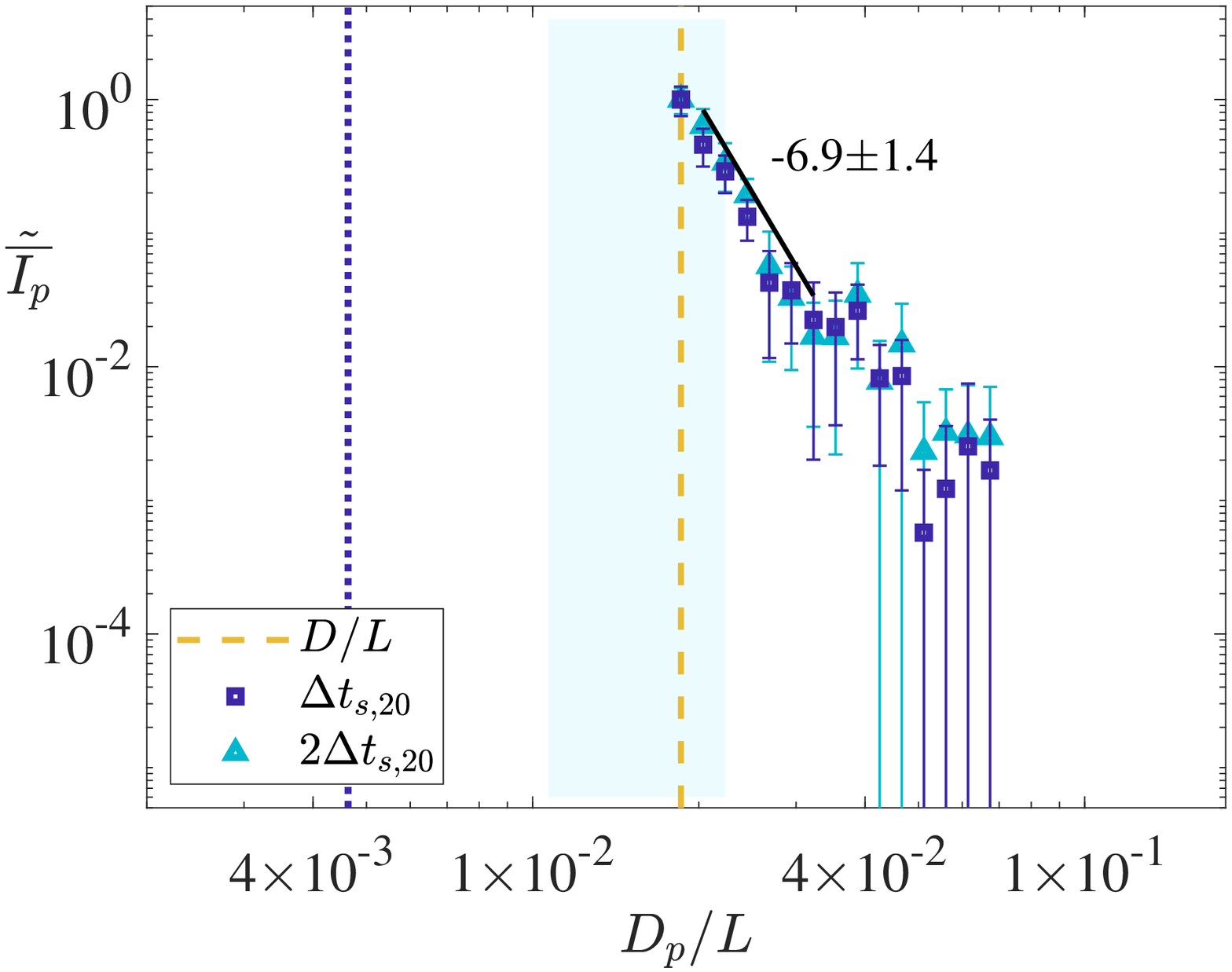}
}
  \caption{The normalized differential incoming transfer rate $\tilde{\ov{I_p}}\left(D_{p,j};t|D\right)$ from parent bubbles as a function of non-dimensional parent bubble size $D_p/L$ for the 20-realization baseline ensemble subset at the non-dimensional cut-off size $D/L = 1.86\times10^{-2}$, and at the times (a) $t=2.51$ and (b) $t=2.99$. These statistics are averaged over the non-dimensional sampling interval $16 \Delta t_{s,20} = 9.6\times10^{-2}$ used in figure~\ref{fig:breakup_gbfe}. For a description of the normalization, shaded regions, vertical lines, excluded events, error bars, the bubble-size subrange and dataset for the power-law fits, and the fit-exponent uncertainties, refer to the caption of figure~\ref{fig:breakup_i}.}
\label{fig:breakup_i_instant}
\end{figure}

Figure~\ref{fig:breakup_i} plots the time-averaged differential incoming $\left(\text{parent, }\ov{I_p}^T\right)$ and outgoing $\left(\text{child, }\ov{I_c}^T\right)$ contributions to the break-up flux for the two time intervals identified in \S~\ref{sec:waveevol}, \S~\ref{sec:sizedistevol}, and \S~\ref{sec:trac-gbfe}, and an intermediate cut-off bubble size near $D/L = 2\times10^{-2}$. This choice of $D/L$ is representative of a bubble size within the intermediate size subrange in figures~\ref{fig:bsd_compare_powerlaw}--\ref{fig:breakup_gbfe_timeavg} in which the break-up cascade scalings were recovered during the first time interval of interest. Other choices of $D/L$, which are not shown here, display a similar behaviour. The differential incoming contributions from parent bubbles strongly decay with increasing parent bubble size, while the differential outgoing contributions to children bubbles strongly decay with decreasing child bubble size, indicating that the break-up flux is strongly infrared and ultraviolet local, as suggested in Part 1. Power-law fits of these decay rates exhibit exponents near $-7$ and $5$ for $\ov{I_p}^T$ and $\ov{I_c}^T$, respectively. The theory in Part 1 predicts these exponents for a uniform child bubble volume distribution, $q_b$, that remains invariant with parent bubble size. Slight deviations from these idealized values may be accounted for by the mild deviations of $\ov{\check{q}_b}^T$ from uniformity and parent-bubble-size invariance observed in \S~\ref{sec:trac-qb}.  Note that reasonable agreement with the exponents from Part 1 is obtained for both time intervals even though the $D^{-10/3}$ scaling is absent from the size distribution in the second interval. This suggests that cascade-like behaviour persists in this interval, even as the dissipation rate decays. 

One might observe that the decay rate of $\ov{I_p}^T$ becomes gentler at very large parent bubble sizes, especially during the first time interval. This is reminiscent of the occurrence of large-size-ratio cascade-bypassing break-up events in large parent bubbles in figure~\ref{fig:breakup_qb}(e). Two remarks are in order here. First, the contributions from these events to the break-up flux are orders of magnitude smaller than the transfers from events of highest locality. Second, these contributions arise from intermittent break-up events as evidenced in figure~\ref{fig:breakup_i_instant}, which illustrates the instantaneous differential incoming contributions from parent bubbles at two time instances during the first interval. Notwithstanding these caveats, these results suggest that the break-up dynamics are well approximated as local in bubble-size space in both time intervals of interest.

\subsubsection{Size variation and time evolution of the break-up flux}\label{sec:trac-wb-evolution}

\begin{figure}
  \centerline{
(a)
\includegraphics[width=0.42\linewidth,valign=t]{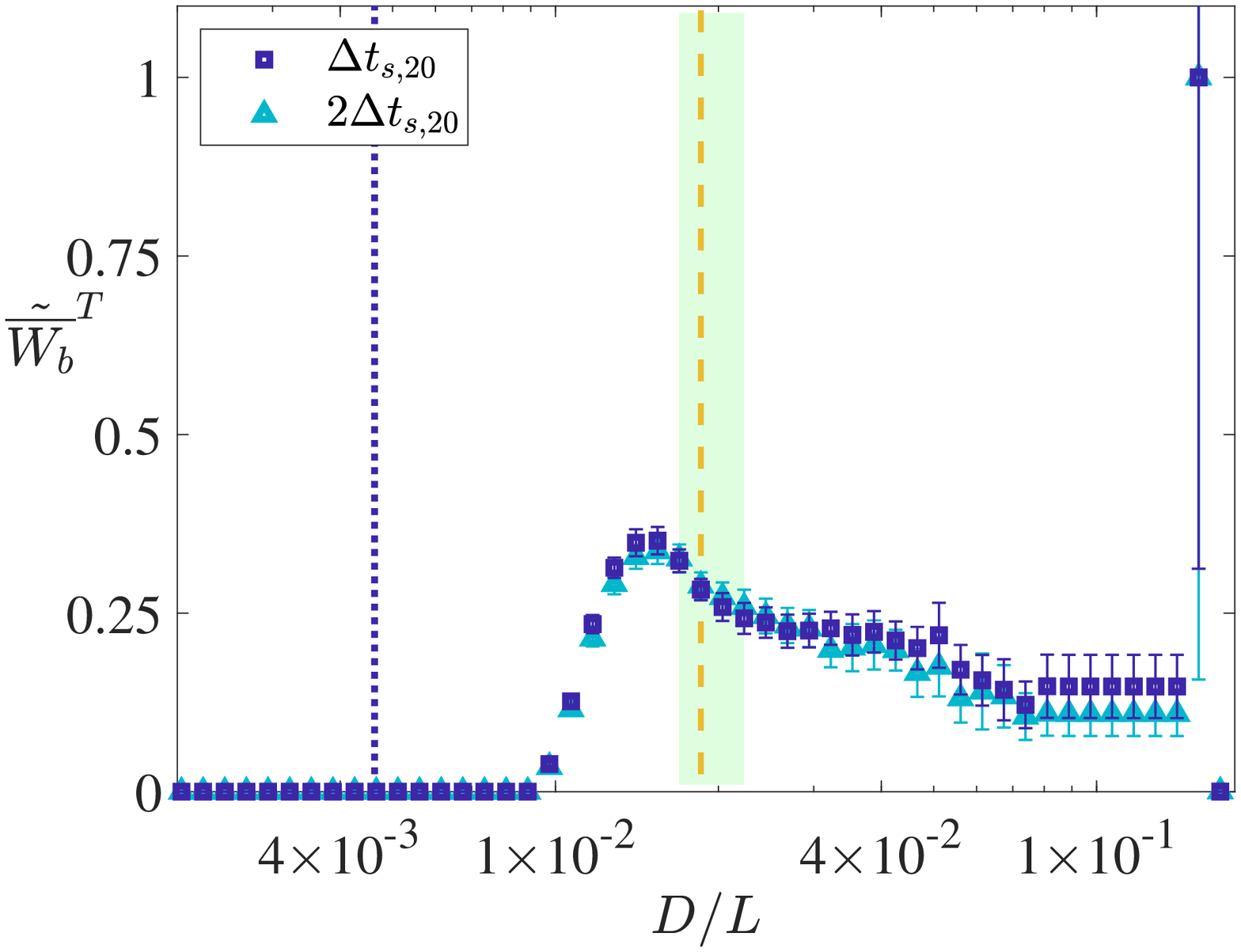}
\quad
(b)
\includegraphics[width=0.42\linewidth,valign=t]{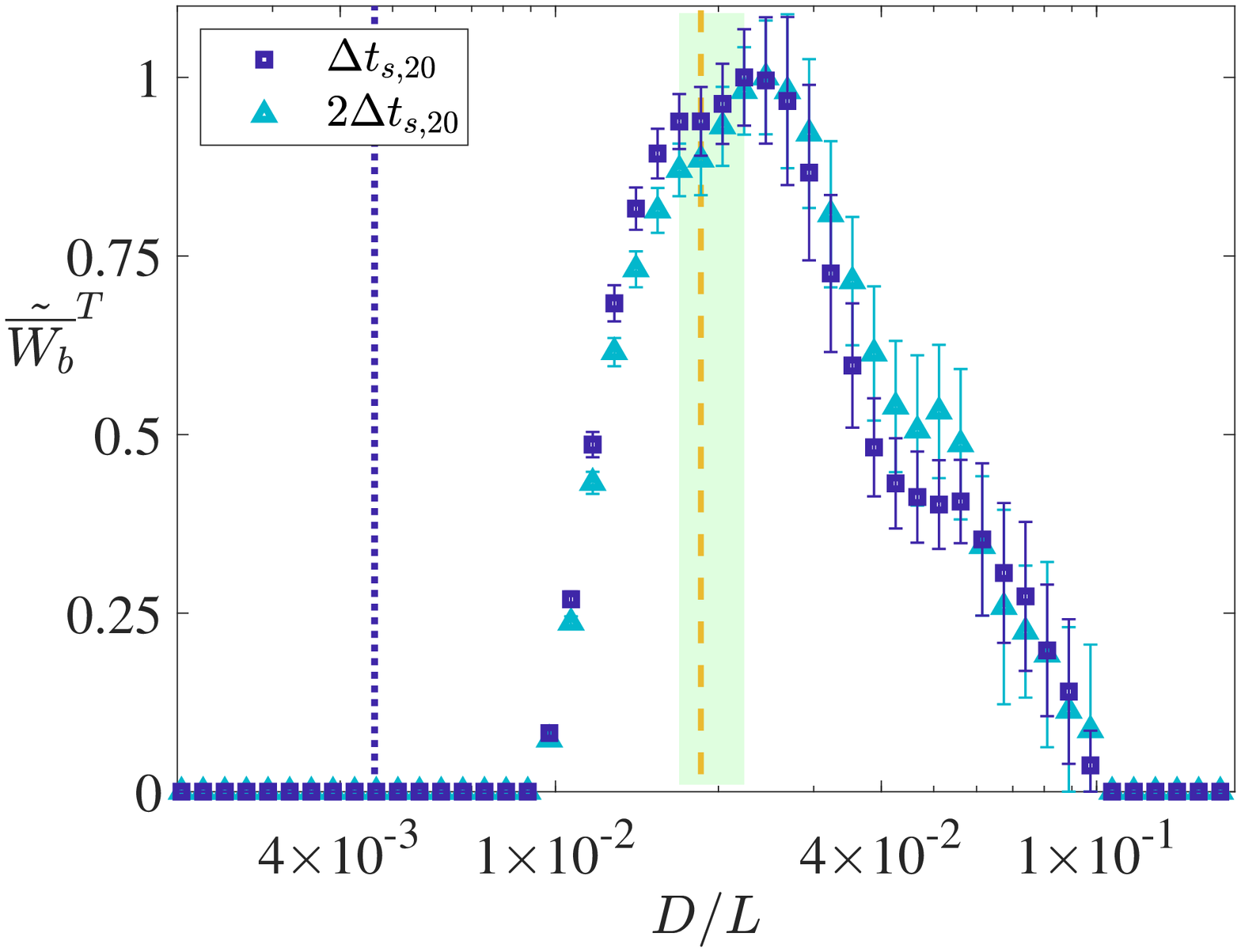}
}
  \caption{The normalized time-averaged break-up flux $\tilde{\ov{W_b}}^T(D/L)$ against non-dimensional size $D/L$ for the 20-realization baseline ensemble subset. These statistics are time averaged over the (a) early ($t=2.30$--$3.14$) and (b) late ($t=3.45$--$3.95$) wave-breaking stages. For a description of the normalization, vertical lines, excluded events, and error bars, refer to the caption of figure \ref{fig:breakup_i}. The shaded regions depict the bubble-size subrange over which the size-averaged break-up flux in figure~\ref{fig:breakup_w_t} was obtained.}
\label{fig:breakup_w_d}
\end{figure}

\begin{figure}
  \centerline{
\includegraphics[width=0.55\linewidth]{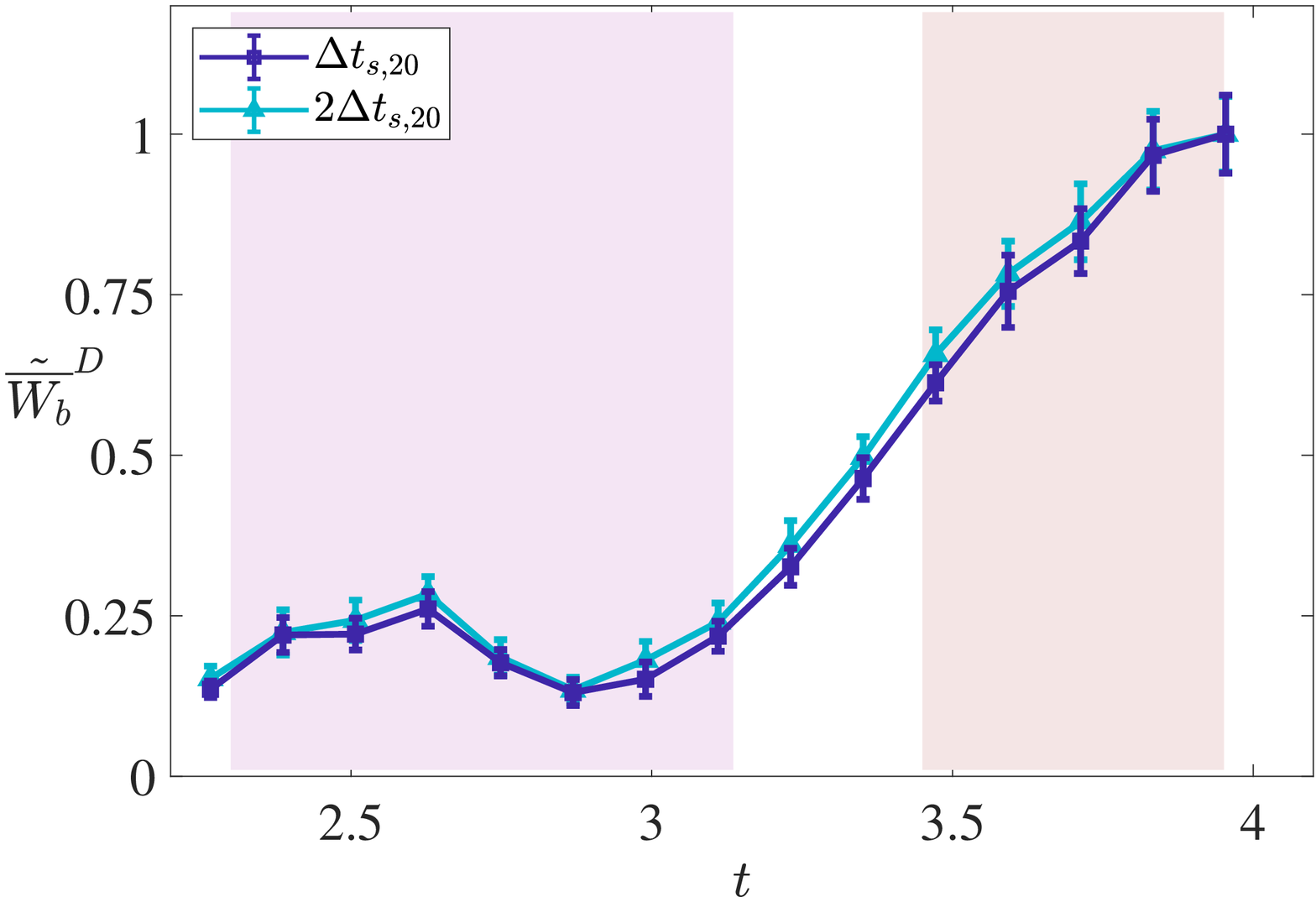}
}
  \caption{The normalized size-averaged break-up flux $\tilde{\ov{W_b}}^D(t)$ against non-dimensional time $t$ for the 20-realization baseline ensemble subset. The flux is averaged over the bubble-size subrange spanning non-dimensional diameters between $1.69\times10^{-2}$ and $2.23\times10^{-2}$, as indicated by the shaded regions in figure~\ref{fig:breakup_w_d}. For a description of the normalization, excluded events, and error bars, refer to the caption of figure \ref{fig:breakup_i}. The shaded regions span the same time intervals as those in figures~\ref{fig:energy}(b), \ref{fig:bsd_powerlaw_time}, and \ref{fig:breakup_gbfe_time}.} 
\label{fig:breakup_w_t}
\end{figure}

Figure~\ref{fig:breakup_w_d} plots the variation of the time-averaged break-up flux $\ov{W_b}^T(D)$ with cut-off bubble size $D$ for the two time intervals identified in \S~\ref{sec:waveevol}, \S~\ref{sec:sizedistevol}, and \S~\ref{sec:trac-gbfe}. Observe that there is some degree of size invariance in the break-up flux in both time intervals, even though the scale separation in the simulated waves is not exceedingly large as remarked in \S~\ref{sec:sizedistevol} and \S~\ref{sec:trac-gbfe}. This size invariance is expected in the theory from Part 1 if $I_p \sim D_p^{-7}$ and $I_c \sim D_c^5$, or more generally if the magnitude of the power-law exponent for $I_p$ is larger than that of $I_c$ by 2. These exponents were approximately recovered for both time intervals in \S~\ref{sec:trac-wb-locality}. Note that the size invariance of the break-up flux, $\ov{W_b}$, is also consistent with approximate stationarity in the size distribution, $\ov{f}$, as implied by \eqref{eqn:pbe_reduced}, as well as the usage of the break-up flux as a proxy for the entrainment flux, as discussed in \S~\ref{sec:bridging_pbe}. Further implications of the consistency of quasi-self-similarity and quasi-stationarity in the break-up dynamics are explored in appendix~\ref{app:averaging} with reference to the discussion on time averaging in \S~\ref{sec:sizedistevol}. Finally, observe that the ratio of the flux at intermediate sizes to the flux at large sizes differs in the two time intervals of interest. The ratio is close to unity in the first interval, with variations due to large-scale inhomogeneities, but increases by an order of magnitude in the second interval. This suggests that active entrainment at the large scales directly drives bubble-mass transfer from large to small bubble sizes in the first interval. The entrainment appears to have ceased in the second interval as mentioned in \S~\ref{sec:sizedistevol}, and bubble-mass transfer seems to be chiefly the result of the fragmentation of already-entrained large cavities and bubbles. 

Figure~\ref{fig:breakup_w_t} plots the variation of the flux $\ov{W_b}^D(t)$ with time, averaged over a narrow intermediate range of bubble sizes enveloping the cut-off size considered in \S~\ref{sec:trac-wb-locality}. Because of the quasi-size-invariance observed in figure~\ref{fig:breakup_w_d}, the results of figure~\ref{fig:breakup_w_t} are representative of the intermediate-size dynamics considered in this work. The flux has distinct characteristics in the two time intervals identified in \S~\ref{sec:waveevol}, \S~\ref{sec:sizedistevol}, and \S~\ref{sec:trac-gbfe}. The first time interval with a relatively constant rate of energy dissipation appears to be concomitant with an approximately constant flux, closing the loop on the presence of a quasi-steady turbulent bubble break-up cascade in this interval. The oscillations in the flux mirror the oscillations in the dissipation rate in figure~\ref{fig:energy}(b), and appear to be related to the earlier discussion in \S~\ref{sec:sizedistevol} on the oscillating power-law exponent for the size distribution. The second time interval with a decreasing dissipation rate appears to be associated with a flux that linearly increases with time. As discussed in \S~\ref{sec:trac-wb-locality}, the flux was observed to be strongly size local during this interval, which is indicative of the persistence of cascade-like behaviour. However, the temporally increasing flux does not drive a uniform increase in the magnitude of the size distribution at all sizes, as depicted in figures~\ref{fig:bsd_compare_powerlaw} and \ref{fig:bsd_timeavg}. Instead, it was observed in \S~\ref{sec:sizedistevol} that the power-law scaling of the size distribution transitions from $D^{-10/3}$ to $D^{-8/3}$. This suggests that small bubbles may be depleted more rapidly at these late times. The continuation of the flux over several characteristic times suggests that the volume of gas present in small bubble sizes gradually increases over time. The assumption in Part 1 that one may neglect this accumulation should eventually break down towards the late wave-breaking stages. The increased importance of coalescence due to this accumulation, and its potential role in the aforementioned depletion of small bubbles, is addressed in appendix~\ref{app:coalescence}.

\section{Conclusions}\label{sec:conclusions}

\begin{table}
  \begin{center}
\def~{\hphantom{0}}
  \begin{tabular}{c|ccccc}
      \multirow{2}{*}{Stage} & Nature of & Nature of & Scaling & Locality of & Description of \\
      & flux $\ov{W_b}$ & dissipation $\ov{\varepsilon}$ & of $\ov{f}$ & flux $\ov{W_b}$ & bubble-mass transfer \\[6pt]
       Early & Quasi-steady & Quasi-steady & $D^{-10/3}$ & Strong & Bubble break-up cascade \\[6pt]
       \multirow{2}{*}{Late} & \multirow{2}{*}{Increasing} & \multirow{2}{*}{Decreasing} & \multirow{2}{*}{$D^{-8/3}$} & \multirow{2}{*}{Strong} & Cascade-like behaviour \\
       &&&&& ${}+{}$ other dynamics \\
  \end{tabular}
  \caption{Summary of bubble-mass transfer dynamics in the active wave-breaking phase.}
  \label{tab:summary}
  \end{center}
\end{table}

This paper investigates the evolution of the bubble size distribution and other bubble statistics beneath a breaking wave during the active wave-breaking phase using ensembles of numerical simulations. A large ensemble of simulations enhances statistical convergence in these bubble statistics, which is particularly important for the time-dependent behaviour explored in this paper. Two time intervals of interest are identified, with comparison and contrast between them throughout the paper. This analysis goes beyond the size distribution by identifying and characterizing individual break-up events in order to directly observe characteristics associated with the bubble-mass cascade phenomenology that were identified in Part 1. The evolving dynamics in the bubble-mass transfer process are summarized in Table~\ref{tab:summary}. The results of this paper (Part 2) suggest that cascade-like behaviour is present in both time intervals, even when the equilibrium $D^{-10/3}$ power law is not observed in the size distribution at intermediate sizes. They highlight that different bubble generation and evolution mechanisms emerge at different length and time-scales during the wave-breaking process, even within the relatively limited range of length scales captured in these simulations. Future ensembles of higher $\RR_L$ and $\We_L$ simulations with resolution of a larger range of scales may bring about more clarity on these mechanisms.

The evolution of the bubble size distribution and other associated bubble statistics in the present simulation ensembles may be summarized as follows. In the early wave-breaking stages, large air cavities are entrained and successively fragmented. The corresponding rate of bubble-mass transfer from large to small bubble sizes is quasi-steady, as is the energy dissipation rate, which also reaches a maximum value in this time interval. These are the ideal characteristics of a forward bubble-mass cascade driven by turbulent fragmentation. As originally proposed by~\citet{Garrett1}, this is accompanied by a $D^{-10/3}$ power-law scaling in the size distribution. The break-up statistics of Part 2 provide strong support for the theoretical results of Part 1 in this early time interval. The time-averaged size distribution matches the $D^{-10/3}$ scaling in agreement with~\citet{Deane1},~\citet{Wang1}, and~\citet{Deike1}. Deviations of the instantaneous size distribution from this scaling are observed but are reduced by time averaging over the entire interval. The $D^{-4}$ power-law scaling derived in Part 1 for the differential break-up rate is also recovered. Notwithstanding several intermittent non-local events for large parent bubbles, the probability distribution of child bubble volumes is roughly uniform in the range of parent bubble sizes where the $D^{-10/3}$ scaling is observed in the time-averaged size distribution. The resulting break-up flux remains strongly infrared and ultraviolet local. The incoming differential flux from parent bubbles approaches a $D_p^{-7}$ power-law scaling over a range of parent bubble sizes, while the outgoing differential flux to children bubbles approaches a $D_c^5$ power-law scaling over a range of child bubble sizes. These scaling exponents were predicted in Part 1 for a uniform child bubble volume distribution. The observed scalings suggest that the underlying bubble-mass transfer process is indeed size local and self-similar, although a more robust demonstration of self-similarity may only be evident in larger waves of higher $\RR_L$ and $\We_L$. Together, these observations provide strong support for the presence of a quasi-steady turbulent bubble break-up cascade at intermediate bubble sizes during the early wave-breaking stages.

In the late wave-breaking stages, large entrained bubbles continue to break up into smaller bubbles. The corresponding rate of bubble-mass transfer from large to small sizes accelerates while the dissipation rate begins to decay. Two distinct power-law scalings appear in the size distribution in agreement with~\citet{Tavakolinejad1} and~\citet{Masnadi1} but in deviation from the $D^{-10/3}$ scaling. The increased magnitude of the size distribution suggests that small and intermediate bubbles are more populous in this time interval. Vestiges of locality, stationarity, and self-similarity are nonetheless present in the corresponding break-up statistics, indicating that the break-up dynamics still exhibit cascade-like behaviour. The steepening of the size distribution at large bubble sizes is indicative of buoyant degassing. A shallower size distribution appears at intermediate bubble sizes. The data appears to be consistent with a more rapid depletion of small bubbles, potentially via coalescence.

Several concluding remarks are made in view of these observations. The theoretical framework developed in Part 1 enables one to go above and beyond the traditional $D^{-10/3}$ power-law scaling in the size distribution in diagnosing the presence of a break-up cascade. The results of Part 2 demonstrate that the access to detailed spatial information provided by numerical simulations allows a more precise evaluation of various components of the theoretical framework. The multifaceted nature of this analysis enables a more nuanced description of the underlying break-up dynamics even when the $D^{-10/3}$ scaling is absent. For example, a number of key physical aspects of the cascade phenomenology, such as locality, were determined to be present even in the late wave-breaking stages. The elements of the formalism in Part 1 could thus be seen as an analytical toolkit that may be used to selectively probe the characteristics of break-up and coalescence dynamics in multiscale two-phase flows. This toolkit may be used to advance understanding of cascading processes, and more generally to inform the validation and development of model kernels for the population balance equation.

Knowledge of the physical mechanisms underlying bubble generation enables the quantification of interfacial fluxes and radiation scattering, which are of practical importance in maritime and climate studies, such as carbon sequestration and the persistent wake signatures of seafaring vessels. The rich temporal evolution of the bubble-mass transfer process observed in this work reflects the complexity in modelling these mechanisms. Elucidation of sustained cascade-like behaviour in the bubble-mass transfer dynamics presents opportunities for modelling of subgrid bubble break-up in large eddy simulations.

\section*{Acknowledgments} 

This investigation was funded by the Office of Naval Research, Grant \#N00014-15-1-2726, and is also supported by the Advanced Simulation and Computing programme of the U.S. Department of Energy's National Nuclear Security Administration via the PSAAP-II Center at Stanford University, Grant \#DE-NA0002373. W.~H.~R. Chan is also funded by a National Science Scholarship from the Agency of Science, Technology and Research in Singapore. The authors acknowledge computational resources from the U.S. Department of Energy's INCITE Program. The authors are grateful to A. Mani for fruitful discussions on the bubble size distribution, and M.~S. Dodd for joint work on algorithms to retrieve statistics from numerical simulations that inspired this work.

\section*{Declarations}

The authors report no conflict of interest.

\appendix


\section{Averaging operations for computing bubble statistics}\label{app:averaging}

The ensemble-averaged bubble size distribution $f\left(\bs{x},D;t\right)$ was defined in Part 1 at every location $\bs{x}$, for every bubble size $D$, and at some time $t$ by ensemble averaging $\left(\langle\cdot\rangle\right)$ the number density function of a bubble population, which is constructed by summing a contribution from each bubble in the population $i=1,\ldots,N_b(t)$ having a centroid location $\bs{x}_i$ and an equivalent size $D_i$. $f$ may then be expressed as
\begin{equation}
f\left(\bs{x},D;t\right) = \left\langle \sum_{i=1}^{N_b(t)} \delta\left(\bs{x}-\bs{x}_i(t)\right) \delta\left(D-D_i(t)\right) \right\rangle.
\end{equation}
It was alluded to in \S~\ref{sec:bridging} that the volume-averaged and ensemble-averaged size distribution $\ov{f}(D;t)$ is typically reported in breaking-wave simulations and experiments as
\begin{equation}
\ov{f}(D;t) = \f{1}{\mathcal{V}} \int_\Omega \diff\bs{x} \: f\left(\bs{x},D;t\right) = \f{1}{\mathcal{V}} \left\langle \sum_{i=1}^{N_b(t)} \delta\left(D-D_i(t)\right) \right\rangle,
\end{equation}
where $\mathcal{V} = \int_\Omega \diff \bs{x}$ is the volume of the domain $\Omega$ over which the bubble population is integrated. $\Omega$ should be selected such that it always contains all $N_b(t)$ bubbles. In this work, $\mathcal{V}$ is chosen to be the characteristic wave volume, $L^3$. Since this coincides with the computational domain volume, all $N_b(t)$ bubbles are thereby always included. The volume-averaging operation is analogously defined for other bubble statistics like $\ov{g_b f}$. The ensemble-averaging and volume-averaging operations commute when $\mathcal{V}$ and $\Omega$ are identical across all the ensemble realizations. Thus, $\ov{f}$ satisfies the normalization conditions
\begin{equation}
\f{\left\langle N_b(t) \right\rangle}{\mathcal{V}} = \left\langle n_b(t) \right\rangle = \int_0^\infty \diff D \: \ov{f}(D;t),
\end{equation}
where $n_b(t)$ is the number of bubbles in the system per unit domain volume for an ensemble realization, or the global bubble number density in the realization. The population balance equation for $f\left(\bs{x},D;t\right) D^3$ was written in Part 1 as
\begin{multline}
\f{\partial\left[f\left(\bs{x},D;t\right)D^3\right]}{\partial t} + \f{\partial\left[v_i\left(\bs{x},D;t\right) f\left(\bs{x},D;t\right)D^3\right]}{\partial x_i} + \f{\partial\left[v_D\left(\bs{x},D;t\right) f\left(\bs{x},D;t\right)D^3\right]}{\partial D} ={}\\
{}= H\left(\bs{x},D;t\right),
\end{multline}
where $v_i$ and $v_D$ are the velocities of $fD^3$ in the $\bs{x}$-$D$ phase space along the spatial and bubble-size dimensions, respectively, and $H$ is a model term that includes source, sink, and non-local transfer terms for the transport of $fD^3$. Volume averaging the equation over a domain that always includes all $N_b(t)$ bubbles yields \eqref{eqn:pbe}. A similar procedure may be used to obtain \eqref{eqn:pbe_kernel}. Note that the spatial and bubble-size dimensions are orthogonal in the $\bs{x}$-$D$ phase space. Thus, the expected $D$-scaling of a quantity, such as the theoretical variations of $f\left(\bs{x},D;t\right)$ and $\ov{f}(D;t)$ with $D$, should remain invariant under volume averaging provided there is sufficient statistical convergence in the quantity.

Volume averaging aggregates statistics at different locations, and provides a convenient way of achieving statistical convergence in a limited number of realizations, as well as a more concise description of the underlying dynamics~\citep{Hinze1}. It is argued here that volume averaging also achieves a good approximation for low-order bubble statistics, including the size distribution, as if the underlying flow was statistically homogeneous~\citep{Deike1}. More specifically, it averages these statistics over small, localized regions that may each be treated as statistically homogeneous in the spirit of local isotropy. This may be motivated by the arguments behind Kolmogorov's refinements to his original energy cascade hypotheses~\citep{Kolmogorov2}. The original hypotheses~\citep{Kolmogorov1} are best applied in regions where the assumption of spatial homogeneity is robust. In a statistically inhomogeneous flow, diffusion fluxes across large-eddy length and time-scales may become important, thereby manifesting as large-scale variations in the turbulent kinetic energy, as well as intermittency in the localized production and dissipation rates. The refinements of~\citet{Kolmogorov2} attempt to characterize these variations systematically through an assumption on their statistical distribution. To enable this, volume averaging is performed in small, localized regions that are each approximately statistically homogeneous. The original hypotheses are then recovered for locally volume-averaged flow statistics, with constants of proportionality that account for large-scale inhomogeneities. The original and refined hypotheses do not yield significantly different results when applied to low-order flow statistics~\citep[\S~6.7.4]{Pope1}, such as the second-order velocity structure functions. The global volume-averaging procedure adopted in this work eliminates these diffusion fluxes by averaging flow and bubble statistics over a volume where no net transport in or out is expected. This assumes each small, localized region in the global averaging volume is subject to the same characteristic large-eddy length and time-scales, in the spirit of the original hypotheses of~\citet{Kolmogorov1}. This is thus likely to be reasonable for low-order flow and bubble statistics, with additional corrections necessary for higher-order statistics when, for example, $\ov{\varepsilon^n}$ is no longer well-approximated by $\left(\ov{\varepsilon}\right)^n$ for large $n$. Hence, volume averaging is a physically relevant way of computing low-order bubble statistics in addition to its expedience. Nevertheless, knowledge of the spatial distribution of bubbles may at times be important, and two representative snapshots of the ensemble-averaged and spanwise-averaged liquid volume fraction are provided in appendix~\ref{app:voidfraction} to provide a general sense of this spatial variation in the context of breaking waves.

As an end-note, recall the remark in \S~5.1 of Part 1 that the analysis for the bubble-mass cascade resembles the monofractal analysis of~\citet{Eyink1} for the energy cascade. The identification of a single power-law scaling in the bubble size distribution under a given set of conditions also suggests monofractality in the bubble break-up dynamics~\citep{Turcotte1}. For example, the $D^{-10/3}$ power-law scaling corresponds to a fractal dimension of $7/3$, or that of a convoluted surface [see also~\citet{Sreenivasan1},~\citet{Sreenivasan2},~\citet{Sreenivasan3}, and~\citet{Vassilicos1}]. It appears that the act of averaging out the aforediscussed large-scale fluctuations yields monofractal dynamics that provide an average sense of any underlying multifractal dynamics. This applies to both the volume-averaging procedure discussed above and the time-averaging procedure introduced in \S~\ref{sec:sizedistevol}. The latter implicitly assumes that the dynamics in these small, localized, and quasi-homogeneous regions are also quasi-stationary. The additional observation of quasi-self-similarity in these break-up dynamics in an intermediate range of bubble sizes, as discussed in \S~\ref{sec:trac-wb-evolution}, brings to mind the five-dimensional turbulent cascade in the three spatial dimensions, time, and scale discussed by~\citet{Cardesa1}.


\section{Estimating the energy dissipation rate $\ov{\varepsilon}$}\label{app:dissipation}

Recall from Part 1 and \S~\ref{sec:bridging} that the wavelength and deep-water-wave phase velocity of a characteristic wave are respectively used to estimate $L$ and $u_L$ in the context of breaking waves. These characteristic scales may then be used to estimate the characteristic energy dissipation rate $\ov{\varepsilon} \sim u_L^3/L$, and thereafter the global Kolmogorov and Hinze scales in \eqref{eqn:kolmodim} and \eqref{eqn:hinzedim}. However, some experiments~\citep{Rapp1,Na1} have suggested that the largest eddies and gaseous cavities have characteristic sizes and velocities that are an order of magnitude smaller. Other experiments have further suggested that most of the turbulence that originates from wave breaking initially resides in a region whose thickness is on the order of one wave height from the surface~\citep{Rapp1,Terray1,Thomson1}. These have motivated the choice of other quantities for the estimation of characteristic wave scales. For example, typical surface trajectories during wave breaking may have motivated the development of a ballistic scaling by~\citet{Drazen1} that uses the wave height to estimate $L$ and the corresponding ballistic velocity to estimate $u_L$. The corresponding ballistic time can be smaller than the wave period by an order of magnitude depending on the wave slope. The ballistic scaling was demonstrated to be adequate for single breaking events with moderate wave slope $S$, where $S=\sum_{i=1}^{N_m} a_i k_i$ may be interpreted as the maximum slope of the initial waveform. The waveform is assumed to be a wave packet with $N_m$ component modes, and $a_i$ and $k_i$ are the initial amplitude and wavenumber, respectively, of the $i$-th mode. However, it has been observed~\citep{Loewen3,Melville2,Loewen2,Drazen1,Deane7,Deane6} that the dissipation rate eventually saturates with increasing wave slope due to the presence of multiple breaking events. This necessitates a different choice of the characteristic scales for steeper waves or more general breaking patterns. In this work, for example, the characteristic maximum slope of the initial waveform is $S = \sum_{i=1}^3 a_i k_i = 0.76$, where $i=2$ and $i=3$ refer respectively to the first and second harmonics. This value of $S$ is in the saturated regime identified by~\citet{Drazen1}. The forthcoming discussion suggests that in the absence of additional information, the wavelength and phase velocity remain the most generic choices for the characteristic length and velocity scales, respectively, in breaking waves.

Two canonical scenarios are considered for energy conversion during wave breaking. First, consider the average energy transfer rate from coherent to turbulent motion in a single travelling wave of height $h$ and wavelength $\lambda$. Before breaking occurs, the wave energy per unit width and wavelength $\squart \rho_l g h^2$ is carried forward at the group velocity $\sqrt{(g\lambda)/(8\upi)}$. Suppose that the action of breaking is to transfer this energy into a volume with cross-sectional area $h$ by $h$~\citep{Lamarre2,Loewen1,Deane1,Drazen1,Rojas2}. The resulting average energy transfer rate per unit mass is $\ov{\varepsilon} \sim 0.8c_p^3/\lambda$, where $c_p$ is the wave phase velocity. Second, consider the average energy transfer rate from large to small waves in a travelling wave-packet with central wavelength $\lambda$ and a corresponding central phase velocity $c_p$. \citet{Gemmrich1} and \citet{Babanin1} reported wave and velocity spectra from individual wave-breaking events indicating the momentary upshifting of the peak wavenumber towards smaller scales before breaking, and then downshifting towards larger scales after breaking. This hints at the accumulation and then release of energy at some small limiting scale. A similar focusing process was observed in physical space during the pre-breaking phase in the two-dimensional numerical study of a spilling breaker by~\citet{Iafrati2}, and in the experimental study of a plunging breaker by~\cite{Bonmarin1}. Inspired by these observations, as well as the arguments by~\citet{Kitaigorodskii1}, one may estimate $\ov{\varepsilon}$ as a ratio of the characteristic energy transported by the wave $c_p^2$ to the time-scale associated with the limiting breaking scale $t_b \sim L_b / u_b$. Assuming $L_b$ is a function of only $\ov{\varepsilon}$ and $g$, and taking $u_b$ to be the phase velocity of a wave of length $L_b$, one eventually obtains the estimate $\ov{\varepsilon} \sim 4c_p^3/\lambda$. Both estimates for $\ov{\varepsilon}$ are of the order $O(c_p^3/\lambda)$. In the absence of additional constraints to restrict this estimate further, it is proposed that $\ov{\varepsilon} \sim u_L^3 / L$ be estimated by assuming $L \sim \lambda$ and $u_L \sim c_p$.


\section{Bubble identification}\label{app:bubbleident}

The error incurred by the bubble identification algorithm introduced in \S~\ref{sec:ident} is briefly discussed for a number of simple test cases. The reader is referred to the discussion by~\citet{Chan4} for a more detailed comparison of the employed grouping criterion with other criteria. First, the error in computing the volume of a single drop was determined by ballistically advecting the drop in a quiescent gas and by advecting it with a constant imposed velocity. In both cases, a uniform mesh was adopted, and there were $O(10)$ grid cells spanning the drop diameter. The volume error normalized by the grid-cell volume $(\Delta D)^3$ was of the order $O(10^{-2}$--$10^{-1})$, while the centroid error normalized by the grid-cell spacing $\Delta D$ was of the order $O(10^{-4})$. Next, these errors were used to estimate the ability of the algorithm to distinguish between a closely spaced large drop--small drop pair and fluctuations in the volume of the large drop from one time-step to another. The performance of the algorithm on this task is crucial in determining the accuracy of the tracking algorithm in \S~\ref{sec:tracking}. It is first assumed that the normalized volume error of a drop, $\Delta \mathcal{V}_\text{err}/(\Delta D)^3$, is proportional to its normalized surface area, $\upi D^2/(\Delta D)^2$
\begin{equation}
\f{\Delta \mathcal{V}_\text{err}}{(\Delta D)^3} = M \left[ \upi \f{D^2}{(\Delta D)^2} \right].
\end{equation}
The proportionality constant, $M$, is assumed to depend only on the grouping criterion. Next, suppose $d$ and $D$ are the diameters of the small and large drops, respectively. Then, in the limit that the two scenarios cannot be distinguished, one may write
\begin{equation}
M \left[ \upi \f{D^2}{(\Delta D)^2} \right] \simeq \f{\pi d^3}{6 (\Delta D)^3}.
\end{equation}
This yields the limiting size ratio
\begin{equation}
r = \f{D}{d} \simeq \sqrt{\f{d}{6M\Delta D}}.
\end{equation}
For the grouping criterion in this work, $M \sim O(10^{-4})$ and $r/\sqrt{N} \simeq 20$--$40$ for a given limiting bubble resolution $N = d/(\Delta D)$, so the algorithm can distinguish between these two scenarios for size ratios of the order $O(10^2)$ even when the small bubble is not well resolved. Note that this estimate for $M$ leads to the non-dimensional bubble diameter error for the breaking-wave baseline ensemble $\Delta D_\text{err}/L \sim O(10^{-6})$.

It is worth noting that many identification schemes are afflicted by the corner case of distinguishing two bubbles spaced a grid cell apart from a dumbbell-shaped bubble where two gaseous masses are connected by a thin gaseous bridge with the dimensions of a grid cell. The occurrence of this corner case does not necessarily suggest a deficiency in any of these schemes. Instead, it is representative of an inherent limitation of a sharp-interface field with finite resolution: when a thin gaseous bridge is numerically indistinguishable from a small under-resolved bubble or a small gaseous protrusion on the surface of a larger bubble, none of the geometries necessarily represent reality more accurately in the absence of additional information. Consequently, any decision by any scheme to favour any of the geometries is essentially arbitrary to a certain degree. The grouping criterion introduced in this work collectively identifies the gaseous masses and the gaseous bridge as a single bubble if the bridge is connected to cells in both gaseous masses with sufficiently large $1-\phi$. Otherwise, it returns two large bubbles and a small under-resolved bubble.


\section{Event detection}\label{app:eventdetect}

The constraints required to track a bubble from one simulation snapshot to another in the event detection algorithm introduced in \S~\ref{sec:tracking} are discussed here. As a bubble is advected, its mass\textemdash and volume in an incompressible setting\textemdash remains constant between the snapshots even if it deforms, within the error identified in appendix~\ref{app:bubbleident}. The principle of mass conservation is necessarily satisfied even if one bubble breaks up into two or if two bubbles coalesce into one. Suppose bubble \#0 breaks up into bubbles \#1 and \#2, or \#1 and \#2 coalesce to form \#0. In the case of break-up, one may write
\begin{equation}
\left| \mathcal{V}_0^n - \left[ \mathcal{V}_1^{n+1} + \mathcal{V}_2^{n+1} \right] \right| < \Delta \mathcal{V}_\text{err},
\label{eqn:consmassDB}
\end{equation}
while in the case of coalescence, one may write
\begin{equation}
\left| \mathcal{V}_0^{n+1} - \left[ \mathcal{V}_1^n + \mathcal{V}_2^n \right] \right| < \Delta \mathcal{V}_\text{err},
\label{eqn:consmassDC}
\end{equation}
where $\mathcal{V}_i^j$ is the volume of bubble \#$i$ in a flow snapshot $j$, $n$ denotes a generic flow snapshot, and $\Delta \mathcal{V}_\text{err}$ is the volume error discussed in appendix~\ref{app:bubbleident}. Also, if the simulation satisfies the CFL condition, then each fluid--fluid interface cannot traverse more than a single cell width every time-step, multiplied by the maximum permissible Courant number $C'$ for the employed advection scheme. Thus, the centroid of a bubble remains stationary between snapshots insofar as the permissible distance error is the product of the local grid spacing $\Delta x$, the number of time-steps between the snapshots $N_t$, and $C'$. This condition is also necessarily satisfied during break-up and coalescence. Denote the centroid of a bubble $i$ in snapshot $j$ as $\bs{x}_i^j$. In the case of break-up, one may write
\begin{equation}
\left|\left| \bs{x}_0^n - \f{ \bs{x}_1^{n+1} \mathcal{V}_1^{n+1} + \bs{x}_2^{n+1} \mathcal{V}_2^{n+1} }{\mathcal{V}_1^{n+1} + \mathcal{V}_2^{n+1}} \right|\right| < C' N_t \Delta x,
\label{eqn:centroidDB}
\end{equation}
while in the case of coalescence, one may write
\begin{equation}
\left|\left| \bs{x}_0^{n+1} - \f{ \bs{x}_1^n \mathcal{V}_1^n + \bs{x}_2^n \mathcal{V}_2^n }{\mathcal{V}_1^n + \mathcal{V}_2^n} \right|\right| < C' N_t \Delta x.
\label{eqn:centroidDC}
\end{equation}
Note that the identification algorithm also generates spatial errors in the centroid. However, as noted in appendix \ref{app:bubbleident}, these errors are typically much smaller than the grid spacing, and will be neglected. The constraints \eqref{eqn:consmassDB}--\eqref{eqn:centroidDC} are sufficient for the identification of break-up and coalescence events, as well as continuing bubbles, using the bubble volumes and centroids. These constraints assume that all break-up and coalescence events are binary. Note also that \eqref{eqn:consmassDB} and \eqref{eqn:consmassDC} imply a critical size ratio $r$ above which events involving a small bubble and a large bubble cannot be distinguished from fluctuations in the volume of the large bubble between snapshots. This ratio was discussed in appendix \ref{app:bubbleident}. Choosing an identification scheme with a lower volume error increases the critical size ratio keeping the resolution of the smallest bubble constant, allowing more relevant events to be captured accurately.


\section{The role of coalescence}\label{app:coalescence}

\begin{figure}
  \centerline{
\includegraphics[width=0.55\linewidth]{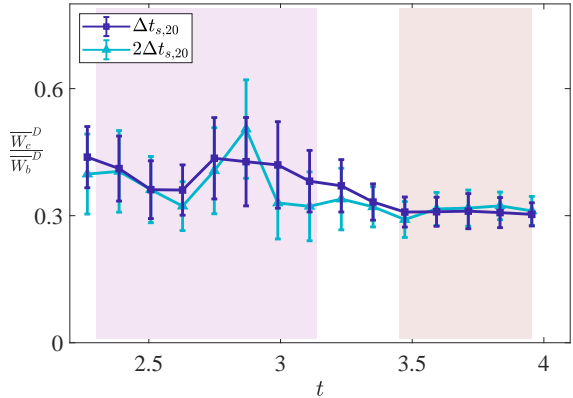}
}
  \caption{The ratio of the size-averaged coalescence flux to the size-averaged break-up flux, $\ov{W_c}^D/\ov{W_b}^D(t)$, against non-dimensional time $t$ for the 20-realization baseline ensemble subset. Coalescence events involving subgrid parent (ancestral) bubbles are excluded. For a description of the shaded regions, excluded break-up events, error bars, and the bubble-size subrange over which the fluxes were averaged, refer to the caption of figure \ref{fig:breakup_w_t}.} 
\label{fig:coalebreak_w_t}
\end{figure}

\begin{figure}
  \centerline{
(a)
\includegraphics[width=0.42\linewidth,valign=t]{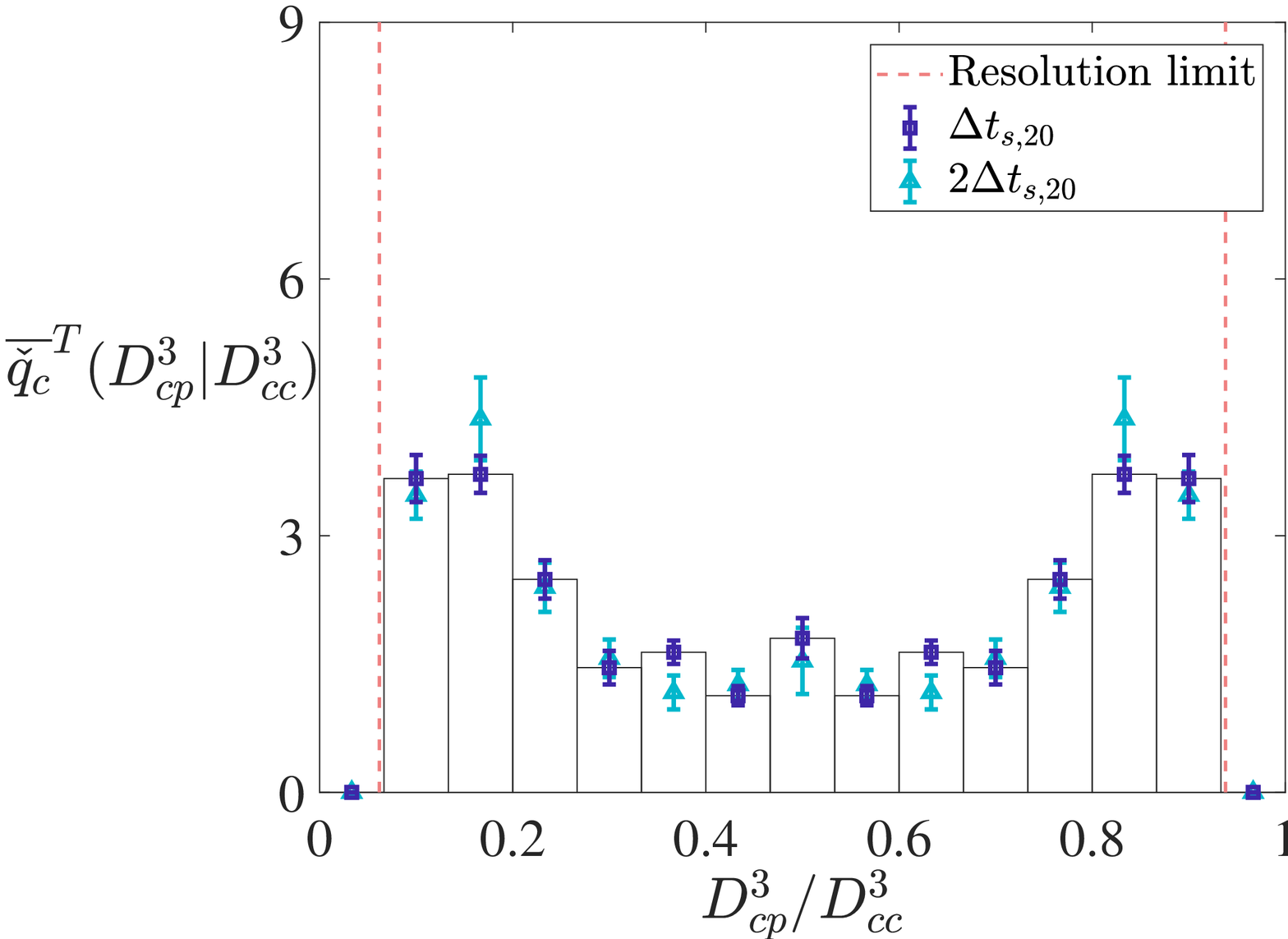}
\quad
(b)
\includegraphics[width=0.42\linewidth,valign=t]{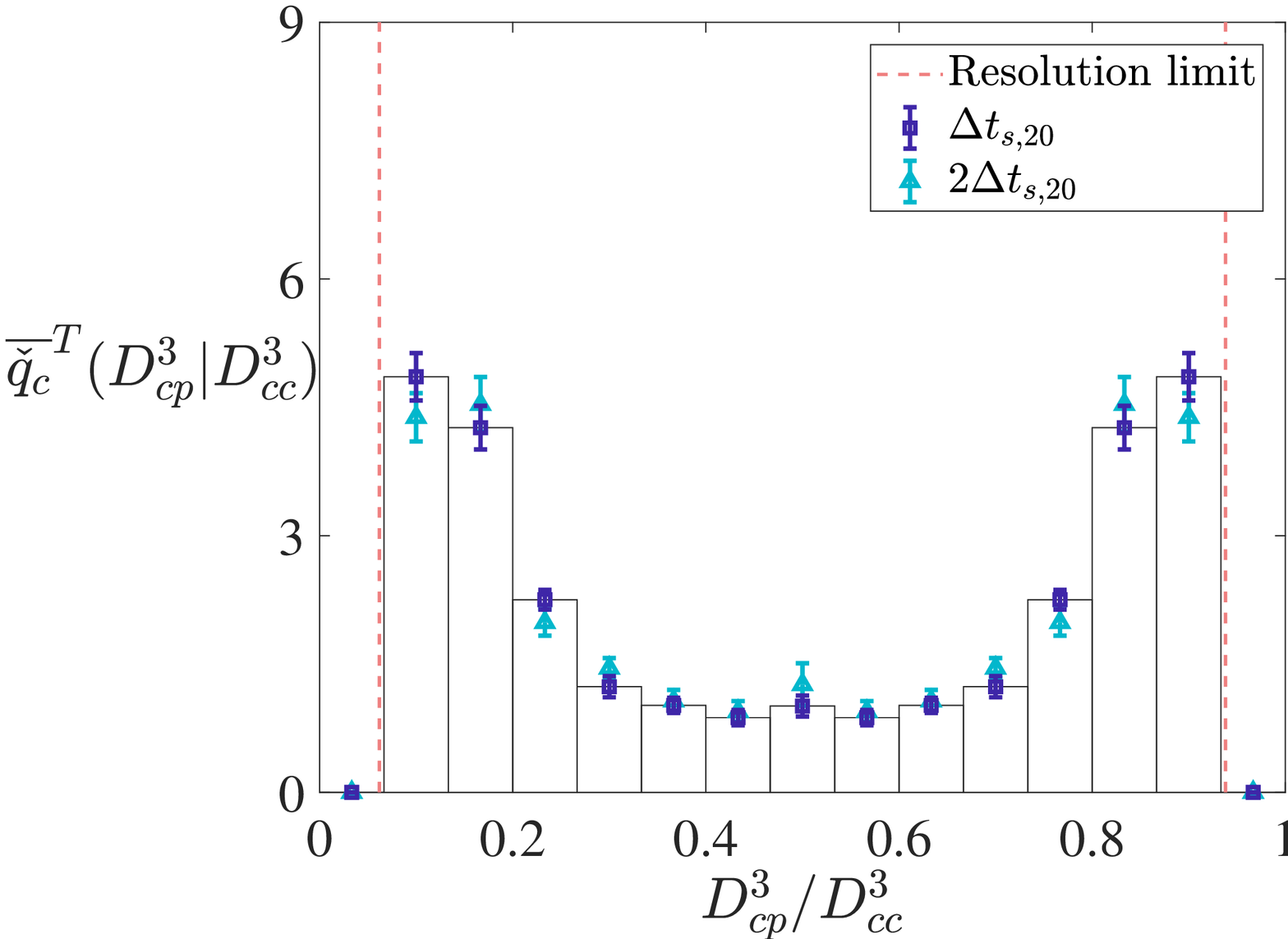}
}
  \centerline{
(c)
\includegraphics[width=0.42\linewidth,valign=t]{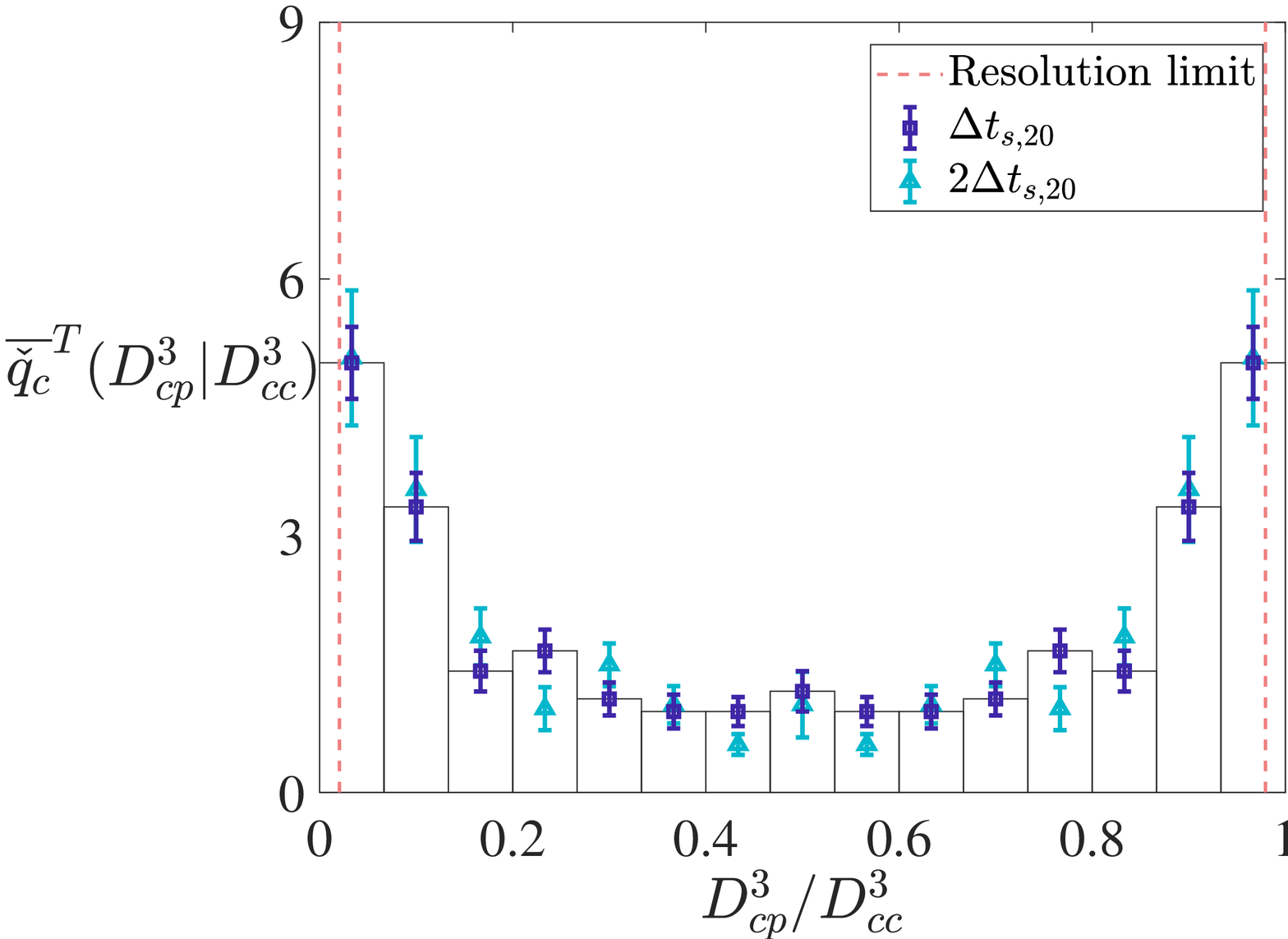}
\quad
(d)
\includegraphics[width=0.42\linewidth,valign=t]{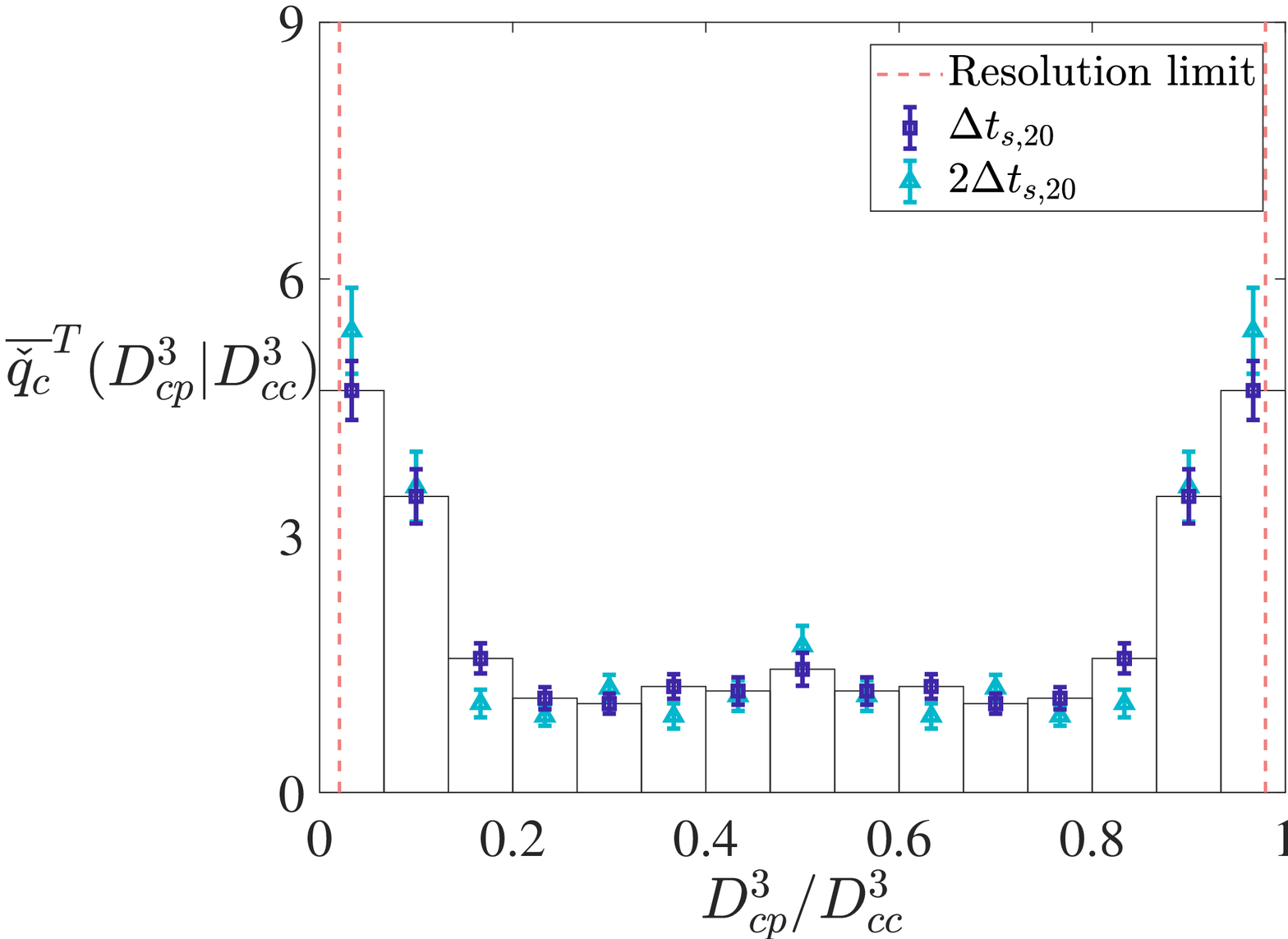}
}
  \centerline{
(e)
\includegraphics[width=0.42\linewidth,valign=t]{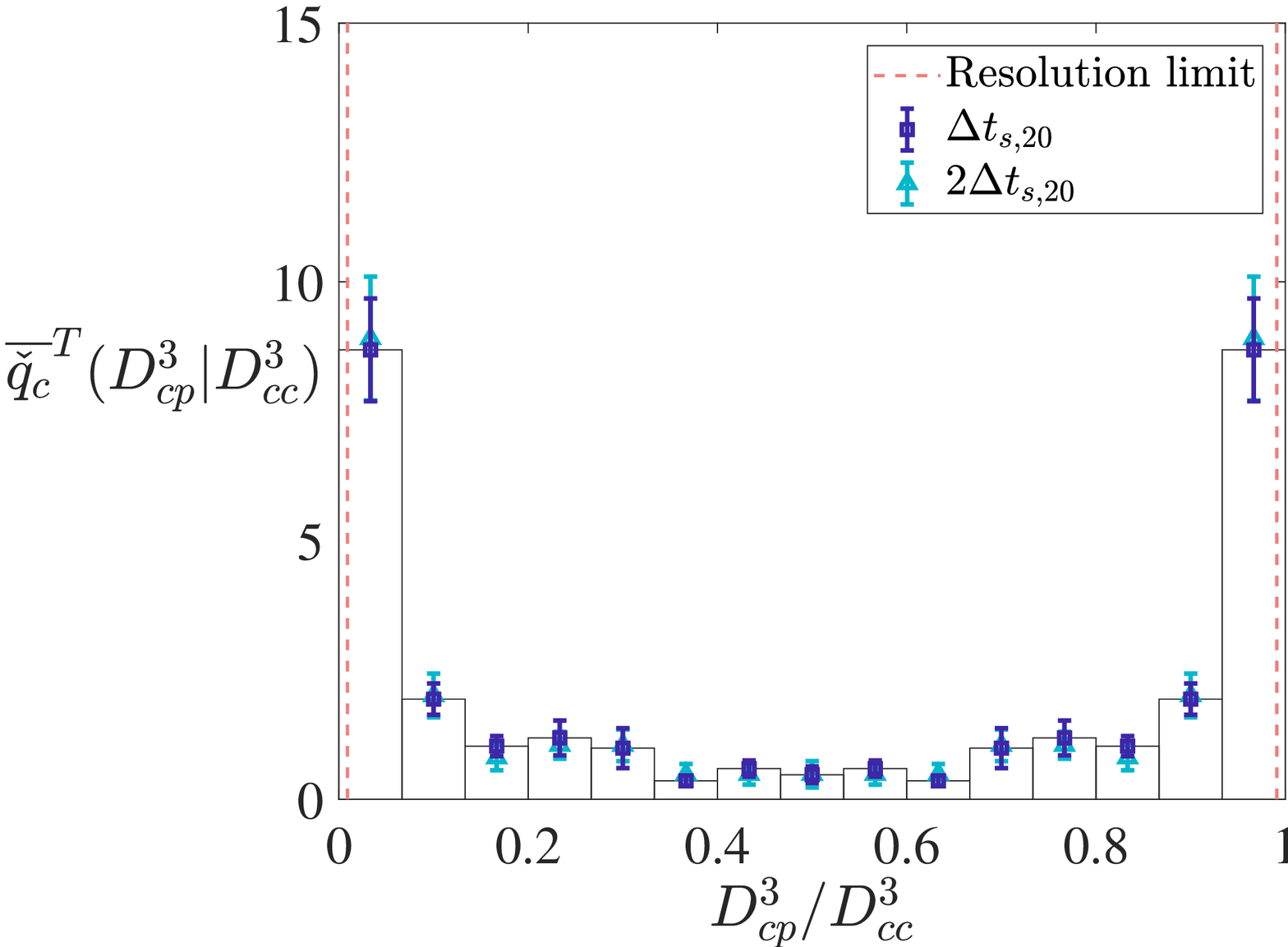}
\quad
(f)
\includegraphics[width=0.42\linewidth,valign=t]{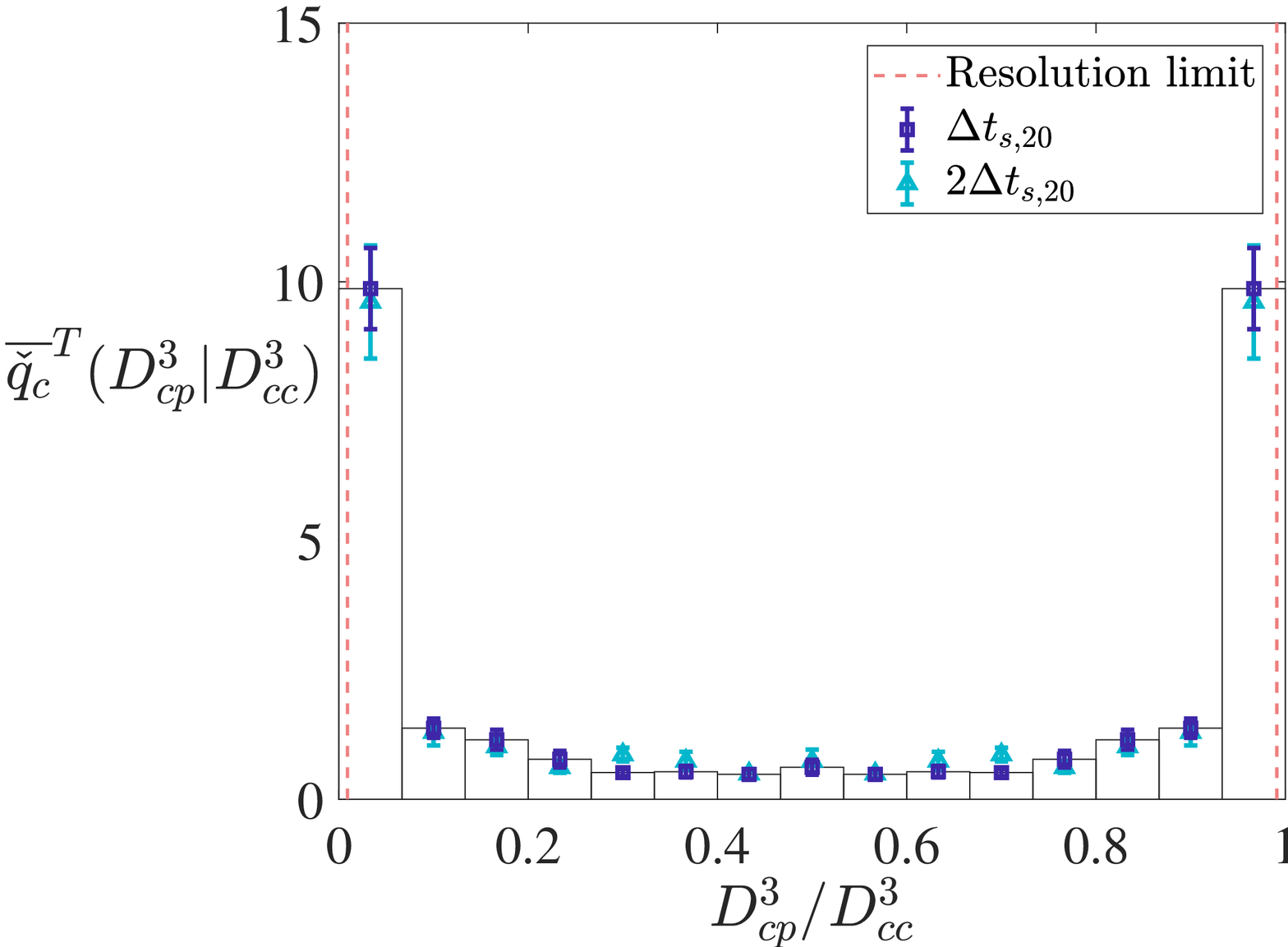}
}
  \caption{The time-averaged coalescence probability $\ov{\check{q}_c}^T\left(D_{cp,j}^3|D_{cc}^3\right)$ against normalized parent (ancestral) bubble volume $D_{cp}^3/D_{cc}^3$ for the 20-realization baseline ensemble subset during the (a,c,e) early ($t=2.30$--$3.14$) and (b,d,f) late ($t=3.45$--$3.95$) wave-breaking stages. The non-dimensional child (descendant) bubble sizes considered are (a,b) between $1.86 \times 10^{-2}$ and $2.23 \times 10^{-2}$, (c,d) between $2.68 \times 10^{-2}$ and $3.23 \times 10^{-2}$, and (e,f) between $3.54 \times 10^{-2}$ and $4.25 \times 10^{-2}$. Only parent bubbles of radii larger than the mesh resolution are considered. The vertical dashed lines demarcate the parent bubble volumes in the coalescence event where one of the parent bubble radii corresponds to this resolution limit. The error bars denote one standard error over the baseline ensemble.}
\label{fig:coalescence_qc}
\end{figure}

Given that the importance of coalescence events is expected to increase with the passage of time and the increase in the number density of small bubbles, coalescence statistics were investigated to determine the role of coalescence in the late wave-breaking stages. Note that coalescence events driven by large-scale dynamics such as buoyancy and turbulence are governed by the corresponding large-scale characteristic time-scales~\citep[see, e.g.,][]{RodriguezRodriguez1}. This means that while the precise moment of coalescence may not be captured accurately since most full-scale numerical simulations cannot feasibly resolve the film drainage that precedes coalescence, macroscopic coalescence statistics remain reasonably reliable unless the system is devoid of these large-scale external influences. There is, however, a possibility that numerical coalescence elevates the total observed coalescence flux. The coalescence flux $\ov{W_c}(D;t)$ across a cut-off size $D$ was computed in an analogous fashion to $\ov{W_b}(D;t)$ using the event detection algorithm in \S~\ref{sec:tracking}. Figure~\ref{fig:coalebreak_w_t} plots the time evolution of $\ov{W_c}^D/\ov{W_b}^D(t)$ where size averaging is performed over the same bubble-size subrange used in figure~\ref{fig:breakup_w_t}. The plot indicates that $\ov{W_c}^D$ is consistently smaller than $\ov{W_b}^D$, indicating that coalescence is possibly not a major player in the dynamics even towards $t=4$. A closer look at the statistics suggests that the complete story may be more nuanced. Figure~\ref{fig:coalescence_qc} plots the time-averaged coalescence probability $\ov{\check{q}_c}^T\left(D_{cp}^3|D_{cc}^3\right)$, which is the probability that a coalescence event generating a child (descendant) bubble of size $D_{cc}$ involves a parent (ancestral) bubble of size $D_{cp}$ and another parent bubble such that the total gaseous volume remains constant, for the two time intervals identified in \S~\ref{sec:waveevol}, \S~\ref{sec:sizedistevol}, and \S~\ref{sec:trac-gbfe}. The distributions reveal that for most children bubbles, large-size-ratio events involving one small parent bubble and one large parent bubble dominate the dynamics. The dominance of large-size-ratio coalescence events may be analytically supported from kinetic theory arguments, such as the one posited by~\citet{Chan5}, which argues that large-size-ratio collisions are favoured when the size distribution decays with increasing bubble size. In short, the dynamics of coalescence appear to be dominated by parent bubbles that are beyond the reach of the current simulations. It is possible, then, that $\ov{W_c}$ has been under-estimated in these simulations, and the jury is still out on the importance of coalescence in the late wave-breaking stages. A study of coalescence is deferred to later work where the capability to computationally capture these small bubbles is included. One possible approach is to replace bubbles under-resolved by the current Eulerian treatment with Lagrangian point particles, as alluded to in Part 1 and to be discussed in forthcoming work.


\section{Spatial distribution of the liquid volume fraction}\label{app:voidfraction}

\begin{figure}
  \centerline{
(a)
\includegraphics[trim={0pt 20pt 20pt 0pt},clip=true,width=0.42\linewidth,valign=t]{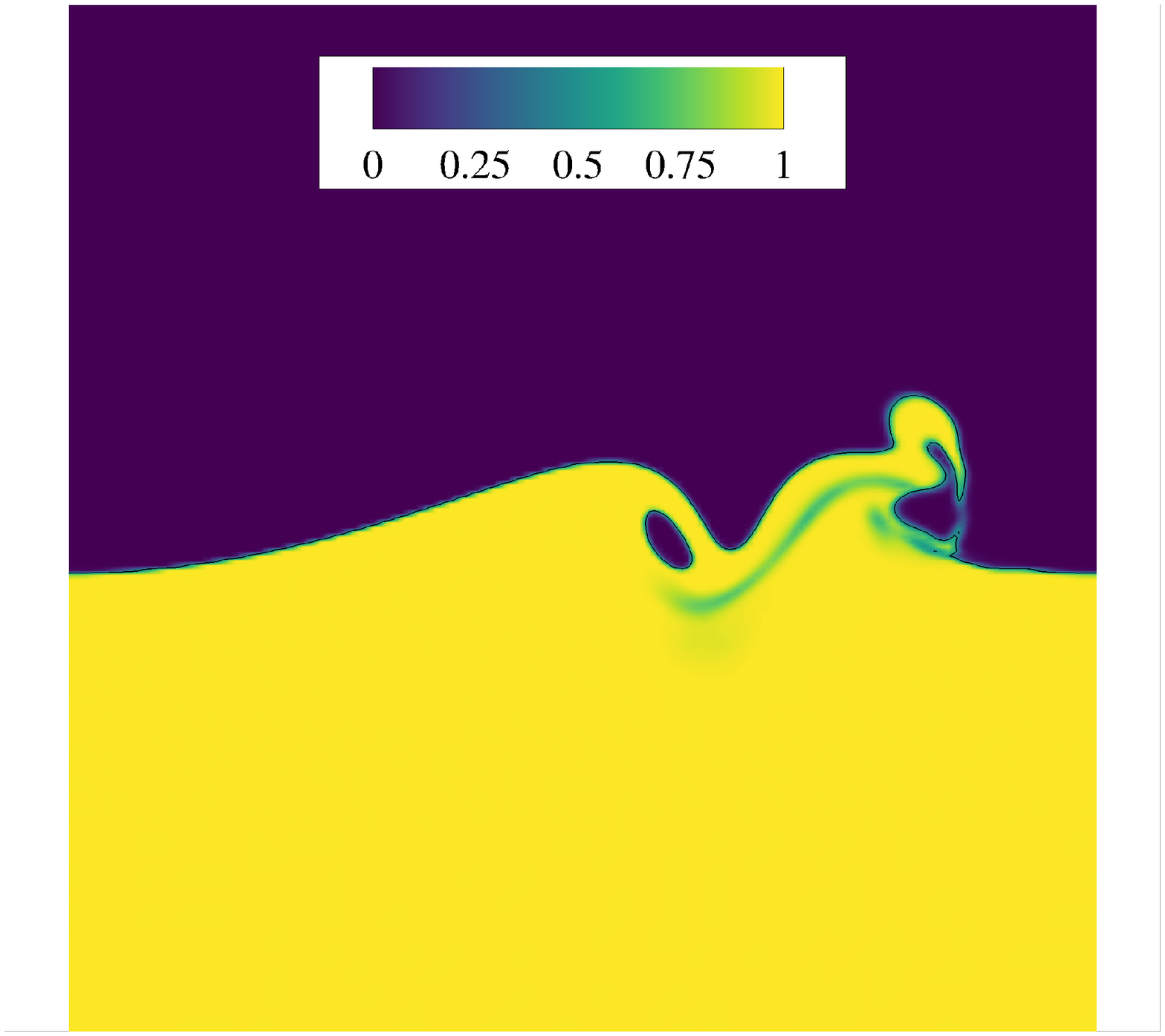}
\quad
(b)
\includegraphics[trim={0pt 20pt 20pt 0pt},clip=true,width=0.42\linewidth,valign=t]{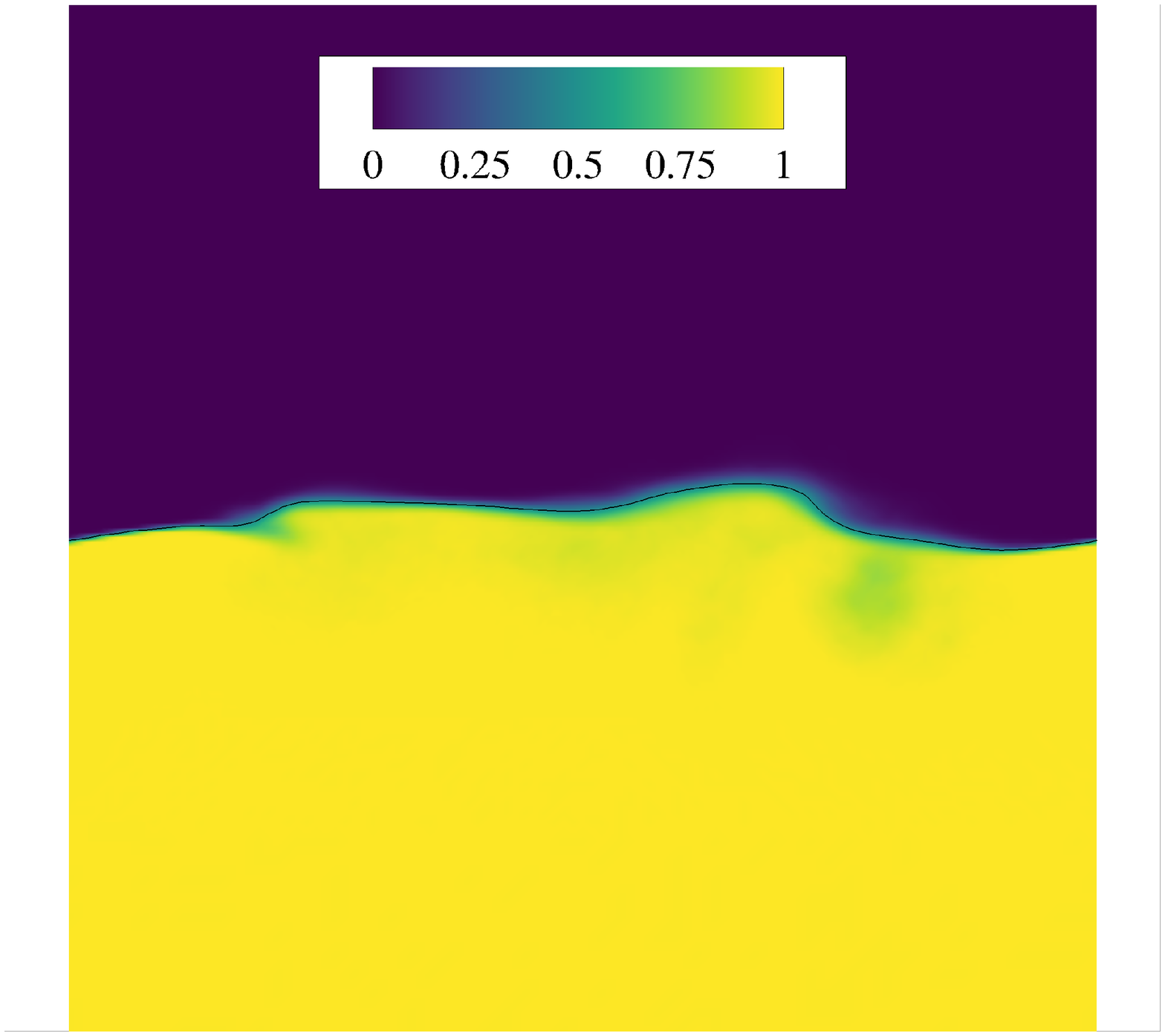}
}
  \caption{The ensemble-averaged and spanwise-averaged liquid volume fraction $\phi$ for the 30-realization baseline ensemble, computed at (a) $t=2.23$ and (b) $t=4.04$. The solid line denotes the $\phi=0.5$ isocontour.}
\label{fig:voidfraction}
\end{figure}

In order to provide a general sense of how the bubbles are spatially distributed during the wave-breaking process, figure~\ref{fig:voidfraction} plots the ensemble-averaged and spanwise-averaged liquid volume fraction, $\phi$, at two representative time instances, one just before the first characteristic time interval and one just after the second characteristic time interval identified in Table~\ref{tab:summary}. Note that the void fraction may be correspondingly computed as $(1-\phi)$. In general agreement with the other works discussed in \S~\ref{sec:waveevol} and \S~\ref{sec:sizedistevol}, high-void-fraction mixed-phase regions are initially concentrated near the areas with rapid topology change as evidenced in figure~\ref{fig:voidfraction}(a), and then gradually evolve into a number of diffuse bubble clouds near the wave surface as evidenced in figure~\ref{fig:voidfraction}(b).


\bibliographystyle{jfm}
\bibliography{wavebreakup_R1}

\end{document}